\documentstyle[preprint,aps,epsf,floats]{revtex}

\newcommand{\nn}{\nonumber}
\newcommand{\ply}{ {\rm Li}_2 }

\begin{document}
\setlength\baselineskip{20pt}

\preprint{\tighten \vbox{\hbox{UCSD/PTH 99-13} \hbox{CALT-68-2243}
\hbox{UTPT-99-16}
}}

\title{NNLO Corrections to Nucleon-Nucleon Scattering \\ and Perturbative Pions
\\[20pt]
}
\author{Sean Fleming$^a$, Thomas Mehen$^b$, and Iain W. Stewart$^{b,c}$\\[10pt]}
\address{\tighten $^a$Department of Physics, University of Toronto, Toronto,
Ontario, M5S 1A7 \\[5pt] $^b$California Institute of Technology, Pasadena, CA
91125\\[5pt] $^c$Department of Physics, University of California at San
Diego,\\[2pt] 9500 Gilman Drive, La Jolla, CA 92099}

\maketitle

{\tighten
\begin{abstract}

The $^1S_0$, $^3S_1$, and $^3D_1$ nucleon-nucleon scattering phase shifts are
calculated at next-to-next-to-leading order (NNLO) in an effective field theory.
Predictions for the $^1P_1$, $^3P_{0,1,2}$, $^1D_2$, and $^3D_{2,3}$ phase shifts at
this order are also compared with data.  The calculations treat pions perturbatively
and include the NNLO contributions from order $Q_r^3$ and $Q_r^4$ radiation pion
graphs. In the $^3S_1$, $^3D_1$, and $^3P_{0,2}$ channels we find large disagreement
with the Nijmegen partial wave analysis at NNLO. These spin triplet channels have
large corrections from graphs with two potential pion exchange which do not vanish
in the chiral limit.  We compare our results to calculations within the Weinberg
approach, and find that in some spin triplet channels the summation of potential
pion diagrams seems to be necessary to reproduce the observed phase shifts.  In the
spin singlet channels the nonperturbative treatment of potential pions does not
afford a significant improvement over the perturbative approach.

\end{abstract}
}
\vspace{0.7in}

\newpage

\section{Introduction}

Understanding how nuclear forces emerge from the fundamental theory of Quantum
Chromodynamics (QCD) remains an outstanding problem in theoretical physics. To
study the physics of hadrons at scales where QCD is strongly coupled, it is
useful to employ the technique of effective field theories.  Model independent
predictions for low energy nuclear phenomena can be made by using an effective
Lagrangian which includes nucleons and pions as explicit degrees of freedom and
all possible interactions that are consistent with the symmetries of the
underlying QCD theory. This method, known as chiral perturbation theory, has
been successfully applied to processes involving 0 and 1 nucleons (see e.g.
\cite{dgh,mj,bkm}).

Weinberg \cite{weinberg} originally proposed using effective field theory for
few body problems in nuclear physics. Weinberg's procedure applies ordinary
chiral perturbation theory power counting to the nucleon-nucleon potential and
then solves the Schr\"{o}dinger equation using this potential. Phenomenological
studies of NN scattering phase shifts and deuteron properties which use this
technique can be found in Refs.~\cite{orefs,ork,Ulf}.

Application of effective field theory to two nucleon systems is complicated by
the existence of a shallow bound state in the spin triplet channel and the
large scattering length in the spin singlet channel. In Refs.~\cite{ksw1,ksw2},
Kaplan, Savage, and Wise (KSW) proposed a new power counting which accounts for
these effects. This approach is more like ordinary chiral perturbation theory
in that power counting is applied to the amplitude rather than the potential.
All observables are expanded in powers of $Q/\Lambda$, where $Q$ is either
$m_\pi$ or $p$ (the nucleon momentum), and $\Lambda$ is the range of the
effective field theory.  Because the S-wave scattering lengths (denoted by $a$)
are large, powers of $p a$ must be summed to all orders \cite{bira}. This
requires a nonperturbative treatment of the leading 4 nucleon operators with no
derivatives. Higher derivative operators and pions are treated perturbatively.
The perturbative treatment of pions makes it possible to obtain analytic
expressions for amplitudes.  One theoretically appealing aspect of the KSW
power counting is that all ultraviolet divergences appearing in loop graphs are
cancelled by contact operators appearing at either the same or a lower order in
the expansion. This is in contrast with Weinberg's approach in which
unsubtracted divergences introduce cutoff dependence which is cancelled at
higher orders in the expansion.  The residual dependence on the cutoff gives an
estimate of the size of higher order corrections.

It is clear from naive dimensional analysis that the KSW expansion will
converge slowly.  To see this, compare the contribution to the amplitude from
single pion exchange and the pion box diagram:
\begin{eqnarray}\label{pilps}
  \begin{picture}(25,20)(1,1)
   \put(-10,-6){\line(1,0){20}} \put(-10,14){\line(1,0){20}}
     \multiput(0,-5)(0,7){3}{\line(0,1){4}}
  \end{picture}\!\!\!\!\!\! = {g_A^2 \over 2 f^2}\ A\! \left( {p \over m_\pi}
  \right), \qquad\qquad
 \begin{picture}(25,20)(1,1)
   \put(-15,-6){\line(1,0){30}} \put(-15,14){\line(1,0){30}}
   \multiput(-8,-6)(0,7){3}{\line(0,1){4}} \multiput(8,-6)(0,7){3}{\line(0,1){4}}
  \end{picture}\!\!\! =
  \left( {g_A^2 \over 2 f^2}\right)^2 {Mm_\pi\over 4 \pi}\ B\! \left( {p \over
   m_\pi} \right )\,,
\end{eqnarray}
where $A,B$ are dimensionless functions. The factor of $M$, the nucleon mass,
comes from performing the energy integral by contour integration and taking a
pole from one of the nucleon propagators.   The factor of $1/(4 \pi)$ is an
estimate of the size of the loop correction. (If a pion pole is taken the
contribution is smaller by $m_\pi/M$ \cite{kaiser}.)  From
Eq.~(\ref{pilps}) one expects an expansion parameter of order $(g_A^2 m_\pi M)/(8
\pi f^2) \equiv m_\pi/\Lambda_{NN} \simeq 0.5$ \cite{ksw2}. This suggests that
perturbative pions will converge, albeit slowly.

Many processes involving two nucleons have been computed to next-to-leading
order (NLO) in the KSW expansion \cite{oNLO,ChenWk}. The results of some of
these calculations are reviewed in \cite{martin}. Typically, one finds
30\%--40\% errors at leading order (LO) and 10\% errors at NLO. These results
suggest an expansion parameter $Q/\Lambda \sim 1/3$ or $\Lambda \approx
400\,{\rm MeV}$. This is consistent with the estimate of the expansion
parameter given above. Obviously, it is important to extend existing
calculations to higher orders to see if the convergence of the expansion
persists.

At the present time, few NNLO calculations\footnote{\tighten For the NN
scattering amplitude LO is $Q^{-1}$, NLO is $Q^0$, and NNLO is order $Q$. In
this paper this terminology will be used even for cases where the LO
contribution vanishes.} are available in the theory with pions. The deuteron
quadrupole moment is calculated to NNLO in Ref.~\cite{binger}. The result of the
NNLO calculation of the $^1S_0$ phase shift has been presented in
Ref.\cite{rupak1,rupak2} and independently in Ref.\cite{msconf}. However, these
$^1S_0$ calculations are incomplete because the full order $Q$ contributions
from radiation pion graphs were not included.  The ${}^3S_1 - {}^3D_1$ mixing
parameter is calculated to NNLO in Ref.\cite{fms}, where it is demonstrated that
the expansion is converging for $p \leq 140\,{\rm MeV}$. For these momenta, the
error is comparable to that of calculations within the Weinberg
approach~\cite{ork}.

In this paper, we present NNLO calculations of the ${}^1S_0$, ${}^3S_1$, and
${}^3D_1$ phase shifts in nucleon-nucleon scattering, including contributions
from radiation pions. At this order we find that the radiation pion diagrams
have trivial momentum dependence and their effect cannot be distinguished from
the contributions of a local operator. In the $^1S_0$ channel, the NNLO fit
agrees with data to $<1$\% accuracy for $p \simeq m_\pi$.  In this channel, the
KSW expansion works as expected. However, in the spin triplet channel we find
that the expansion breaks down at NNLO.  For the $^3S_1$ and $^3D_1$ phase
shifts, the NNLO calculation actually does worse at fitting the data than the
NLO prediction. In the $^3S_1$ channel, the NNLO corrections are as large as the
NLO corrections for $p = m_\pi$. In the $^3D_1$ channel there is no sign
of convergence for any value of $p$. We find that the failure of the EFT
expansion in these two triplet channels is due to large non-analytic corrections
that grow with $p$ coming from graphs with two potential pions.  These terms do
not appear in the spin singlet channel.  The reason for the difference in the
quality of the perturbative expansions in the two channels is that the potential
between nucleons arising from pion exchange is much more singular in the spin
triplet channel than in the singlet channel.  We elaborate on this point in
section \ref{discuss} of the paper.

Next, we examine the NNLO predictions for the $P$ and $D$ wave phase shifts drawing
on results from Ref.~\cite{kaiser}. At LO these phase shifts vanish.  In these
channels the only contributions at NLO and NNLO come from potential pion exchange.
Contact interaction and radiative pion contributions do not enter until higher
order.  Thus predictions for these phase shifts contain no free parameters, and it
is possible to unambiguously test the perturbative treatment of pions. In the spin
singlet channels (${}^1P_1$, ${}^1D_2$) corrections from two potential pion exchange
are small, and the errors at $p = m_\pi$ are (13\%, 33\%). At $p = m_\pi$, the NNLO
predictions for the ${}^3P_1$, ${}^3D_2$ channels have errors of the expected size
(15\%, 8\%). In the ${}^3P_0$, ${}^3P_2$ channels errors are bigger than expected
(170\%, 52\%). Like the $^3S_1$ and $^3D_1$ channels, these spin triplet channels
have non-analytic contributions that grow with $p$.

Our final section includes a comparison of our calculations with those of
Refs.\cite{ork,Ulf} which use the Weinberg approach. In spin singlet channels the
corrections obtained by summing perturbative potential pion exchange to all orders
are negligible. In particular, in the $^1P_1$ and $^1D_2$ channels, single pion
exchange gives the same answer as the LO Weinberg calculation which treats potential
pions nonperturbatively. Here corrections from soft and radiation pion graphs as
well as contact interactions appear to be much more important. In the KSW expansion,
these effects appear at one higher order than the results presented in this paper.

In some spin triplet channels ($^3S_1$, $^3P_1$, $^3D_1$) the summation of potential
pions gives significant improvement relative to the calculation which treats the
pion perturbatively. There are also spin triplet channels where nonperturbative
potential pions seem to be less important than soft pion graphs and four nucleon
operators.  This is true in the $^3P_0$ and $^3P_2$ channels, where the LO
calculation in the Weinberg scheme does no better than the LO term in the KSW expansion.
Finally, in the $^3D_{2,3}$ channels, the KSW expansion at NNLO gives predictions
that are as accurate as the NNLO Weinberg calculations and so a nonperturbative
treatment of pions does not seem to be necessary in these channels.

The rest of the paper will be organized as follows. In section II, the formalism
relevant for our calculation is introduced. We define all operators appearing in
the Lagrangian to the order we are working and discuss the solution to their
renormalization group equations (RGE). Solving the RGE perturbatively ensures
that observables are renormalization scale independent, as in pion chiral
perturbation theory. We also discuss our method for fitting the constants at
each order in the expansion. In section III, expressions are presented for the
$^1S_0$, $^3S_1$ and $^3D_1$ amplitudes up to NNLO. Detailed comparison of the
theoretical phase shifts with the Nijmegen phase shift analysis \cite{Nij}
appears in this section.  In Section IV, we look at NLO and NNLO contributions
to nucleon-nucleon scattering in the ${}^1P_1$, ${}^3P_{0,1,2}$, ${}^1D_2$, and
${}^3D_{2,3}$ waves. In the final section, we discuss our results and their
implications for the perturbative treatment of pions.  Details of the
calculations are contained in the Appendices.  In Appendix~\ref{AppProj} we
describe a trace formalism for projecting partial wave amplitudes from Feynman
diagrams.  In Appendix~\ref{AppQ} we give explicit expressions for all
individual graphs at NNLO, except for graphs involving radiation pions. We also
describe a general strategy for analytically evaluating massive non-relativistic
multi-loop Feynman diagrams.  In Appendix~\ref{AppRad}, the S-wave radiation
pion contribution is discussed in detail. The power counting for radiation and
soft pions is reviewed and the complete order $Q$ contribution is evaluated.

\section{Formalism}

In this paper, we will follow the notation in Refs.\cite{ksw2,fms,ms1}. The
relevant Lagrangian for NN scattering at NNLO is
\begin{eqnarray}  \label{Lpi}
 {\cal L} &=& \frac{f^2}{8} {\rm Tr}\,( \partial^\mu\Sigma\: \partial_\mu
 \Sigma^\dagger )+\frac{f^2\omega}{4}\, {\rm Tr} (m_q \Sigma+m_q \Sigma^\dagger)
 + N^\dagger \bigg( i D_0+\frac{\vec D^2}{2M} \bigg) N  \nn \\
 &+& \frac{ig_A}2\, N^\dagger \sigma_i (\xi\partial_i\xi^\dagger -
  \xi^\dagger\partial_i\xi) N -{C_0^{(s)}}
  {\cal O}_0^{(s)} +\frac{C_2^{(s)}}{8}{\cal O}_2^{(s)}
  -{D_2^{(s)}} \omega {\rm Tr}(m^\xi ) {\cal O}_0^{(s)}  \nn\\[5pt]
 &-& \frac{{C_4^{(s)}}}{64} {\cal O}_4^{(s)}
   + \frac{E_4^{(s)}}{8} \omega {\rm Tr}(m^\xi ) {\cal O}_2^{(s)}
   - \frac{D_4^{(s)}}{2} \omega^2 \Big\{ {\rm Tr}^2(m^\xi )+
   2{\rm Tr}[(m^\xi)^2] \Big\} {\cal O}_0^{(s)}\nn \\[3pt]
  &-& C_2^{(SD)}\, {\cal O}_2^{(SD)} + \ldots \,.
\end{eqnarray}
Here $g_A=1.25$ is the nucleon axial-vector coupling, $\Sigma = \xi^2 = \exp(2
\,i\, \Pi/f)$ where
\begin{eqnarray}
 \Pi = \left( \begin{array}{cc} \pi^0/\sqrt{2} & \pi^+ \\
    \pi^- & -\pi^0/\sqrt{2} \end{array} \right) \,,
\end{eqnarray}
$f=131\, {\rm MeV}$ is the pion decay constant, the chiral covariant derivative
is $D_\mu=\partial_\mu+\frac12 (\xi\partial_\mu\xi^\dagger +
\xi^\dagger\partial_\mu\xi)$, and $m^\xi=\frac12(\xi m_q \xi + \xi^\dagger m_q
\xi^\dagger)$, where $m_q={\rm diag}(m_u,m_d)$ is the quark mass matrix.  At
the order we are working $\omega {\rm Tr}(m^\xi )=\omega (m_u+m_d) = m_\pi^2 =
(137\, {\rm MeV})^2$.  In Eq.~(\ref{Lpi}), $s={}^1S_0$ or $^3S_1$. Below this
superscript will be dropped when it is clear from the context which channel is
being referred to or when the reference is to both channels.  The two-body
nucleon operators are:
\begin{eqnarray} \label{ops}
 {\cal O}_0^{(s)} &=& ( N^T P^{(s)}_i N)^\dagger ( N^T P^{(s)}_i N)
     \nn \,, \\
 {\cal O}_2^{(s)} &=& ( N^T P^{(s)}_i N)^\dagger ( N^T P^{(s)}_i \:
     \tensor{\nabla}^{\,2} N) + h.c. \nn\,, \\
 {\cal O}_4^{(s)} &=& ( N^T P^{(s)}_i N)^\dagger ( N^T P_i^{(s)}\:
  \tensor{\nabla}^{\,4} N) + h.c. +  2( N^T P_i^{(s)} \tensor{\nabla}^{\,2}
  N)^\dagger ( N^T P_i^{(s)} \:\tensor{\nabla}^{\,2} N) \nn \,,\\
 {\cal O}_2^{(SD)} &=& ( N^T P^{(^3S_1)}_i N)^\dagger ( N^T
     P^{(^3D_1)}_i N) + h.c.  \,,
\end{eqnarray}
where the projection matrices are
\begin{eqnarray} \label{proj}
 && P_i^{({}^1\!S_0)} = { (i\sigma_2) \, (i\tau_2 \tau_i) \over 2\sqrt{2} }\,,
   \qquad
  P_i^{({}^3\!S_1)} = { (i\sigma_2 \sigma_i  ) \, (i\tau_2) \over 2\sqrt{2}}\,,
    \nn \\
&& P_i^{({}^3\!D_1)} =  \frac{n}{4\sqrt{n-1}} \: \Big( \tensor{\nabla}_i
  \tensor{\nabla}_j -  \frac{\delta_{ij}}{n}\:\tensor{\nabla}^{\,2}  \Big) \:
  P_j^{({}^3\!S_1)} \,,
\end{eqnarray}
and $\tensor{\nabla} = \overleftarrow{\nabla}- \overrightarrow{\nabla}$. The
derivatives in Eqs.~(\ref{ops}) and (\ref{proj}) should really be chirally
covariant, however, only the ordinary derivatives are needed for the
calculations in this paper.

Ultraviolet divergences are regulated using dimensional regularization. All
spin and isospin traces are done in $n$ dimensions, where $d=n+1$ is the
space-time dimension. Regulating the theory in this way preserves the chiral
and rotational symmetry of the theory as well as the Wigner symmetry
\cite{Wigner,msw} of the leading order Lagrangian, as discussed in
Ref.\cite{fms}.

The KSW power counting is manifest in renormalization schemes such as power
divergence subtraction (PDS) \cite{ksw1,ksw2} or off-shell momentum subtraction
(OS) \cite{gegelia,ms0,ms1}. (In this paper the PDS scheme will be used.) In
these schemes the coefficients of the S-wave operators in Eq.~(\ref{ops}) scale
as $C_{2n}^{(s)} \sim 1/(M\Lambda^n \mu^{n+1})$, where $\mu$ is the
renormalization scale, and $\Lambda$ is the range of the effective field
theory. The renormalization scale is chosen to be on the order of the nucleon
momentum $p$ which is of order $m_{\pi}$. Letting $\mu \sim p \sim m_\pi \sim
Q$ the scaling of the coefficients in Eq.~(\ref{Lpi}) is:
\begin{eqnarray} \label{Qscale}
  {\rm LO} &:& \qquad C_0^{(s)}(\mu)\sim 1/Q \\
  {\rm NLO} &:&\qquad p^2\,C_2^{(s)}(\mu) \sim Q^0, \,\,\, m_\pi^2 \,
    D_2^{(s)}(\mu) \sim Q^0 \nn \\
  {\rm NNLO} &:& \qquad p^4\, C_4^{(s)}(\mu)\sim Q, \,\,\, m_\pi^2 p^2 \,
    E_4^{(s)}(\mu)\sim Q, \,\,\,m_\pi^4 \, D_4^{(s)}(\mu)\sim Q \nn \,,
    \,\,\, p^2 C_2^{(SD)}(\mu) \sim Q \,.
\end{eqnarray}
Note that from simple dimensional analysis one would expect these coefficients
to scale as $C_{2n} \sim 1/(M\Lambda^{2n+1})$. However, these coefficients are
larger than naive dimensional analysis predicts because the theory flows to a
non-trivial fixed point for $a\to \pm\infty$. (See Refs.~\cite{ksw2,birse,pths}
for a more detailed explanation.) Since $C_0^{(s)}(\mu)\sim 1/Q$, and each
nucleon loop gives a factor of $Q$, power counting demands that graphs with
$C_0$'s be summed to all orders.  This sums all powers of $ap$. Operators with
derivatives or insertions of the quark mass matrix scale as $Q^n$, $n \geq 0$,
and are treated perturbatively.

In Eq.~(\ref{Lpi}) we have not included four nucleon operators for partial
waves with $L\ge 1$ because these operators enter at order $Q^2$ or higher. For
example, the coefficients of the four P-wave operators with two derivatives are
not enhanced by the renormalization group flow near the fixed point and
therefore scale as $1/(M \Lambda^3)$. Thus, these P-wave terms in the
Lagrangian are order $Q^2/(M \Lambda^3)$. As a result the order $Q$ predictions
for partial waves with $L\ge 1$ come completely from pion exchange and have no
free parameters.

There is another term in the Lagrangian in Eq.~(\ref{Lpi}) with an S-wave
four-derivative operator distinct from ${\cal O}_4^{(s)}$, ${\cal L} =
\tilde C_4^{(s)}\: \tilde {\cal O}_4^{(s)}$ where
\begin{eqnarray} \label{C4t}
\tilde {\cal O}_4^{(s)} &=& \Big[  ( N^T P^{(s)}_i N)^\dagger ( N^T P_i^{(s)}\:
  \tensor{\nabla}^{\,4} N) + h.c. -  2( N^T P_i^{(s)} \tensor{\nabla}^{\,2}
  N)^\dagger ( N^T P_i^{(s)} \:\tensor{\nabla}^{\,2} N) \Big] \,.
\end{eqnarray}
For the process $N(p_1) N(p_2) \to N(p_3) N(p_4)$ this operator vanishes on-shell
since energy-momentum conservation gives $(\vec p_1- \vec p_2)^2 = (\vec p_3-
\vec p_4)^2$. In deriving the RGE's only on-shell amplitudes are relevant.  In
fact the off-shell Green's functions do not have to be $\mu$ independent, as
illustrated by the off-shell $C_2$ amplitude given in Eq.~(\ref{offC2}) of
Appendix~\ref{AppRad}.   Thus, to derive an RGE for $\tilde C_4^{(s)}(\mu)$ it is
necessary to consider an on-shell process in which this coefficient gives a
non-zero contribution.  Although $\tilde C_4^{(s)}(\mu)$ does not contribute to
NN scattering, it may contribute to interactions with photons when the operator
in Eq.~(\ref{C4t}) is gauged. Diagrams with two ${\cal O}_2^{(s)}$ operators
renormalize ${\cal O}_4^{(s)}$ making $C_4^{(s)}(\mu)\sim 1/\mu^3$. The fact that
${\cal O}_4^{(s)}$ rather than ${\cal O}_4^{(s)} - \tilde {\cal O}_4^{(s)}$
has an enhanced coefficient differs from the conclusion in Ref.~\cite{nopi}.

Relativistic corrections contribute at order $Q$ to the S-wave amplitudes. They
are suppressed relative to the leading order amplitude by $(Q/M)^2$ rather than
$(Q/\Lambda)^2$. In Ref.\cite{nopi} these corrections are computed and found to
be negligible relative to other order $Q$ contributions. Therefore they are left
out of our analysis.

For momenta $p\gtrsim m_\pi$ pions should be included in the theory. There are
three types of contributions from pions: radiation, potential, and soft. In
evaluating non-relativistic loop diagrams the energy integrals are performed
using contour integration.  When the residue of a nucleon pole is taken the pion
propagators in the loop are potential pions.  When the residue of a pion pole is
taken the pion will be either radiation or soft. Potential pion exchange scales
as $Q^0$, and is therefore perturbative.  Radiation and soft pions begin to
contribute at order $Q$ and $Q^2$ respectively. The power counting for pions is
discussed in detail in Appendix~\ref{AppRad}. Because pion exchange is treated
perturbatively the dominant scaling of the $C_{2n}(\mu)$ coefficients is the
same as in the theory without pions.

The $Q$ scaling in Eq.~(\ref{Qscale}) can be determined by computing the beta
functions for the four nucleon couplings appearing in Eq.~(\ref{Lpi}) to the
order we are working.  The procedure used for computing beta functions in the
PDS scheme is described briefly in Appendix B and in detail in Ref.~\cite{ms1}.
Our results are slightly different than Ref.\cite{ms1} because all spin and
isospin traces are performed in $n$ dimensions rather than 3 dimensions.  For
the  $^1S_0$ channel, the beta functions to NNLO are:
\begin{eqnarray} \label{C0RGE}
 \mu {\partial \over \partial \mu} C_0 &=& {M \mu \over 4 \pi} (C_0)^2
    \left[ 1 + 2 {g_A^2 \over 2 f^2}{M \mu \over 4 \pi}+ 3 \left({g_A^2 \over
    2 f^2}{M \mu \over 4 \pi} \right)^2 \right] \,, \\
 \mu {\partial \over \partial \mu} C_2 &=& 2{M \mu \over 4 \pi} C_0 C_2 \left(
    1+2 {g_A^2 \over 2 f^2}{M \mu \over 4 \pi} \right) \,, \nn \\
\mu {\partial \over \partial \mu} D_2 &=& 2{M \mu \over 4 \pi} C_0 D_2 \left(
   1+2 {g_A^2 \over 2 f^2}{M \mu \over 4 \pi}\right)
   + {g_A^2 \over 2 f^2}\left( {M \over 4 \pi} \right)^2 (C_0)^2 \nn \\
   && +2 \left( {M g_A^2 \over 8 \pi f^2} \right)^2 C_0  \left( 1+ { M
    \mu \over 4 \pi} C_0 \right) + \beta_{D_2}^{{\rm rad}} \,, \nn \\
\mu {\partial \over \partial \mu} C_4 &=& {M \mu \over 4 \pi} \Big[ 2\, C_0 C_4
   + (C_2)^2 \Big] \,, \nn \\
\mu {\partial \over \partial \mu} E_4 &=& {M \mu \over 4 \pi} \Big[ 2\, C_0 E_4
   + 2 D_2\, C_2 \Big] + 2\, C_2\, C_0\, {g_A^2 \over 2 f^2}\left( {M \over
   4 \pi} \right)^2 \,, \nn \\
\mu {\partial \over \partial \mu} D_4 &=& {M \mu \over 4 \pi} \Big[ 2\, C_0 D_4
   + (D_2)^2 \Big] + 2\, D_2\, C_0\, {g_A^2 \over 2 f^2}\left( {M \over 4 \pi}
   \right)^2  \nn \,,
\end{eqnarray}
where the contribution $\beta_{D_2}^{{\rm rad}}$ from radiation pions is given
in Eq.~(\ref{bd2rad}). All coupling constants are functions of $\mu$. In the
$^3S_1$ channel, the beta functions for $C_0$, $C_2$, and $D_2$ are:
\begin{eqnarray} \label{C0RGE2}
\mu {\partial \over \partial \mu} C_0 &=& {M \mu \over 4 \pi} (C_0)^2 \left(
  1 + 2 {g_A^2 \over 2 f^2}{M \mu \over 4 \pi}\right) + 4 {M \mu \over 4 \pi}
  \left( {g_A^2 \over 2 f^2}\right)^2 \,, \\
\mu {\partial \over \partial \mu} C_2 &=& 2{M \mu \over 4 \pi} C_0 C_2 \left(
  1+2 {g_A^2 \over 2 f^2}{M \mu \over 4 \pi} \right) +12 \left({M g_A^2 \over 8
  \pi f^2}\right)^2 C_0 \left(1+{M \mu \over 4 \pi} C_0 \right) \nn \,, \\
\mu {\partial \over \partial \mu} D_2 &=& 2{M \mu \over 4 \pi} C_0 D_2 \left(
   1+2 {g_A^2 \over 2 f^2}{M \mu \over 4 \pi}\right)
   + {g_A^2 \over 2 f^2}\left( {M \over 4 \pi} \right)^2 (C_0)^2 \nn \\
   && +7 \left( {M g_A^2 \over 8 \pi f^2} \right)^2 C_0  \left( 1+ { M
    \mu \over 4 \pi} C_0 \right) + \beta_{D_2}^{{\rm rad}} \,, \nn
\end{eqnarray}
and the beta functions for $C_4$, $E_4$ and $D_4$ are identical to those in the
$^1S_0$ channel. (We have corrected a sign error in the $C_2^{(^3S_1)}$ beta
function computed in Ref.~\cite{ms2}.)  The running of $C_2^{SD}(\mu)$ is
discussed in Ref.~\cite{fms}.  Terms in the beta functions that vanish as $\mu
\to 0$ are from linear power divergences and are renormalization scheme
dependent. These terms are necessary for a consistent power counting near the $a
\to \pm\infty$ fixed points. Taking $g_A \to 0$ gives the dominant power
contributions, and these terms are the same in renormalization schemes with a
manifest power counting like PDS or OS.  Finally, terms that do not vanish as
$\mu \to 0$ correspond to logarithmic divergences and are scheme independent.

It is desirable that the amplitude, and hence all physical quantities, like the
scattering length, be $\mu$ independent at each order in the expansion. This can
be accomplished by expanding the coupling constants in $Q$ \cite{ms0}:
\begin{eqnarray}\label{pc}
C_0 &\rightarrow& C_0 + C_{0}^{(0)} + C_{0}^{(1)} \nn\\ C_2 &\rightarrow& C_2 +
C_{2}^{(-1)} \nn \\ D_2 &\rightarrow& D_2 + D_{2}^{(-1)} \,.
\end{eqnarray}
The first piece of $C_0$ is treated nonperturbatively (i.e. $C_0 \sim Q^{-1}$),
while $C_{0}^{(0)} \sim Q^0, C_{0}^{(1)} \sim Q$. Because of the perturbative
expansion of the couplings in Eq.~(\ref{pc}) there are ten constants of
integration that appear in the calculation of the NNLO S-wave phase shifts.
However, the NNLO amplitude depends only on six independent linear combinations
of these constants. The coupling constants are also subject to two further
constraints:
\begin{enumerate}
\item At this order, $C_4$, $E_4$ and $D_4$ are determined entirely in terms of
   lower order couplings as a consequence of solving the RGE's and applying the
   KSW power counting.
\item Spurious double and triple poles in the NLO and NNLO amplitudes must
   be cancelled in order to obtain a good fit at low momentum.
\end{enumerate}

An example of constraint 1 is provided by the solution of the RGE for $C_4$
given in Eq.~(\ref{C0RGE})~\cite{ksw2}:
\begin{eqnarray} \label{fixC4}
   C_4 = {(C_2)^2 \over C_0} + \rho {M \over 4 \pi} (C_0)^2 \,,
\end{eqnarray}
where $\rho$ is a constant of integration. In the theory without pions, $\rho$ is
proportional to the shape parameter, which is $\sim Q^0$ in the KSW power counting.
In the theory with pions $\rho \sim Q^0$ too, since its size is determined by the
scale $\Lambda$.  Therefore, $(C_2)^2/C_0 \sim Q^{-3}$, while $\rho (C_0)^2 M/(4
\pi) \sim Q^{-2}$. The second term is subleading in the $Q$ expansion, and should be
omitted at NNLO, so $C_4=(C_2)^2/C_0$. Solving Eq.~(\ref{C0RGE}) gives similar
relations for $E_4$, and $D_4$:
\begin{eqnarray} \label{fixED4}
  E_4 = {2 C_2 D_2 \over C_0} + {\cal O}(Q^{-2})\,, \qquad\quad
  D_4 = {(D_2)^2 \over C_0} + {\cal O}(Q^{-2}) \,,
\end{eqnarray}
assuming that the constants of integration are order $Q^0$. The beta functions
for $E_4$ and $D_4$ have contributions from chiral logarithms, which are
determined by the $\ln(\mu)$ in $D_2$.

\begin{figure}[!t]
  \centerline{\epsfxsize=15 cm \epsfbox{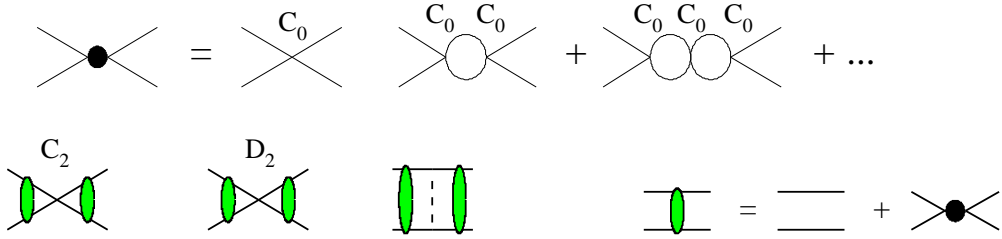}  }
{\tighten \caption[1]{Order $1/Q$ (first row) and $Q^0$ (second row) diagrams
for NN scattering.  The solid lines are nucleons and the dashed line is a
potential pion.} \label{figQ0} }
\end{figure}
Constraint 2 is due to the nonperturbative treatment of $C_0$, which
gives rise to spurious poles at higher orders in the expansion. The
leading order amplitude ${\cal A}_{-1}$ has a simple pole at $p= i
\gamma$.  The NLO amplitude is proportional to ${\cal A}_{-1}^2$,
and therefore has a double pole, while the NNLO amplitude has terms
proportional to ${\cal A}_{-1}^2$ and ${\cal A}_{-1}^3$. To obtain a
good fit at low momentum, parameters need to be fixed so that the
amplitude has only a simple pole at each order in the expansion. This
requires that ${\cal A}_{-1}$ have its pole in the correct location
and that the residues of the spurious double and triple poles
vanish. This requirement leads to the following good fit conditions
\cite{ms0}:
\begin{eqnarray}\label{goodfit}
  \left. {1\over {\cal A}_{-1}}\right|_{p= p^*} \!\!\!\! = 0 \,, \qquad
  \left. {{\cal A}_0\over ({\cal A}_{-1})^2}\right|_{p=p^*}\!\!\!\! = 0\,,\qquad
  \left. {{\cal A}_1\over {(\cal A}_{-1})^2}\right|_{p=p^*}\!\!\!\!= 0 \,,\qquad
\end{eqnarray}
where $p^*$ is the location of the pole.  The second condition first appears at
NLO, the third at NNLO.  The residue of the triple pole in ${\cal A}_1$
vanishes by the second equation in Eq.~(\ref{goodfit}). The first equation
results in $\gamma = -i p^*$, while the other equations give constraints which
eliminate two of the remaining parameters. In order to solve the constraints in
Eq.~(\ref{goodfit}) we must allow the coupling constants $C_0^{(0)}$ and
$C_0^{(1)}$ to have non-analytic dependence on $m_\pi$. Ideally, all $m_\pi$
dependence should be explicit in the Lagrangian and the coupling $C_0$ should
only depend on short distance scales.  However, the fine tuning that results in
the large scattering lengths is a consequence of a delicate cancellation
between long and short distance contributions, and in order to put the pole in
the physical location, one must induce explicit $m_\pi$ dependence in the
perturbative parts of $C_0$ \cite{rupak1,sk}. Eq.~(\ref{goodfit}) will be applied
to both S-wave channels. After imposing these conditions, there is one free
parameter at NLO and two free parameters at NNLO.

\section{Amplitudes and Phase Shifts}


\subsection{$^1S_0$ channel}

\begin{figure}[!t]
  \centerline{\epsfysize=8.0 cm \epsfbox{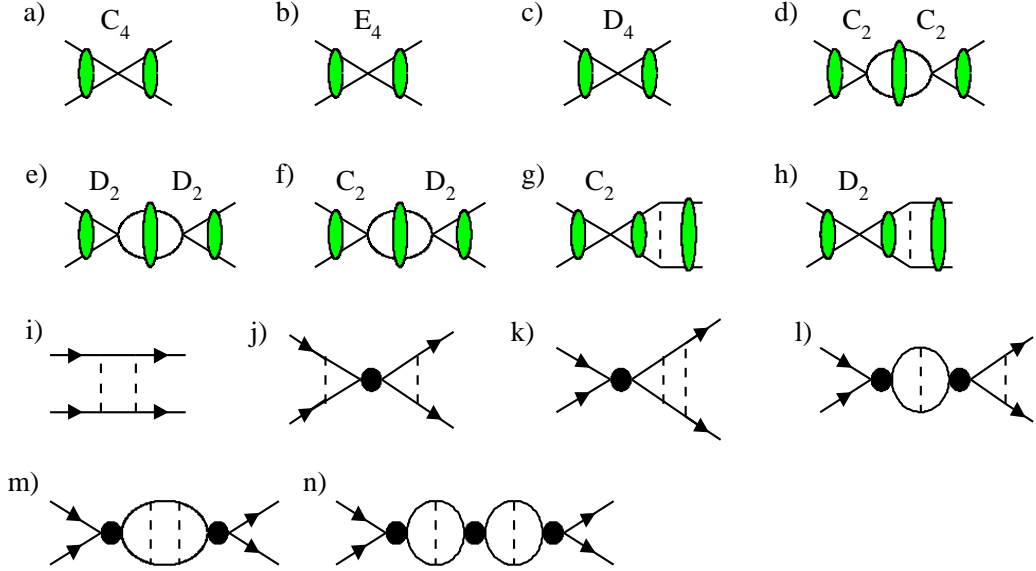}  }
{\tighten \caption[1]{Order $Q$ contact interaction and potential pion graphs
for the $^1S_0$ and ${}^3S_1$ channels. At this order the first three graphs do
not introduce new parameters as explained in the text. Radiation pion diagrams
with order $Q$ contributions are shown in Appendix~\ref{AppRad}.} \label{figQ}}
\end{figure}
In this section, we present the NNLO calculation of the $^1S_0$ phase shift. At
NLO the amplitude involves the diagrams in Fig.~\ref{figQ0} calculated in
Ref.~\cite{ksw2}.  Graphs contributing to the NNLO amplitude include those with
one insertion of an order $Q$ operator and those with two insertions of either a
potential pion or order $Q^0$ operator. These graphs are shown in Fig.\ref{figQ}.
A discussion of the techniques used to evaluate these graphs and explicit
expressions for each individual graph are given in Appendix~\ref{AppQ}. The NNLO
amplitude also receives contributions from graphs with radiation pions which are
discussed in Appendix~\ref{AppRad}.

By expanding $\exp(2 i \delta)=1+ ipM{\cal A}/(2\pi)$ in powers of $Q$ we
obtain expressions for the $^1S_0$ phase shift $\delta = \delta^{(0)} +
\delta^{(1)} + \delta^{(2)}$ (where $\delta^{(n)}\sim Q^n$) in terms of the
amplitudes ${\cal A}= {\cal A}_{-1}+{\cal A}_0+ {\cal A}_1$ (where $A_n\sim
Q^n$),
\begin{eqnarray}
 {\delta}^{(0)} &=&  {1 \over 2 i}\, {\rm ln}\left(1+{i p M \over 2
  \pi} {\cal A}_{-1} \right)  \,, \qquad {\delta}^{(1)} = { p M
  \over 4 \pi} {{\cal A}_{0} \over 1+ {i p M \over 2 \pi} {\cal
  A}_{-1}}   \,,\nn \\[5pt]
  {\delta}_0^{(2)} &=& { p M \over 4 \pi} {{\cal A}_{1} \over 1+ {i
  p M \over 2 \pi}{\cal A}_{-1} } - i\left( { p M \over 4 \pi}
  \right)^2 \left( {{\cal A}_{0} \over 1+ {i p M \over 2
  \pi}{\cal A}_{-1} } \right)^2   \,.
\end{eqnarray}
Our final result for the amplitude at NNLO is quite simple:
\begin{eqnarray} \label{SinAmp}
{\cal A}_{-1} &=& -{4\pi\over M}{1\over \gamma + ip}\,,  \nn \\[5pt] { {\cal A}_0  }
&=& -  {\cal A}_{-1}^2 (\zeta_1 \, p^2 + \zeta_2 \, m_\pi^2)  \\
   &&+ {g_A^2 \over 2 f^2} {\cal A}_{-1}^2 \Big({M m_\pi \over 4 \pi}\Big)^2 \bigg[
  { (\gamma^2 - p^2) \over 4 p^2}{\rm ln}\Big(1 + {4 p^2 \over m_\pi^2}\Big) - \,
  {\gamma \over p} {\rm tan}^{-1}\left({2  p \over m_\pi} \right) \bigg] \nn\,,\\[5pt]
{ {\cal A}_{1} } &=& {{\cal A}_0^2 \over {\cal A}_{-1}} -  {\cal A}_{-1}^2
  \Big(\zeta_3 \, m_\pi^2 +\zeta_4 \,p^2 + \zeta_5 {p^4 \over m_\pi^2} \Big)
  + { {\cal A}_0 }{M g_A^2 \over 8 \pi f^2}\,  {m_\pi^2\over p}
  \bigg[ { \gamma \over 2 p}\: {\rm ln}\Big(1 + {4 p^2 \over m_\pi^2}\Big)  \nn \\
&& - {\rm tan}^{-1}\Big({2 p \over m_\pi}\Big) \bigg]
  +{M {\cal A}_{-1}^2 \over 4 \pi}  \Big({ M g_A^2 \over 8\pi f^2}\Big)^2
   {m_\pi^4 \over 4 p^3}  \Bigg\{ 2(\gamma^2-p^2)\, {\rm Im}\, {\rm Li_2} \Big(
   {-m_\pi \over m_\pi-2 i p} \Big)  \nn \\
&&  - 4 \gamma\, p\: {\rm Re}\, {\rm Li_2}\Big({-m_\pi \over m_\pi-2 i p}
   \Big) - { \gamma\, p\, \pi^2 \over 3} -(\gamma^2+p^2) \bigg[ {\rm Im} \,
  {\rm Li_2} \Big({m_\pi+2 i p \over -m_\pi+2 i p} \Big) \nn \\
&& + {\gamma \over 4 p} \ln^2\Big(1+\frac{4p^2}{m_\pi^2}\Big) - \tan^{-1}
   \Big(\frac{2p}{m_\pi} \Big) \ln\Big(1+\frac{4p^2}{m_\pi^2}\Big) \bigg] \Bigg\} \nn
   \,.
\end{eqnarray}
Using Eq.~(\ref{SinAmp}) it is easy to verify that the S-matrix is unitary to the
order we are working. The six linearly independent constants appearing in the
amplitude are $\gamma,\zeta_1, \zeta_2,\zeta_3,\zeta_4,\zeta_5$:
\begin{eqnarray} \label{zeta1S0}
\gamma&=&\,\,{4\pi\over M C_0} + \mu \ \,, \qquad\quad \zeta_1 = \,\left[ {C_2
\over (C_0)^2} \right] \,, \nn \\
 \zeta_2 &=& \left[ {D_2 \over (C_0)^2 } -{g_A^2 \over 4 f^2} \Big(
   {M\over 4 \pi} \Big)^2 {\rm ln}\Big({\mu^2 \over m_\pi^2}\Big) \right] +
   \frac{1}{m_\pi^2} \left[ {C_{0}^{(0)} \over (C_0)^2} + {g_A^2 \over 2f^2}
   \Big({M\over 4 \pi}\Big)^2 (\gamma^2\!-\!\mu^2) \right] \,, \nn \\
\zeta_3 &=& -{g_A^2 \over 2 f^2}  {M m_\pi \over 4 \pi}
    \left[ {C_2 \over (C_0)^2}\right] +{1 \over m_\pi^2} \left[ {C_0^{(1)}
    \over (C_0)^2} -{(C_0^{(0)})^2 \over (C_0)^3} -\left({g_A^2 \over 2f^2}
    \right)^2 \left({M \over 4 \pi}\right)^3(\mu^3-\gamma^3)\right]  \nn \\
 && - {2\gamma \over m_\pi^2} \,{M g_A^2 \over 8 \pi f^2}\left[{C_0^{(0)}
    \over (C_0)^2} +{g_A^2 \over 2 f^2}\left({M \over 4\pi}\right)^2(-\mu^2+
    \gamma^2)\right] -2 {M\gamma\over 4\pi} \Big( \frac{M g_A^2}{8\pi f^2}
    \Big)^2 \Big(\ln 2-\frac32 \Big)  \nn \\
 && + m_\pi^2 \left\{ {D_4 \over (C_0)^2} \!\!-\! {D_2^2\over (C_0)^3} \right\}
    \!\!+\! \left[{D_2^{(-1)} \over (C_0)^2} - {2 D_2 C_0^{(0)} \over (C_0)^3}
    -{g_A^2 \over f^2}{M\gamma \over 4\pi} {D_2 \over (C_0)^2}\right]
    +\zeta_3^{rad} \,, \nn \\
\zeta_4 &=& \left[{C_{2}^{(-1)} \over (C_0)^2} - {2\, C_2\, C_{0}^{(0)}\over
    (C_0)^3}-{g_A^2 \over f^2} {M\gamma\over 4\pi} {C_2\over (C_0)^2}  \right]
    +m_\pi^2\ \left\{{E_4 \over (C_0)^2}- {2\, C_2 \, D_2 \over  (C_0)^3}
    \right\}\,, \\
\zeta_5 &=& m_\pi^2\ \left\{ {C_4 \over (C_0)^2} - {(C_2)^2\over (C_0)^3}
    \right\} \,. \nn
\end{eqnarray}
$\zeta_1 -\zeta_5$ are dimensionless constants. Note that $\zeta_2 -\zeta_5$
include factors of $m_\pi$ and are not simply short distance quantities. After
solving the RGE's in Eq.~(\ref{C0RGE}) one finds that
all quantities in square and curly brackets are separately $\mu$ independent.
Furthermore, the quantities in curly brackets vanish at NNLO in the $Q$
expansion due to Eqs.~(\ref{fixC4}) and (\ref{fixED4}). In Eq.~(\ref{zeta1S0})
the order $Q$ radiation pion contributions appear in $\zeta_3^{rad}$ given in
Eq.~(\ref{z3rad}) of Appendix~\ref{AppRad}. At order $Q$, the effect of
radiation pions turns out to be indistinguishable from corrections coming from
contact interactions.

For the $^1S_0$ channel, the location of the pole is determined by solving
\begin{eqnarray}\label{pole}
    -{1 \over a} + {r_0 \over 2}(p^*)^2 - i p^* = 0 \,.
\end{eqnarray}
This fixes $\gamma = - 7.88 \,{\rm MeV}$.  Note that adding the shape parameter
correction to Eq.~(\ref{pole}) changes the location of the pole by less than
$0.01\%$.  The NLO good fit condition in Eq.~(\ref{goodfit}) relates the constants
$\zeta_1$ and $\zeta_2$,
\begin{equation}\label{NLOfit}
  \zeta_2 = {\gamma^2 \over m_\pi^2}\, \zeta_1 - {M \over 4\pi}\,
   {g_A^2 M \over 8 \pi f^2}\,  {\rm log}\left(1+{2 \gamma \over m_\pi}\right)
   \, ,
\end{equation}
leaving one new parameter in the fit at NLO. At NNLO, $\zeta_5 = 0$ once we impose
$C_4 = C_2^2/C_0$.  This leaves $\zeta_3$ and $\zeta_4$, which are related by the
NNLO good fit condition
\begin{equation} \label{NNLOfit}
 \zeta_3 = \frac{\gamma^2}{m_\pi^2}\,  \zeta_4 +  \Big(  {M g_A^2 \over 8\pi f^2}
   \Big)^2 {M\over 4\pi}\ {m_\pi^2 \over \gamma} \bigg[ {\rm Re}\,{\rm Li}
   \Big({-m_\pi \over m_\pi+2 \gamma} \Big) + \frac{\pi^2}{12} \bigg] \,.
\end{equation}
Since $\zeta_1$ and $\zeta_4$ are multiplied by $\gamma^2/m_\pi^2$ in
Eqs.~(\ref{NLOfit}) and (\ref{NNLOfit}) these conditions basically fix the
values of $\zeta_2$ and $\zeta_3$. We have chosen to fix $\zeta_1$ and
$\zeta_4$ by performing a weighted least squares fit to the Nijmegen partial
wave analysis\cite{Nij}. The ranges $p=7-80\,{\rm MeV}$ and $p=7-200\,{\rm MeV}$
were used at NLO and NNLO respectively, with low momentum weighted more heavily.
Using $M=939\,{\rm MeV}$, $m_\pi=137\,{\rm MeV}$, $g_A=1.25$, and
$f=131\,{\rm MeV}$ the parameters for the $^1S_0$ channel are:
\begin{eqnarray}
  {\rm NLO}: && \qquad \zeta_1 = 0.216; \qquad \zeta_2 = 0.0318 ; \nn \\*
  {\rm NNLO}: && \qquad \zeta_1 = 0.0777 ; \qquad \zeta_2 = 0.0313 ;
   \qquad \zeta_3 = 0.1831 ; \qquad \zeta_4 = 0.245 \,.
\end{eqnarray}
The value of these parameters depend on the range of momentum used in the fit,
for instance using the range $p=7-150\,{\rm MeV}$ at NLO gives $\zeta_1=0.25$.
From the power counting we expect $\zeta_1\sim M/(4\pi\Lambda)$ at NLO and
$\zeta_1+\zeta_4\sim M/(4\pi\Lambda)$ at NNLO.  For $\Lambda\simeq 300\,
{\rm MeV}$, $M/(4\pi\Lambda)\simeq 0.25$ in reasonable agreement with the fits.

\begin{figure}[!t]
  \centerline{\epsfxsize=7.5truecm \epsfbox{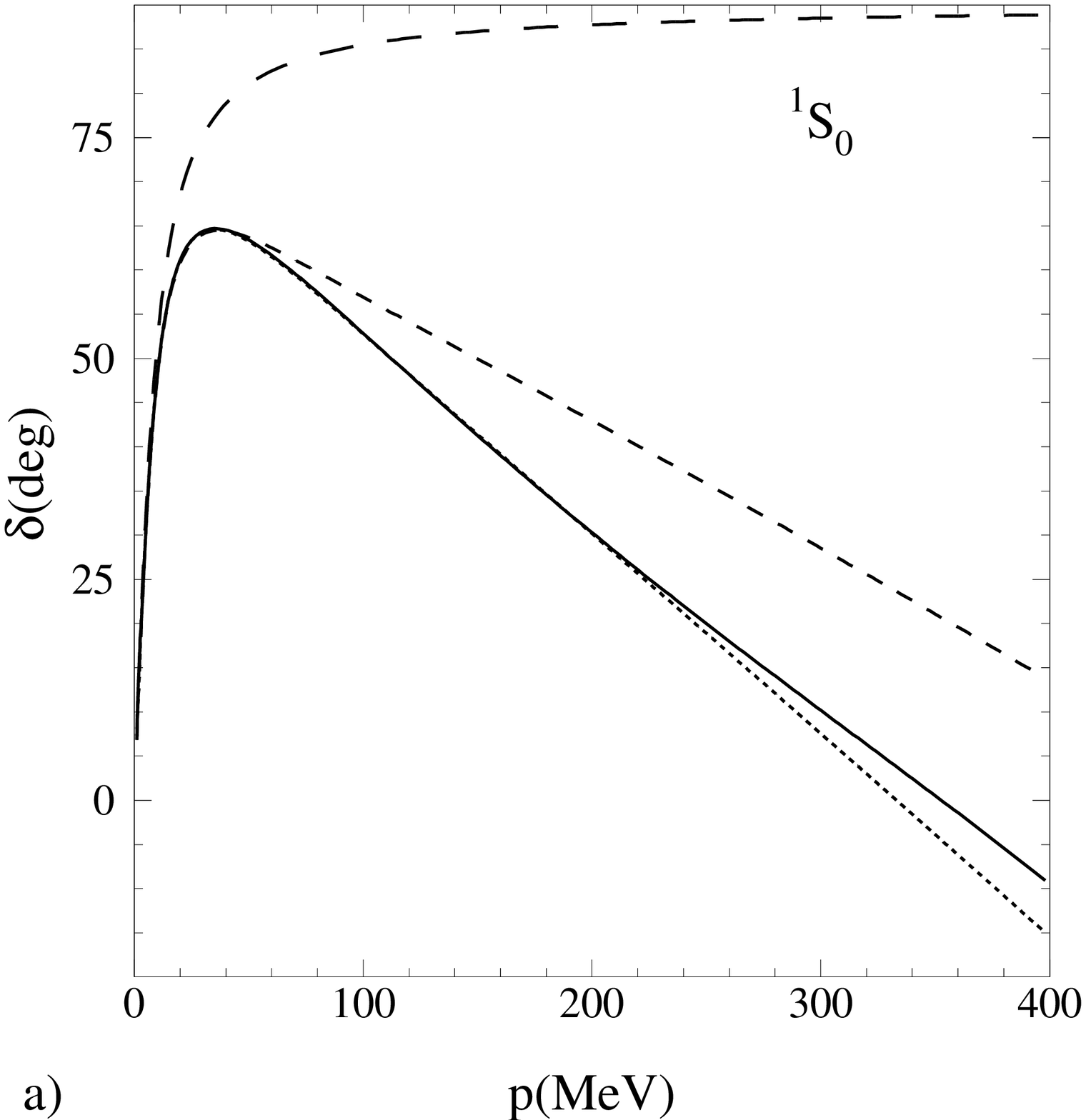}\ \ \ \
  \epsfxsize=7.5truecm \epsfbox{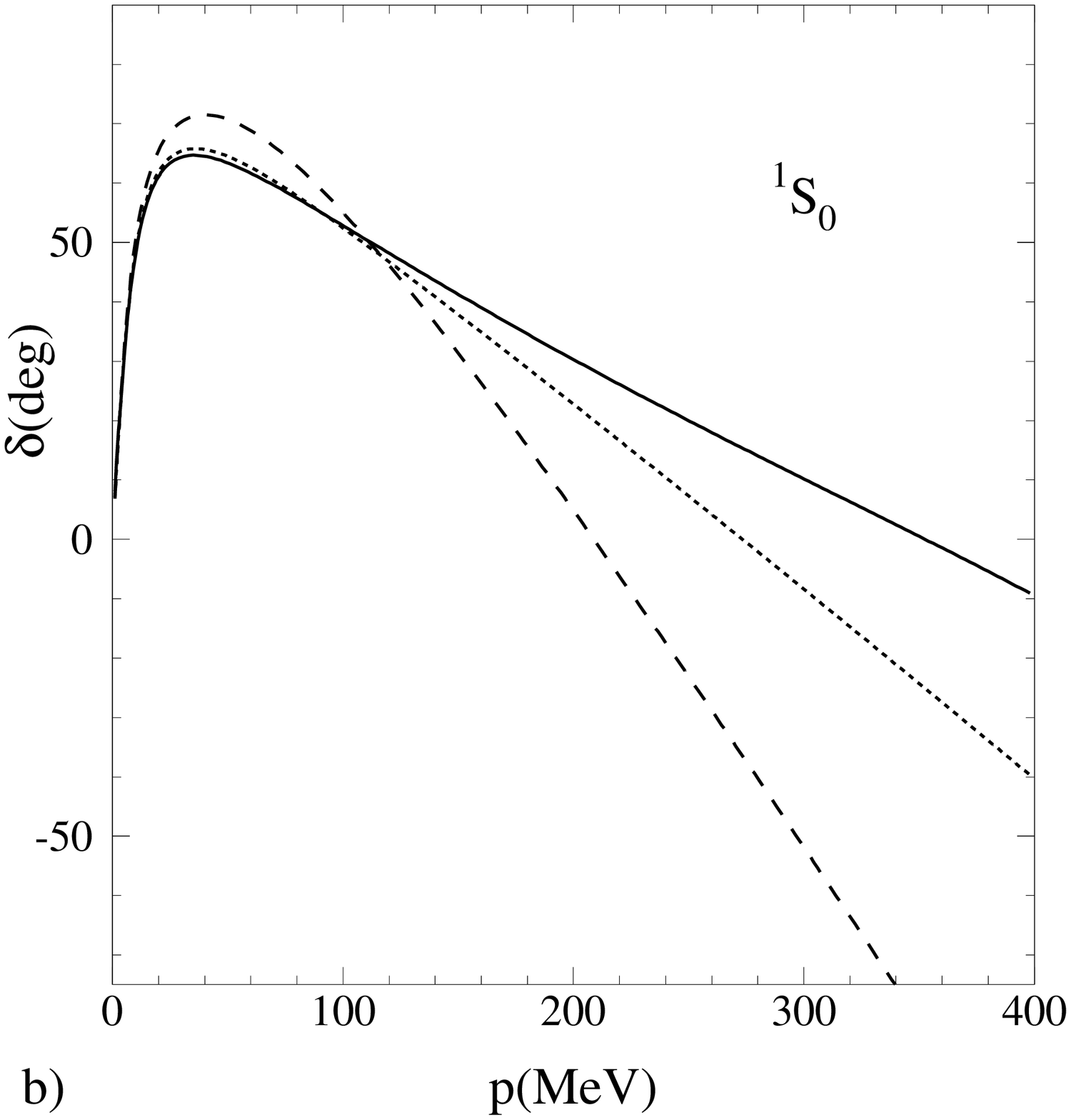} } \vspace{0.3cm}
{\tighten \caption[1]{Fit to the $^1S_0$ phase shift $\delta$.  The solid line
is the Nijmegen fit \cite{Nij} to the data. In a), the long dashed, short
dashed, and dotted lines are the LO, NLO, and NNLO results respectively. In b)
we show two other NNLO fits with a different choice of parameters as described
in the text.} \label{fig_fits} }
\end{figure}
The $^1S_0$ phase shift is shown in Fig.~\ref{fig_fits}a. The solid line is the
result of the Nijmegen phase shift analysis\,\cite{Nij}.  The $^1S_0$ phase
shift has an expansion in powers of $Q$, and we plot the LO, NLO and NNLO
results. The LO phase shift at $p = m_\pi$ is off by $48 \%$.  At NLO, the
error is $17 \%$. At NNLO, the error in the $^1S_0$ channel is less than $1\%$
at $p = m_\pi$, and the NNLO result gives improved agreement with the data even
at $p \sim 400 \,{\rm MeV}$.

Note that $\zeta_3\sim Q$ is larger than $\zeta_2\sim Q^0$ because from
Eqs.~(\ref{NLOfit}) and (\ref{NNLOfit}), $\zeta_3/\zeta_2 \sim m_\pi^2/
(\gamma\Lambda_{NN})$.  The parameter $\zeta_2$ is stable because it is fixed
by the NLO good fit condition.  On the other hand, $\zeta_1$ changes by a
factor of 2.8 going from NLO to NNLO.  One expects the value of coupling
constants to change somewhat at each order in the expansion, but a factor of
three difference is surprising. It is also disturbing that $\zeta_4$ is greater
than $\zeta_1$, since, on the basis of the RGE and KSW power counting, it is
expected that $\zeta_4 < \zeta_1$~\cite{ms1}.  It is possible to do a fit and
impose the constraints that $\zeta_1$ is close to its NLO value and $\zeta_4
\leq \zeta_1$. If this is done the error at $p \simeq m_\pi$ is $\approx 10
\%$, which is still an improvement relative to the NLO calculation and
consistent with an expansion parameter of order $1/2$. This fit is shown as the
dotted line in Fig.~\ref{fig_fits}b.

The potential diagrams for the $^1S_0$ phase shift at NNLO were also computed
by Rupak and Shoresh\, \cite{rupak1}. To fit $\zeta_1$ and $\zeta_4$ they
essentially demand that the experimental value of the effective range is reproduced
at both NLO and NNLO. For the observable $\sin^2\delta$ at $p\simeq m_\pi$, they
find $\simeq 80\%$, $\simeq 65\%$, and $\simeq 5\%$ errors at LO, NLO, and NNLO
respectively \cite{rupak2}.

Kaplan and Steele\,\cite{sk} have proposed that when the perturbative expansion
of coupling constants is made the sub-dominant couplings should not be treated
as new parameters. As an example, in their fitting procedure, $C_2^{(-1)}$ is given by
\begin{equation}  \label{kscond}
    C_2^{(-1)} =   { 2\, C_2\, C_0^{(0)} \over [C_0]^2 } \,.
\end{equation}
Imposing this condition fixes the value of $\zeta_4$ so that there is one less
free parameter at NNLO.  Kaplan and Steele motivated this fitting procedure by
arguing that adding pions should only change long distance physics. Therefore,
the number of free parameters in the theory with pions should be the same as in
the pionless theory.  It is worth pointing out that Ref.~\cite{sk} made use of
toy models in which the pions were represented as a contribution to the
potential which is either a delta-shell removed a finite distance from the
origin or a pure Yukawa. In these models it makes sense to think of the
``pion'' as purely long-distance because the pion effects are cleanly seperated
even in the presence of loop corrections.

In a realistic effective field theory ultraviolet divergences from loops with pions do
not allow a clean separation of long and short distance scales.  As an example
consider $C_2^{(^3S_1)}= C_2 + C_2^{(-1)}+C_2^{(0)}+\ldots$. Here $C_2$ first appears
in the NLO diagrams in Fig.~\ref{figQ0} and introduces a short distance effective
range-like constant. At NNLO the diagram in Fig.~\ref{figQ}k appears and has a
logarithmic ultraviolet divergence that must be absorbed by $C_2^{(-1)}$. This induces
a $\ln(\mu/K)$ dependence into the coupling $C_2^{(-1)}$ (as is clear from
Eq.~(\ref{zeta3S1})). Since the constant $K$ is undetermined it is clear that
$C_2^{(-1)}$ cannot be determined from lower order couplings.  Note that if this
$\ln(\mu)$ is instead absorbed into the leading order $C_2$ then this would induce
additional $\ln(\mu)$ dependence into the part of the NNLO amplitude that depends on
$C_2$.

In the $^1S_0$ channel $C_2^{(-1)}$ does not receive a logarithmic renormaliztion.
However, there is a new logarithmic divergence that must be absorbed into
$C_2^{(0)}$ coming from Fig.~\ref{fig_soft}a \cite{ms2}.  Therefore, $C_2^{(0)}$
must be treated as a parameter.  It is not possible to renormalize the theory in a
$\mu$ independent way without introducing more parameters than exist in the
pionless theory.  The power law sensitivity to the choice of $\mu$ makes the $\mu$
independence of observables an essential criteria. Since, in general, higher order
terms in the expansion of couplings receive ultraviolet renormalizations, we
prefer to treat all $C_{2n}^{(m)}$ as free parameters whose size is only
restricted by their RGE's. This then implies that $\zeta_4$ is a free parameter in
both the $^1S_0$ and $^3S_1$ channels. However, in the $^1S_0$ channel at NNLO
imposing the relation in Eq.~(\ref{kscond}) does give a $\mu$ independent
amplitude. In this case the result of the fit is shown by the dashed line in
Fig.~\ref{fig_fits}b. In general the choice of fit parameters is somewhat
arbitrary, and a true test of the values can only be made by using them to predict
an independent observable.

\begin{table}[t!]
\begin{center} \begin{tabular}{ccccccc} \label{LETs}
 $^1S_0$\hspace{2cm}& & & $r_0$ & $v_2$ & $v_3$ & $v_4$ \\ \hline
&Fit\cite{ch} && $2.73\,{\rm fm}$ & $-0.48\,{\rm fm^3}$ & $3.8\,{\rm fm^5}$ &
    $-17\,{\rm fm^7}$  \\
&NLO && $2.65\,{\rm fm}$ & $-3.3\,{\rm fm^3}$ & $19\,{\rm fm^5}$ &
    $-117\,{\rm fm^7}$  \\
&NNLO && $2.63\,{\rm fm}$ & $-1.2\,{\rm fm^3}$ & $2.9\,{\rm fm^5}$ &
    $-0.7\,{\rm fm^7}$
\end{tabular} \end{center}
\caption{Predictions for terms in the $^1S_0$ effective range expansion.}
\end{table}
Finally, we present NNLO corrections to the higher order terms in the
effective range expansion
\begin{eqnarray}
  p\,{\rm cot}(\delta) = -{1\over a} + {r_0 \over 2}p^2 + v_2 p^4 + v_3 p^4+ v_4 p^4
    + \ldots   \,.
\end{eqnarray}
Using the NLO expression for $p\,{\rm cot}(\delta)$, Cohen and Hansen\,\cite{ch}
obtained predictions for $v_2,v_3$ and $v_4$.  At NLO, the effective field theory
predictions for $v_2$, $v_3$, and $v_4$ disagree with the $v_i$ obtained from a
fit to the Nijmegen phase shift analysis.  The NNLO predictions for the shape
parameters are shown in Table~\ref{LETs}.  The prediction for $r_0$ is not better
at NNLO than at NLO, but is still well within the expected errors. The NNLO $v_i$
predictions depend on $\zeta_1$ and $\zeta_2$.  We see that the NNLO correction
substantially reduces the discrepancy between the effective field theory
prediction and the fit to the Nijmegen phase shift analysis, but the discrepancy
is still quite large. This gives some evidence that the EFT expansion is
converging on the true values of the $v_i$, albeit slowly. Effective field theory
predictions for the shape parameters have been studied in toy models where one is
able to go to very high orders in the $Q$ expansion~\cite{kaplanpc}. In the toy
models, the effective field theory did eventually reproduce the shape parameters,
but the observed convergence is rather slow.


\newpage

\subsection{$^3S_1$ channel}
\label{s3s1}

The S matrix for the $^3S_1$ and $^3D_1$ channels is $2 \times 2$ and
can be parameterized using the convention in Ref.~\cite{Stapp} :
\begin{eqnarray}  \label{S}
  S = {\mathbf{1}} + \frac{i\, M p}{2\pi} \left( \begin{array}{cc} {\cal A}^{SS} &
      {\cal A}^{SD} \\ {\cal A}^{SD} & {\cal A}^{DD} \end{array} \right)
   =   \left(  \begin{array}{cc}
       e^{2i\bar\delta_0} \cos 2\bar\epsilon_1 & i\, e^{i\bar\delta_0+i\bar\delta_2}
       \sin 2\bar\epsilon_1 \\
       i\, e^{i\bar\delta_0+i\bar\delta_2} \sin 2\bar\epsilon_1 & e^{2i\bar\delta_2}
       \cos 2\bar\epsilon_1
     \end{array}   \right) \,.
\end{eqnarray}
The phase shifts and mixing angle are expanded in powers of $Q/\Lambda$:
\begin{eqnarray}  \label{dexpn}
   \bar\delta_0 = \bar\delta_0^{(0)}  + \bar\delta_0^{(1)} + \bar\delta_0^{(2)}
   + \ldots \,,\quad
   \bar\delta_2 = 0  + \bar\delta_2^{(1)}+ \bar\delta_2^{(2)} + \ldots \,,\quad
   \bar\epsilon_1 = 0 + \bar\epsilon_1^{(1)} + \bar\epsilon_1^{(2)} + \ldots
   \,.
\end{eqnarray}
The phase shifts and mixing angles start at one higher order in $Q$ than the
amplitudes because of the factor of $p$ in Eq.~(\ref{S}). In the PDS scheme,
expressions for $\bar\delta_0^{(0,1)}$, $\bar\delta_2^{(1)}$, and
$\bar\epsilon_1^{(1)}$ are given in Ref.~\cite{ksw2}.  The prediction for
$\bar\epsilon_1^{(2)}$ is given in Ref.~\cite{fms} and is discussed in
section~IIIC, and the prediction for $\bar\delta_2^{(2)}$ is given in
section~IIID.

Expressions for the terms in Eq.~(\ref{dexpn}) in terms of the scattering
amplitude are obtained by expanding both sides of Eq.~(\ref{S}) in powers of
$Q$. This gives\footnote{\tighten The branch cut in the logarithm in
Eq.~(\ref{delta0s}) is taken to be on the positive real axis.  This is
consistent with $\bar \delta_0(p\to 0)=\pi$. The sign of our $^3D_1$ state is
the opposite of Ref.~\cite{ksw2}, making $A^{SD}_0$ in Eq.~(\ref{Q0}) have
the opposite overall sign.}
\begin{eqnarray}  \label{delta0s}
  \bar{\delta}_0^{(0)} &=&  {1 \over 2 i}\, {\rm ln}\left(1+{i p M \over 2
  \pi} {\cal A}_{-1}^{SS} \right)  \,, \qquad \bar{\delta}_0^{(1)} = { p M
  \over 4 \pi} {{\cal A}^{SS}_{0} \over 1+ {i p M \over 2 \pi} {\cal
  A}^{SS}_{-1}}   \,,\nn \\
  \bar{\delta}_0^{(2)} &=& { p M \over 4 \pi} {{\cal A}^{SS}_{1} \over 1+ {i
  p M \over 2 \pi}{\cal A}^{SS}_{-1} } - i\left( { p M \over 4 \pi}
  \right)^2 \left[ \left( {{\cal A}^{SS}_{0} \over 1+ {i p M \over 2
  \pi}{\cal A}^{SS}_{-1} } \right)^2 + {({\cal A}^{SD}_{0})^2 \over
  1+ {i p M \over 2 \pi}{\cal A}^{SS}_{-1} }  \right]   \,.
\end{eqnarray}
In $\bar{\delta}^{(2)}_0$ the terms that depend on $A_0^{SS}$ and $A_0^{SD}$
are purely imaginary and cancel the imaginary part of the term proportional to
${\cal A}_1^{SS}$ as required by unitarity. The order $Q^0$
mixing amplitude is\cite{ksw2}
\begin{eqnarray}  \label{Q0}
  {\cal A}_0^{SD} &=& \sqrt{2} \frac{M g_A^2}{8\pi f^2}\: {\cal A}_{-1}^{SS}
  \bigg\{ - \frac{3 m_\pi^3}{4p^2}+ \Big(\frac{m_\pi^2}{2p} +
  \frac{3m_\pi^4}{8p^3} \Big) \tan^{-1}\Big(\frac{2p}{m_\pi}\Big) +
  \frac{3\gamma m_\pi^2}{4p^2} - \frac{\gamma}{2} \nn\\
&&\qquad\qquad\qquad\quad - \Big( \frac{\gamma m_\pi^2}{4 p^2}+
  \frac{3\gamma m_\pi^4}{16p^4}\Big)  \ln\Big(1+\frac{4p^2}{m_\pi^2}\Big) \bigg\}
   \,.
\end{eqnarray}
The diagrams which contribute to the $^3S_1$ amplitude up to NNLO are shown in
Figs.~\ref{figQ0} and \ref{figQ} and give
\begin{eqnarray} \label{TripAmp}
{\cal A}^{SS}_{-1} &=& -{4\pi\over M}{1\over \gamma + ip}\,,  \nn\\[5pt]
 { {\cal A}^{SS}_{0}  } &=&
   -  \Big[{\cal A}^{SS}_{-1}\Big]^2 (\zeta_1 \, p^2 + \zeta_2 \, m_\pi^2)  \\*
  && + \Big[{\cal A}^{SS}_{-1}\Big]^2 {g_A^2 \over 2 f^2} \Big({M m_\pi \over 4 \pi}
   \Big)^2 \bigg[ { (\gamma^2 - p^2) \over 4 p^2}\:{\rm ln}\Big(1 + {4 p^2 \over
   m_\pi^2}\Big) - \, {\gamma \over p}\: {\rm tan}^{-1}\left({2  p \over m_\pi}
  \right) \bigg] \nn\,, \\[5pt]
{ {\cal A}_{1}^{SS} } &=& { [{\cal A}^{SS}_{0}]^2 \over {\cal A}^{SS}_{-1}}
  + {i p M \over 4\pi} \Big[{\cal A}_{0}^{SD} \Big]^2
  + { {\cal A}^{SS}_{0} }\: {M g_A^2 \over 8 \pi f^2}\:  {m_\pi^2\over p}
  \bigg[ { \gamma \over 2 p}\: {\rm ln}\Big(1 + {4 p^2 \over m_\pi^2}\Big) -
  {\rm tan}^{-1}\Big({2 p \over m_\pi}\Big) \bigg] \nn \\
&-&  \Big[{\cal A}^{SS}_{-1}\Big]^2\, \Big(\zeta_3 \, m_\pi^2 +\zeta_4 \,p^2 +
  \zeta_5 {p^4 \over m_\pi^2} \Big) \nn\\
&+& \Big[{\cal A}^{SS}_{-1}\Big]^2\, {M  \over 4 \pi}\, \Big({ M g_A^2 \over
  8\pi f^2} \Big)^2 \Bigg[ {-6\gamma^2m_\pi^3 + 9\gamma m_\pi^4 -3 m_\pi^5 \over
  4 p^2 }+ \ln 2 \Big( \frac{9\gamma m_\pi^6}{4p^4}\!+\! \frac{3\gamma m_\pi^4}
  {2p^2}\!-\! \frac{9 m_\pi^7}{4p^4}\!-\! \frac{3m_\pi^5}{p^2} \Big) \nn\\
&& + \Big( 6\,p^2\!+\!6 m_\pi^2 \!-\!\frac{3 m_\pi^4}{4 p^2}\!-\!\frac{9 m_\pi^6}
   {8p^4} \Big) \bigg[ {p^2\!-\! \gamma^2 \over p } \tan^{-1}\Big(\frac{p}{m_\pi}\Big)
   - \gamma\, \ln\Big(1+\frac{p^2}{m_\pi^2}\Big) \bigg] \nn\\
&&- \Big( \frac{3 m_\pi^5}{p^3}\!+\! \frac{9 m_\pi^7}{4 p^5} \Big) \bigg[ \gamma\,
  \tan^{-1}\Big(\frac{p}{m_\pi}\Big) -\frac{(\gamma^2-p^2)}{4p} \ln\Big(1+\frac{p^2}
  {m_\pi^2}\Big) \bigg] \nn \\
&&+\Big( \frac{9 m_\pi^7}{8 p^5}\!+\! \frac{3 m_\pi^5}{2 p^3}\!-\!
  \frac{9 \gamma m_\pi^6}{8 p^5}\!-\!\frac{3 \gamma m_\pi^4}{4 p^3} \!+\!
 \frac{\gamma m_\pi^2}{p} \Big) \bigg[ \gamma \tan^{-1}\Big(\frac{2p}{m_\pi}\Big)
 +\frac{p}{2} \ln\Big(1+\frac{4p^2}{m_\pi^2} \Big) \bigg] \nn \\
&& + \Big( \frac{9 m_\pi^8}{32 p^7}\!+\!\frac{3m_\pi^6}{4p^5}+\frac{3m_\pi^4}{4p^3}
  \Big) \Bigg\{ 2(\gamma^2\!-\!p^2)\, {\rm Im}\, {\rm Li_2} \Big(
   {-m_\pi \over m_\pi\!-\!2 i p} \Big) \!-\! 4 \gamma\, p\: {\rm Re}\, {\rm Li_2}
  \Big({-m_\pi \over m_\pi\!-\!2 i p}\Big) \!-\! { \gamma\, p\, \pi^2 \over 3}  \nn \\
&&  \ \ \ -(\gamma^2+p^2) \bigg[ {\rm Im} \,
  {\rm Li_2} \Big({m_\pi\!+\!2 i p \over -m_\pi\!+\!2 i p} \Big) + {\gamma \over 4 p}
  \ln^2\Big(1+\frac{4p^2}{m_\pi^2}\Big) - \tan^{-1}
   \Big(\frac{2p}{m_\pi} \Big) \ln\Big(1+\frac{4p^2}{m_\pi^2}\Big) \bigg] \Bigg\} \nn \\
&& + \gamma\, \Big( \frac{9 m_\pi^8}{32 p^6}\!+\!\frac{3m_\pi^6}{4p^4}+
  \frac{m_\pi^4}{2p^2}\Big) \bigg[ \tan^{-1}\Big(\frac{2p}{m_\pi}\Big) -
  \frac{\gamma}{2p} \ln\Big(1+\frac{4p^2}{m_\pi^2} \Big)\bigg]^2 \ \Bigg] \,. \nn
\end{eqnarray}
The six linearly independent constants appearing in Eq.~(\ref{TripAmp}) are:
\begin{eqnarray} \label{zeta3S1}
\gamma&=&\,\,{4\pi\over M C_0} + \mu \ \,, \qquad\quad \zeta_1 = \, \left[
  {C_2 \over (C_0)^2} \right] \,, \\
\zeta_2 &=& \left[ {D_2 \over (C_0)^2 } -{g_A^2 \over 4 f^2} \Big( {M\over
   4 \pi} \Big)^2\, {\rm ln}\Big({\mu^2 \over m_\pi^2}\Big) \right] +
   \frac{1}{m_\pi^2} \left[ {C_{0}^{(0)} \over (C_0)^2} + {g_A^2 \over 2f^2}
   \Big({M\over 4 \pi}\Big)^2 (\gamma^2\!-\!\mu^2) \right] \,, \nn \\
\zeta_3 &=& -{g_A^2 \over 2 f^2}  {M m_\pi \over 4 \pi} \left[ {C_2 \over
  (C_0)^2}\right]- {1 \over m_\pi^2} {M\over 4 \pi} \Big( {Mg_A^2 \over 8\pi f^2}
  \Big)^2 \Big( \gamma^3 -6 m_\pi \gamma^2 -\frac72 m_\pi^2 \gamma + 4 m_\pi^3
  \Big)  \nn\\
 &&+{1 \over m_\pi^2} \left[ {C_{0}^{(1)} \over (C_0)^2} -{(C_0^{(0)})^2 \over
   (C_0)^3}-{g_A^2 \over f^2} {M\gamma\over 4\pi} {C_{0}^{(0)}\over (C_0)^2} -
    {M\over 4 \pi} \Big( {Mg_A^2 \over 8\pi f^2} \Big)^2 \Big( 4\mu \gamma^2
   -6\gamma \mu^2 +\frac43 \mu^3  \Big) \right] \nn\\
 && + m_\pi^2 \left\{ {D_4 \over (C_0)^2} \!\!-\! {(D_2)^2\over (C_0)^3} \right\}
    \!\!+\! \left[{D_{2}^{(-1)} \over (C_0)^2} - {2 D_2 C_{0}^{(0)} \over (C_0)^3}
    - {g_A^2 \over f^2} {M \gamma \over 4\pi} {D_2 \over (C_0)^2} -
    5{M \gamma \over 4\pi} \Big({M g_A^2 \over 8 \pi f^2}\Big)^2
     \ln\Big(\frac{\mu^2}{m_\pi^2}\Big) \right]  \nn \\
 && + \zeta_3^{rad} \,,\nn \\[5pt]
\zeta_4 &=& \left[{C_{2}^{(-1)} \over (C_0)^2} - {2\, C_2 C_{0}^{(0)}\over
    (C_0)^3}- 6{M \gamma \over 4\pi} \Big({M g_A^2 \over 8 \pi f^2}\Big)^2
     \ln\Big(\frac{\mu^2}{m_\pi^2}\Big)  \right] + m_\pi^2\ \left\{{E_4 \over
     (C_0)^2}- {2\, C_2 \, D_2 \over  (C_0)^3}  \right\}  \nn \,, \\
 && -{g_A^2 \over f^2} {M\gamma\over 4\pi} \left[ {C_2\over (C_0)^2} \right]
     - {M\over 4 \pi}\Big( {Mg_A^2 \over 8\pi f^2} \Big)^2 \Big( -3 \gamma +
     6 m_\pi) \,, \nn\\
\zeta_5 &=& m_\pi^2\ \left\{ {C_4 \over (C_0)^2} - {(C_2)^2\over (C_0)^3}
    \right\}\,. \nn
\end{eqnarray}
Solving the beta functions in Eq.~(\ref{C0RGE2}) perturbatively, we find that the
quantities in the square and curly brackets are separately $\mu$ independent,
and the quantities in curly brackets vanish at NNLO. $\zeta^{rad}_3$ includes the
radiation pion contributions to the amplitude.  The expression for
$\zeta_3^{rad}$ in the ${}^3S_1$ channel is obtained from Eq.~(\ref{z3rad}) by
interchanging the spin singlet and spin triplet labels.

\begin{figure}[!t]
  \centerline{\epsfxsize=8.truecm \epsfbox{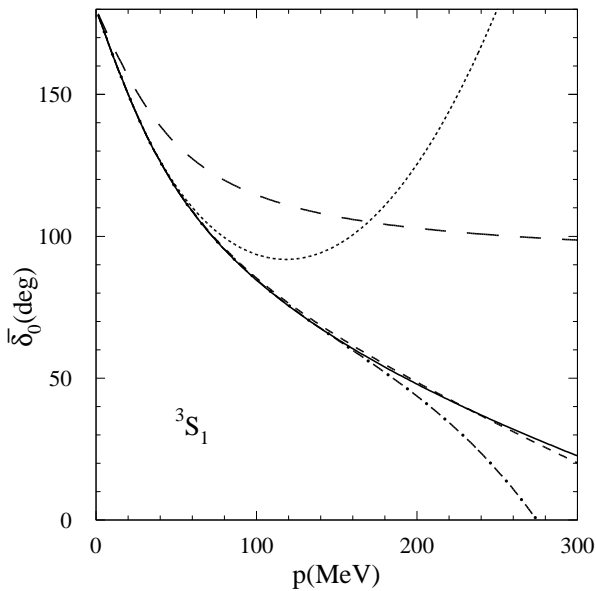}  } \vspace{.5cm}
{\tighten \caption[1]{The $^3S_1$ phase shift for NN scattering.  The solid
line is the Nijmegen multi-energy fit\cite{Nij}, the long dashed line is the LO
effective field theory result, the short dashed line is the NLO result, and the
dotted line is the NNLO result.  The dash-dotted line shows the result of
including the parameter $\zeta_5$ which is \emph{higher order} in the power
counting.} \label{fig_3s1}}
\end{figure}
The LO amplitude ${\cal A}^{SS}_{-1}$ has a pole at $p=i\gamma$ corresponding
to the deuteron bound state. The deuteron has binding energy $B=2.22\,{\rm
MeV}$, so $\gamma= \sqrt{M B} = 45.7\,{\rm MeV}$. The remaining coefficients,
$\zeta_1 - \zeta_4$ are fixed using the same procedure as in the ${}^1S_0$
channel:
\begin{eqnarray} \label{3s1fit}
  {\rm NLO}: && \qquad \zeta_1 = 0.327 ; \qquad \zeta_2 = -0.0936 ;  \\*
  {\rm NNLO}: && \qquad \zeta_1 = 0.432 ; \qquad \zeta_2 = -0.0818 ;
   \qquad \zeta_3 = 0.165 ; \qquad \zeta_4 = 0.399 ; \nn
\end{eqnarray}
The ${}^3S_1$ phase shift is shown in Fig.~\ref{fig_3s1}. The LO phase shift
(long dashed curve) has no free parameters, and at $p=m_{\pi}$ the error is
60\%. The NLO phase shift (short dashed curve) has one free parameter
($\zeta_1$), which is fit to the Nijmegen multi-energy fit (solid curve). The
NLO fit to the data is excellent. However, this agreement is clearly fortuitous
because the NNLO phase shift (dotted line) with two free parameters ($\zeta_1$,
$\zeta_4$) does worse at fitting the data than the NLO phase shift. At $p=
m_{\pi}$ the error is 30\%, exceeding expectations based on an expansion in
$1/2$. The error is even greater for larger values of $p$.  The dash-dotted
line in Fig.~\ref{fig_3s1} shows the result of including the ${\rm N}^3{\rm
LO}$ parameter $\zeta_5=0.26$.  Better agreement with the data is found,
however, including $\zeta_5$ at this order is in violation of the KSW power
counting.

Large NNLO corrections also show up in predictions for the effective range
expansion parameters.  For example, at NLO the effective theory gives an
effective range $r_0=2.2\, {\rm fm}$, which is within 20\% of the experimental
value, $r_0=2.73\, {\rm fm}$.  At NNLO we find $r_0=1.3\,{\rm fm}$.  The NNLO
correction to $r_0$ includes a large negative non-analytic contribution from the
diagrams with two potential pions.

The failure of the EFT at NNLO in the ${}^3S_1$ channel is due to large
corrections from the two pion exchange graphs in Figs.~\ref{figQ}i,k,m. The
term which dominates the NNLO amplitude for large $p$ is
\begin{eqnarray}\label{A1lim}
{\cal A}_1^{SS} \simeq  6\: [{\cal A}_{-1}^{SS}]^2\ {M \over 4 \pi}
 \left({M g_A^2 \over 8 \pi f^2}\right)^2 \: p^3\: {\rm tan}^{-1} \left({p \over
  m_\pi} \right) \,.
\end{eqnarray}
For $p\gg m_\pi$ this term grows linearly with $p$, an effect which can be
clearly seen in Fig.~\ref{fig_3s1} (the growth in Fig.~\ref{fig_3s1} is
quadratic due to the extra $p$ in Eq.~(\ref{delta0s})). The contribution in
Eq.~(\ref{A1lim}) is large because of the coefficient of $6$ which is much
greater than the expansion parameter. For $p\gg m_\pi$ the size of this
contribution relative to the LO amplitude is $3\pi p^2/\Lambda_{NN}^2$. The fact
that this correction survives in the chiral limit indicates that it comes from
the short distance part of potential pion exchange. Large non-analytic
corrections are also found in some of the other spin triplet channels at this
order.


\subsection{$^3S_1-{}^3D_1$ channel}

\begin{figure}[!t]
  \centerline{\epsfxsize=8.truecm \epsfbox{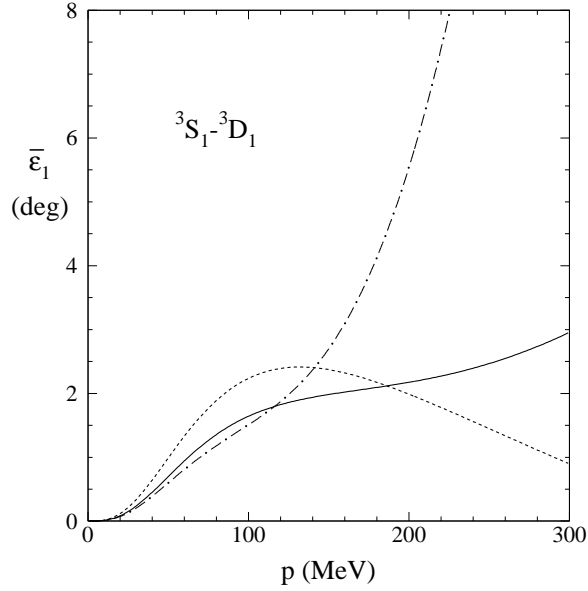}  } \vspace{.5cm}
{\tighten \caption[1]{The $^3S_1-{}^3D_1$ mixing angle for NN scattering. At
LO this phase shift is zero.  The dotted line is the NLO result \cite{ksw2}
and the dash-dotted line is the NNLO result\cite{fms}. The solid line is from
Nijmegen's multi-energy partial wave analysis \cite{Nij}.}
\label{fig_mix}}
\end{figure}

The $^3S_1-{}^3D_1$ mixing amplitude at NNLO was presented in Ref.~\cite{fms}.
The result is briefly summarized here for the sake of completeness. The
prediction is shown in Fig.~\ref{fig_mix}. For $\bar\epsilon_1$ the LO (order
$Q^0$) prediction vanishes and the NLO prediction\cite{ksw2} is parameter free.
At NNLO there is one free parameter $C_2^{(SD)}(\mu)$ which is fit to the data:
$C_2^{(SD)}(m_\pi)=-4.6\,{\rm fm^4}$. This value is consistent with the power
counting estimate which gives $|C_2^{(SD)}(m_\pi)| \sim 4\pi/(M m_\pi^2
\Lambda)\simeq 3.6\,{\rm fm^4}$ for $\Lambda=300\,{\rm MeV}$.  The mixing angle
agrees with expected errors for $p \sim m_\pi$, but for larger values of momentum
there is serious disagreement between theory and experiment.  For $p\sim m_\pi$
this disagreement is comparable to the uncertainty of a calculation of
$\bar\epsilon_1$ within the Weinberg approach\cite{ork}. A more recent
analysis \cite{Ulf} gives a more accurate prediction for $\bar\epsilon_1$, but
an analysis of the uncertainty due to the cutoff dependence is not presented.

At the order we are working ${\cal A}^{SD} = {\cal A}^{SD}_0 + {\cal A}^{SD}_1$
and
\begin{eqnarray}
  \bar \epsilon_1 = \frac{M p}{4\pi} \Big|{\cal A}_{-1}^{SS}\Big|\: {\rm Re}
    \bigg[\, { {\cal A}^{SD}  \over {\cal A}_{-1}^{SS} } \, \bigg] \,.
\end{eqnarray}
The behavior of this mixing angle for $p\gg m_\pi$ can be examined by taking the
$m_\pi\to 0$ limit of the mixing amplitude:
\begin{eqnarray}
 \sqrt{2}\: {\rm Re}\bigg[\, {{\cal A}^{SD} \over {\cal A}_{-1}^{SS} } \,
  \bigg] &=& \frac{-\gamma}{\Lambda_{NN}} + \frac{3\gamma\, \pi\, p}{5\,
  \Lambda_{NN}^2} -p^2 \bigg[ \frac{\sqrt{2}C_2^{(SD)}}{C_0} -
  \frac{4\pi C_2}{M\Lambda_{NN}C_0^2}+\frac{21}{100\Lambda_{NN}^2} +
   \frac{6}{5\Lambda_{NN}^2} \ln\Big(\frac{\mu}{2p}\Big) \bigg] \nn\\
 && + \frac{4\pi}{M\Lambda_{NN}}\: \frac{p^2}{(p^2+\gamma^2)} \bigg[ -
  \frac{C_2\gamma^2}{(C_0)^2} + \frac{C_0^{(0)}}{(C_0)^2} + \frac{g_A^2}{2f^2}
  \Big( \frac{M}{4\pi} \Big)^2 (\gamma^2-\mu^2) \bigg] \,,
\end{eqnarray}
where the $\ln(\mu)$ dependence is cancelled by $C_2^{(SD)}$.  In this
channel the term proportional to $\pi p$ is suppressed by an additional
factor of $\gamma$, and the dominant terms in the NNLO calculation for $p\gg
m_\pi$ are analytic, growing as $p^2$.  The fit to the low energy data in
Ref.~\cite{fms} did not give a value of $C_2^{(SD)}$ that cancelled this
growth as can be seen clearly in Fig.~\ref{fig_mix}.

An interesting way to test the EFT for nucleons is to compare the value of
$C_2^{(SD)}(\mu)$ extracted from our NNLO calculation of $\bar\epsilon_1$ to
the $C_2^{(SD)}(\mu)$ extracted from the NNLO calculation of the deuteron
quadrupole moment \cite{binger}. To make the comparison meaningful the same
renormalization scheme must be used (and the same finite constants must
be subtracted along with the $p^2/\epsilon$ pole).  Ref.\cite{binger}
does not explicitly give $C_2^{(SD)}$ counterterms so it is was not possible to
compare our value for $C_2^{(SD)}(m_\pi)$ with the value extracted there.


\subsection{$^3D_1$ channel}

\begin{figure}[!t]
  \centerline{\epsfysize=2.5truecm \epsfbox{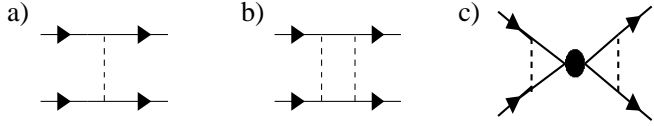}  }
 {\tighten
  \caption[1]{The order $Q^0$ diagram (a) and order $Q$ diagrams (b and c) that
contribute to the P and D wave channels. Only the $^3D_1$ channel gets a
contribution from diagram c).}
\label{figQD} }
\end{figure}
In the KSW expansion, there is no order $1/Q$ contribution to the $^3D_1-
{}^3D_1$ amplitude.  Using Eq.~(\ref{S}) we can derive expressions for the
$^3D_1$ phase shift up to order $Q^2$:
\begin{eqnarray}  \label{delta2s}
  \bar{\delta}_2^{(1)} &=& \frac{Mp}{4\pi} \: {\cal A}_0^{DD} \,, \\
  \bar{\delta}_2^{(2)} &=& \frac{Mp}{4\pi} \: {\cal A}_1^{DD} -i \Big(
     \frac{Mp}{4\pi}
     \Big)^2 \Big(A_0^{DD}\Big)^2 -i \Big( \frac{Mp}{4\pi} \Big)^2
   \Big|{\cal A}_{-1}^{SS}\Big|^2 \bigg( \frac{ {\cal A}_0^{SD}}{A_{-1}^{SS}}
   \bigg)^2 \,. \nn
\end{eqnarray}
The last two terms in $\bar \delta_2^{(2)}$ are purely imaginary and cancel the
imaginary part of $A_1^{DD}$ as required by unitarity.  The NLO contribution
comes entirely from one pion exchange\cite{ksw2},
\begin{eqnarray}
  {\cal A}^{DD}_0 = {g_A^2 \over 2 f^2}
  \left[-{1\over 2} -{3 m_\pi^2 \over 4 p^2} + \left( {m_\pi^2\over 2 p^2}
 + {3 m_\pi^4 \over 16 p^4} \right) {\rm ln}\left(1+{4 p^2 \over m_\pi^2}\right)
  \right]  \,.
\end{eqnarray}
Four nucleon operators which mediate transitions between two $^3D_1$-wave
states must have at least 4 derivatives. Graphs with these operators do not
contribute until order $Q^3$ in the KSW expansion. (The leading operator which
mediates $^3D_1$ wave transitions is renormalized by graphs with two insertions
of the ${\cal O}_2^{(SD)}$ operator. An insertion of ${\cal O}^{(SD)}_2$ is order
$Q$, therefore these graphs are order $Q^3$.) At NNLO, the $^3D_1-{}^3D_1$
amplitude gets contributions from the graphs in Fig.~\ref{figQD}.

The only short distance operator which contributes to this amplitude at
NNLO is ${\cal O}^{(^3S_1)}_0$, whose coefficient is completely
determined by the location of the pole in the spin-triplet channel. Therefore, no
free parameters appear in the calculation of this amplitude.
The NNLO amplitude is:
\begin{eqnarray}
{\cal A}^{DD}_{1} &=& i {M p \over 4 \pi} ({\cal A}^{DD}_0)^2 + {({\cal
   A}^{SD}_0)^2 \over {\cal A}_{-1}} + {3\over 2} \Big({g_A^2 \over 2
   f^2}\Big)^2 {M \over 4 \pi} \Bigg\{ -{2 m_\pi \over 7} + {51 m_\pi^3
   \over 70 p^2} + {3 m_\pi^5 \over 70 p^4} \\
&+& \left( {9 m_\pi^8 \over 32 p^7} + {m_\pi^6 \over p^5} + {m_\pi^4 \over
    p^3} \right) {\rm Im}\left[ {\rm Li}_2\left({-2p^2 +i m_\pi p \over m_\pi^2
    }\right) + {\rm Li}_2\left( {p \over 2 p + i m_\pi} \right) \right] \nn \\
&+& \left( {9 m_\pi^6 \over 8 p^5} + {7 m_\pi^4 \over 4 p^3} + {4 m_\pi^2 \over
    5 p} - {2 p \over 7} \right){\rm tan}^{-1}\left({p \over m_\pi} \right) -
    \left({3 m_\pi^6 \over 8 p^5} + {5 m_\pi^4 \over 4 p^3} + {2 m_\pi^2 \over
    3 p}\right) {\rm tan}^{-1}\left({2 p \over m_\pi}\right) \nn \\
 && + \left({3 m_\pi^7 \over 8 p^6} +{m_\pi^5 \over 2p^4 }\right){\rm ln}
    \left(1 + {4 p^2 \over m_\pi^2}\right) - \left({3 m_\pi^8 \over 16 p^7} +
    {m_\pi^6 \over 2p^5} + {m_\pi^4 \over 3p^3}\right) {\rm tan}^{-1}\left(
    {2 p \over m_\pi}\right) {\rm ln}\left(1 + {4 p^2 \over m_\pi^2}\right)
    \nn \\
 &-& \left( {549 m_\pi^7 \over 560 p^6} +{3 m_\pi^5 \over 4 p^4}\right)
    {\rm ln}\left(1+{p^2 \over m_\pi^2}\right) +{4 \gamma \over 3} \left[{1
    \over 2}- {3 m_\pi^2 \over 4 p^2} + \left( {m_\pi^2 \over 4 p^2} +
    {3 m_\pi^4 \over 16 p^4} \right) {\rm ln}\left(1 + {4 p^2 \over m_\pi^2}
    \right) \right]^2 \Bigg\} \nn\,.
\end{eqnarray}
Values for the individual graphs are given in Eqs.~(\ref{ddb}) and (\ref{ddc}).

\begin{figure}[!t]
  \centerline{\epsfxsize=8.truecm \epsfbox{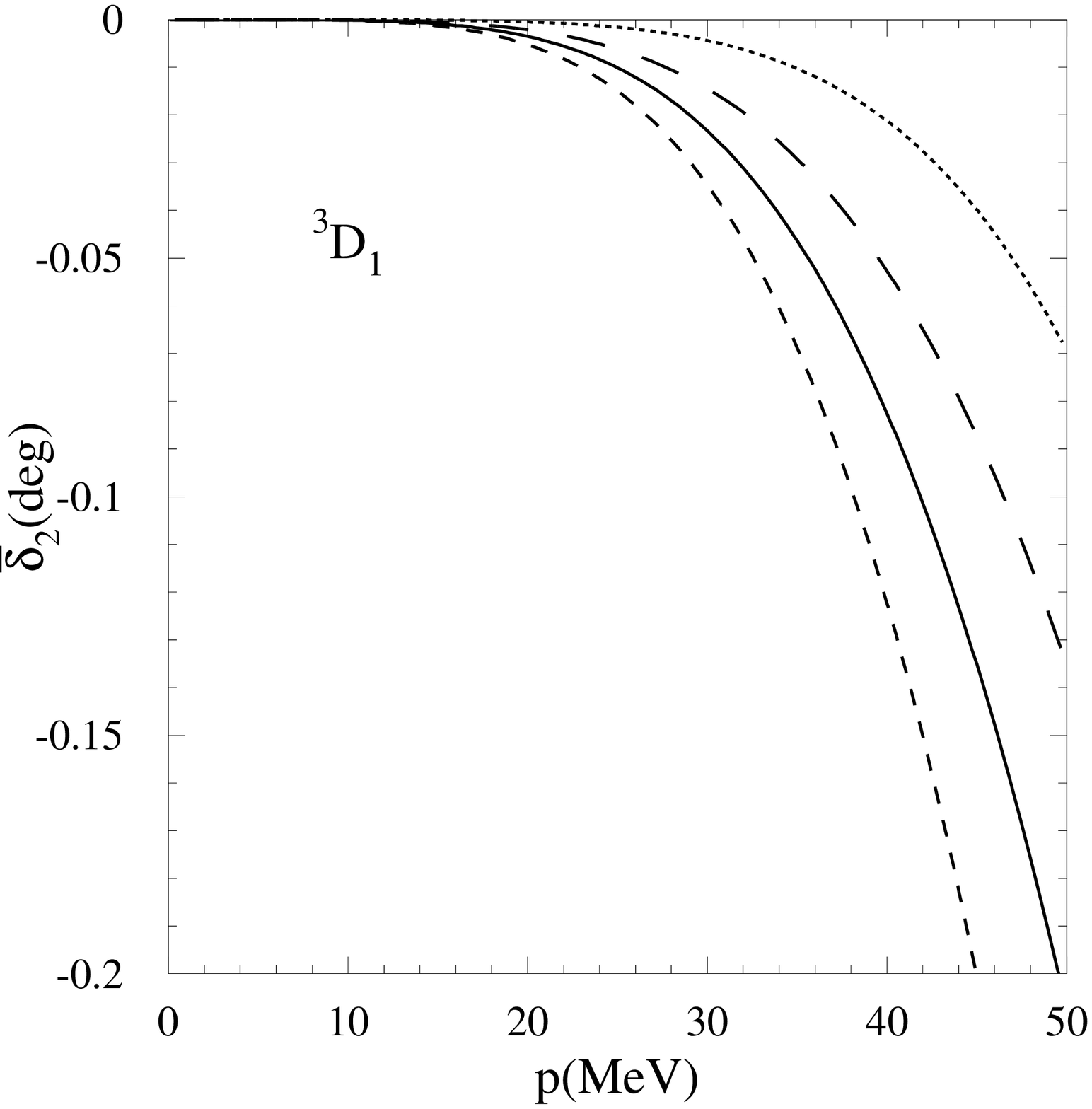}
       \epsfxsize=8.truecm \epsfbox{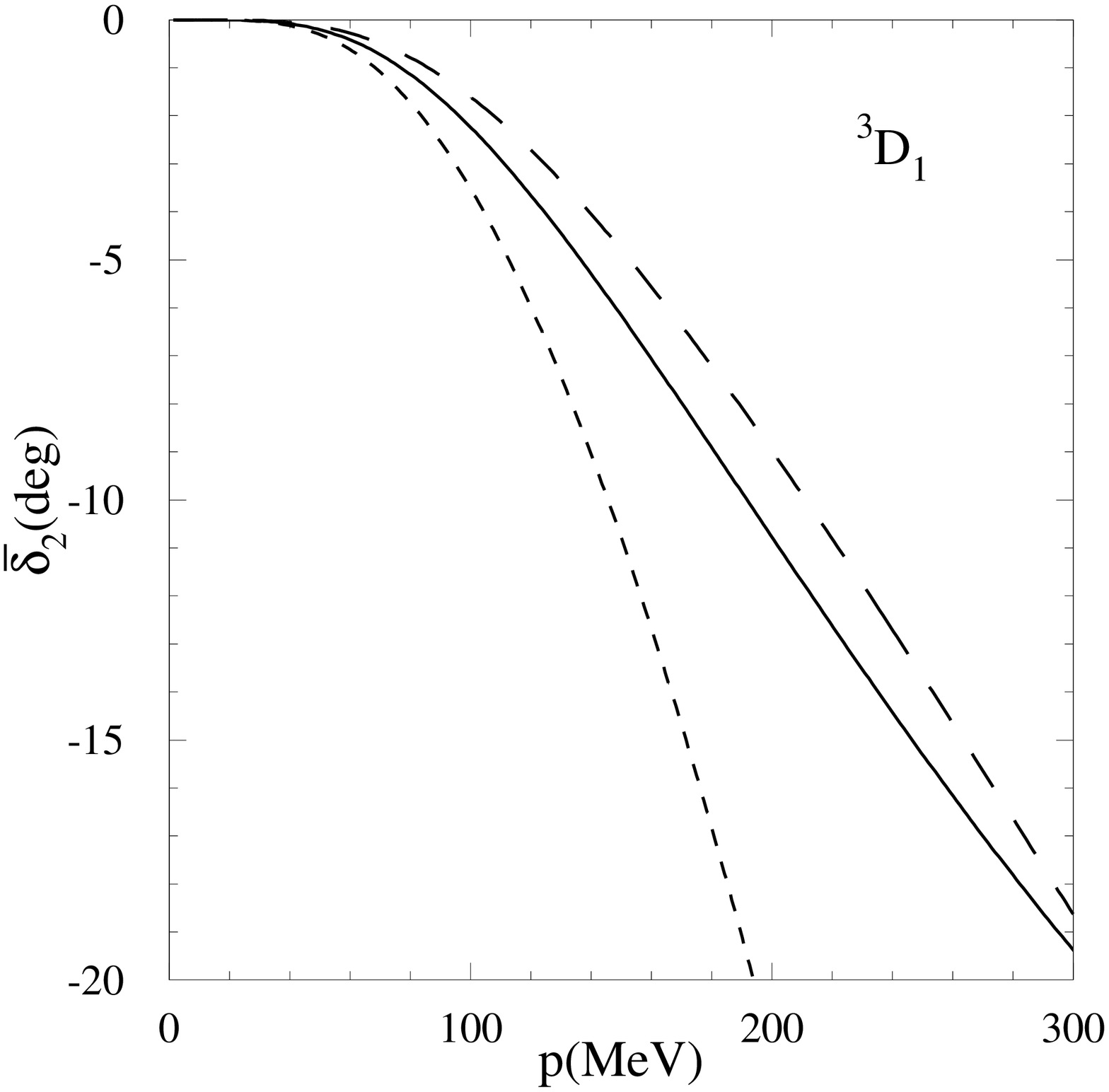}  } \vspace{.5cm}
{\tighten \caption[1]{$^3D_1$ phase shift for NN scattering.  The solid line is
from Nijmegen's multi-energy partial wave analysis\cite{Nij}. At LO
this phase shift is zero.  The long dashed line is the NLO result and the short
dashed line is the NNLO result.  In the first plot the NNLO prediction without
including the graph Fig.~\ref{figQD}c is displayed as a dotted line. The only
parameter entering at this order is $C_0^{(^3S_1)}$, which is fixed by the
location of the pole in the $^3S_1$ amplitude.} \label{fig_3d1}}
\end{figure}
The NLO and NNLO predictions for $\bar{\delta}_2$ are plotted in
Fig.~\ref{fig_3d1}, along with the result of the Nijmegen partial wave
analysis. The NLO result gives satisfactory agreement with data up to
$300\,{\rm MeV}$. The NNLO calculation is less accurate than the NLO
calculation especially for $p>50\,{\rm MeV}$. The error in the NNLO calculation
is always greater than the NLO calculation, so for this observable there is no
sign of convergence of the KSW expansion at any value of $p$.  At NNLO the
prediction for the $^3D_1$ phase shift suffers from the same problem as the
$^3S_1$ phase shift, namely a large term in the amplitude that grows linearly
with $p$ for $p\gg m_\pi$.  Taking $m_\pi\to 0$ we find
\begin{eqnarray}
   {M\over 4\pi} {\cal A}^{DD} &=& \frac{-1}{2\Lambda_{NN}} + \frac{1}
    {\Lambda_{NN}^2} \Big(\, \frac{i p}{4} + \frac{i\gamma p}{2(\gamma+ip)}
     -\frac{3\pi\,p}{14}\, \Big) \,.
\end{eqnarray}
The last term in this equation dominates the phase shift at large momenta.

Note that for low momentum, the inclusion of graph c) in Fig.~\ref{figQD}
improves the agreement over a theory which contains only perturbative pion
exchange. This can be seen in Fig.~\ref{fig_3d1} where the small dashed line
(NNLO with c)) lies closer to the Nijmegen phase shift (solid) than the dotted
line (NNLO without c)).

In this section we have presented calculations of the phase shifts and mixing angles
in the $^1S_0$, $^3S_1$, and $^3D_1$ channels at NNLO. We found that the $^1S_0$
phase shift agrees well with data up to $p \sim 400$ MeV. However, in the spin
triplet channels the effective field theory expansion does not seem to converge. The
$^3S_1-{}^3D_1$ mixing angle $\bar{\epsilon}_1$ agrees with data to within errors
for $p \lesssim m_\pi$. This is not true for the $^3S_1$ and $^3D_1$ phase shifts.
In these channels, two pion exchange graphs give corrections which worsen the
agreement with data. This suggests that the perturbative treatment of pions is
inadequate in spin triplet channels.


\section{$P$ and $D$ wave channels}

In this section we will examine the $^1P_1$, $^3P_{0,1,2}$, $^1D_2$, and
$^3D_{2,3}$ channels in an effort to get a better understanding of perturbative
pions. In these channels there is no order $Q^{-1}$ contribution, the $Q^0$
contribution consists solely of single pion exchange (Fig.~\ref{figQD}a), and
the order $Q$ contribution comes from the potential box diagram in
Fig.~\ref{figQD}b. Four nucleon operators only contribute at higher orders in
$Q$.  Since the coefficients of these operators are not enhanced by the
renormalization group flow near the fixed point they have a scaling determined
by dimensional analysis. In the P waves contact interactions first appear at
order $Q^2$, while in the $^1D_2$ and $^3D_{2,3}$ they first appear at order
$Q^4$.

\begin{figure}[!t]
  \centerline{\epsfysize=3.0truecm \epsfbox{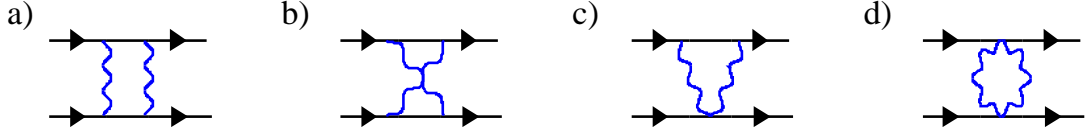}  }
 {\tighten
\caption[1]{Order $Q^2$ soft pion loop graphs for the $P$ and $D$ wave
channels.} \label{fig_soft} }
\end{figure}
In Ref.~\cite{kaiser} the phase shifts with $L\ge 2$ were calculated using
perturbative pion exchange. In this calculation, the one loop potential box,
soft diagrams, and a subset of order $Q^3$ corrections were included
simultaneously. The potential box in Fig.~\ref{figQD}b is order $Q$, while the
soft diagrams in Fig.~\ref{fig_soft} are order $Q^2$. At order $Q^2$ there are
also relativistic corrections and radiation pion contributions. The latter can
be absorbed by using the physical value of $g_A$ in the one-pion exchange
diagram. However, Ref.~\cite{kaiser} did not include the double potential box
\begin{eqnarray}
  \begin{picture}(25,20)(1,1)
   \put(-15,-6){\line(1,0){45}} \put(-15,14){\line(1,0){45}}
   \multiput(-8,-6)(0,7){3}{\line(0,1){4}}
   \multiput(8,-6)(0,7){3}{\line(0,1){4}} \multiput(24,-6)(0,7){3}{\line(0,1){4}}
  \end{picture} \qquad \sim\ \  \Big( \frac{g_A^2}{2 f^2} \Big)^3 \Big(
  \frac{M}{4\pi} \Big)^2 p^2 \,,
\end{eqnarray}
which is also order $Q^2$. Since a complete order $Q^2$ amplitude is not yet
available, no diagrams of order $Q^2$ or higher will be included in our
analysis.

The order $Q$, $Q^2$ phase shifts are given in terms of the amplitude by:
\begin{eqnarray}  \label{deltaos}
  {\delta}^{(1)} =\frac{Mp}{4\pi} \: {\cal A}_0^{(s)} \,, \qquad\qquad
  {\delta}^{(2)} =\frac{Mp}{4\pi}\:{\rm Re} \Big[ {\cal A}_1^{(s)} \Big]\,.
\end{eqnarray}
Projecting Fig.~\ref{figQD}a onto the various P and D waves using the projection
technique discussed in Appendix~\ref{AppProj} gives the results in
Eq.~(\ref{pd1pi}) which agree with Ref.~\cite{kaiser}. In these channels, the box
graph in Fig.~\ref{figQD}b can be evaluated analytically using the techniques
discussed in Appendix B. We have instead chosen to calculate the partial wave
amplitudes by using the expression for the box graph given in Ref.~\cite{kaiser},
and doing the final angular integration numerically.

\begin{figure}[!t]
  \centerline{\epsfxsize=8.truecm \epsfbox{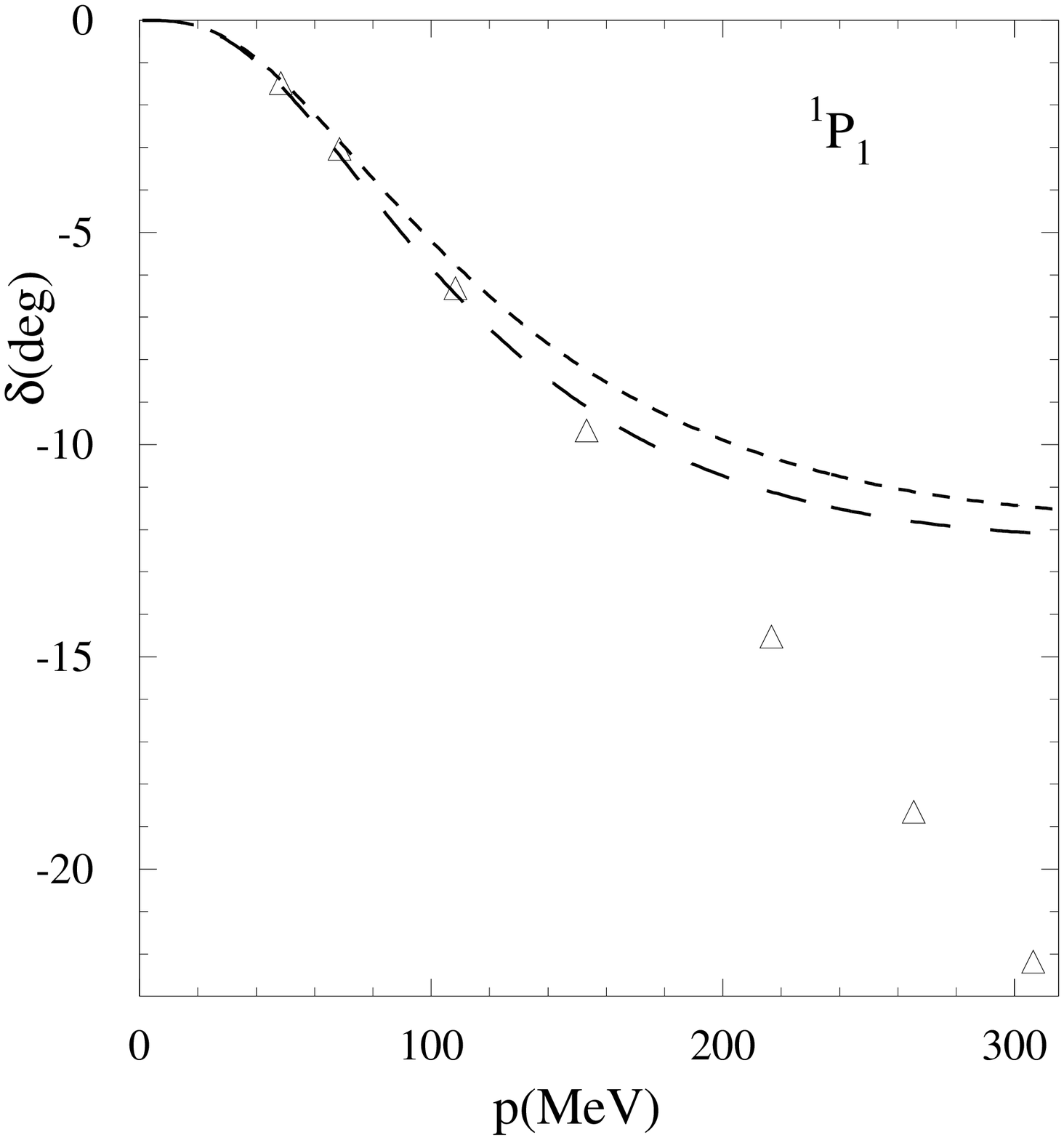}
       \epsfxsize=8.truecm \epsfbox{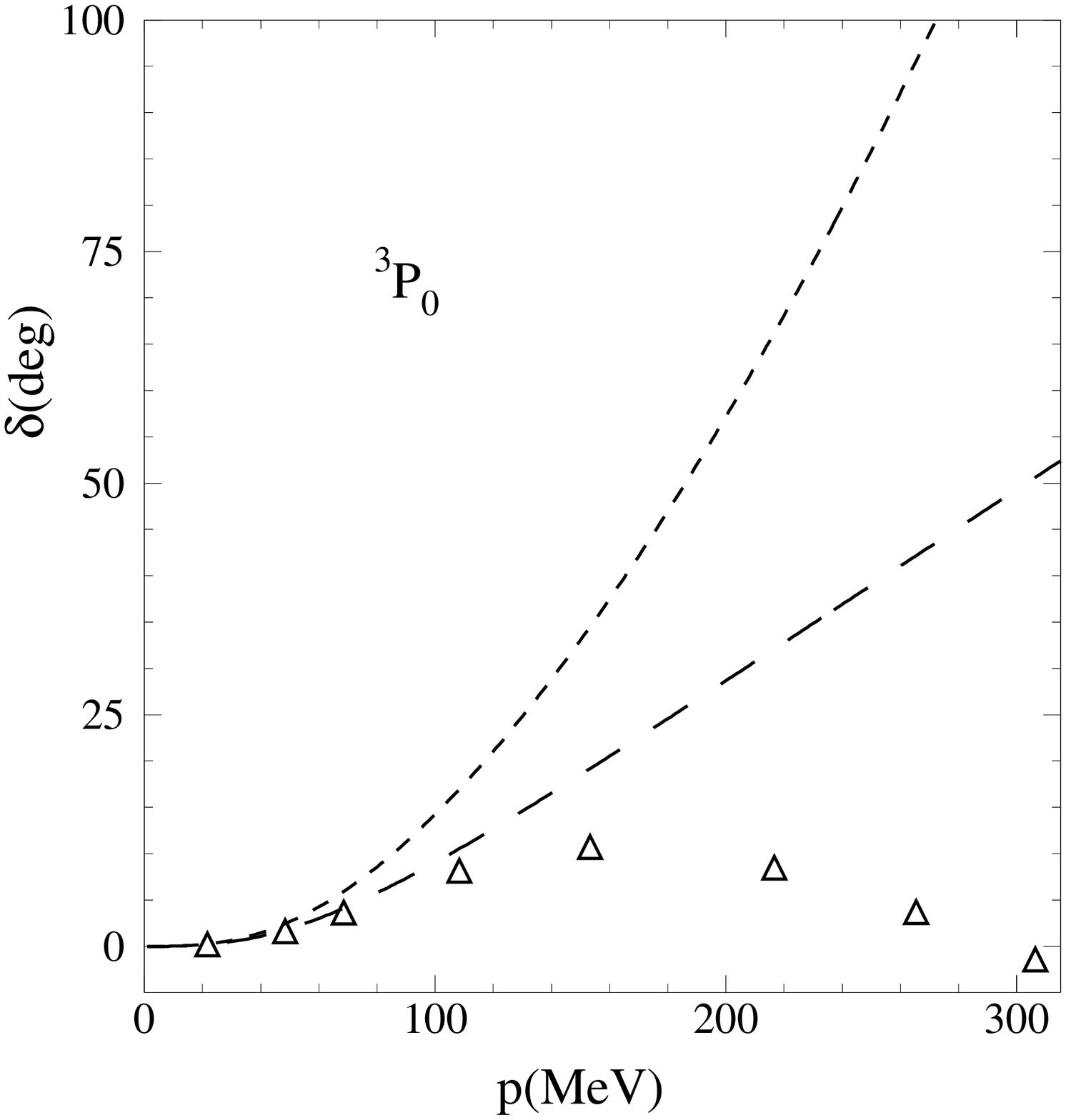}  } \vspace{.5cm}
  \centerline{\epsfxsize=8.truecm \epsfbox{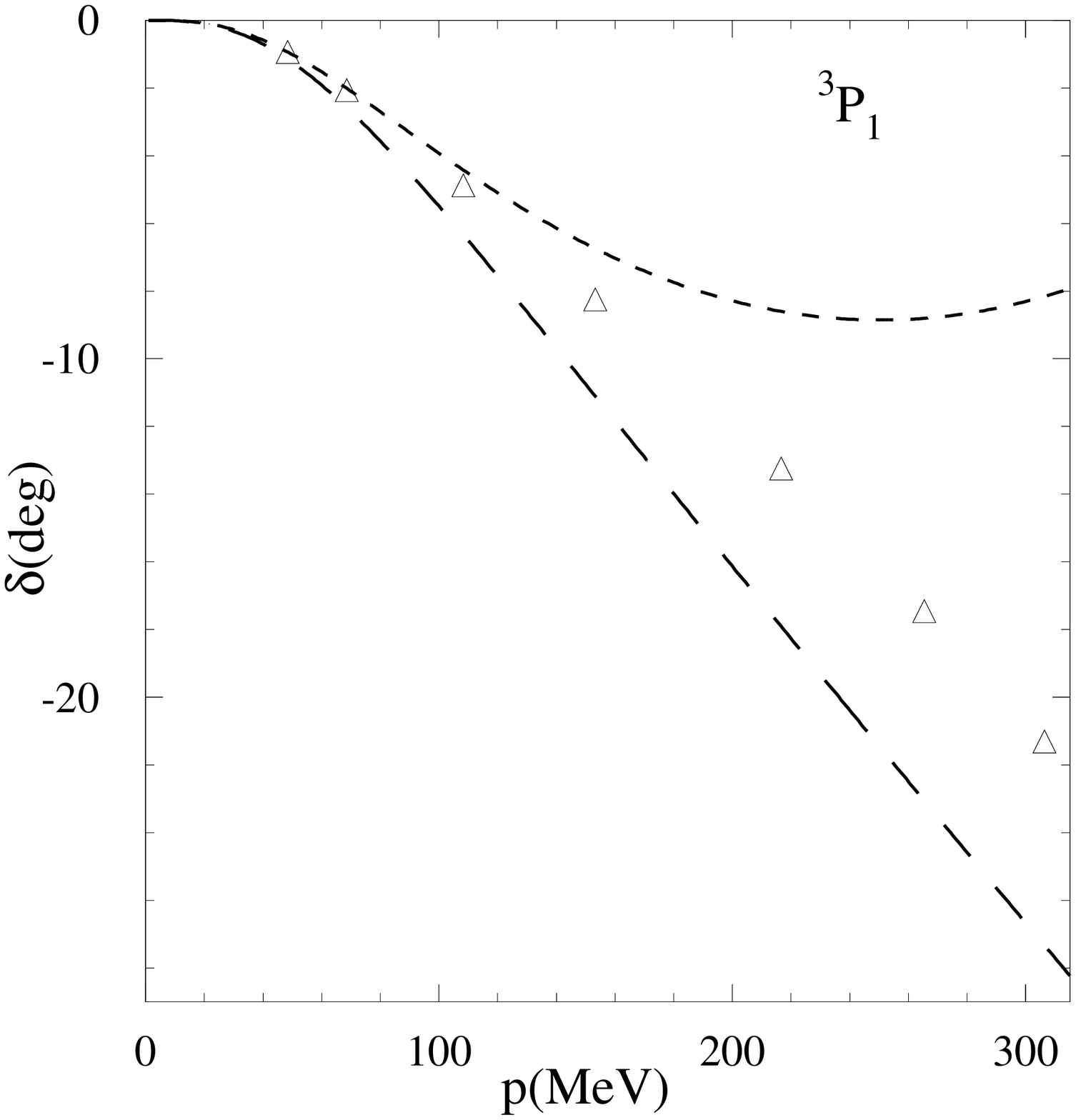}
       \epsfxsize=8.truecm \epsfbox{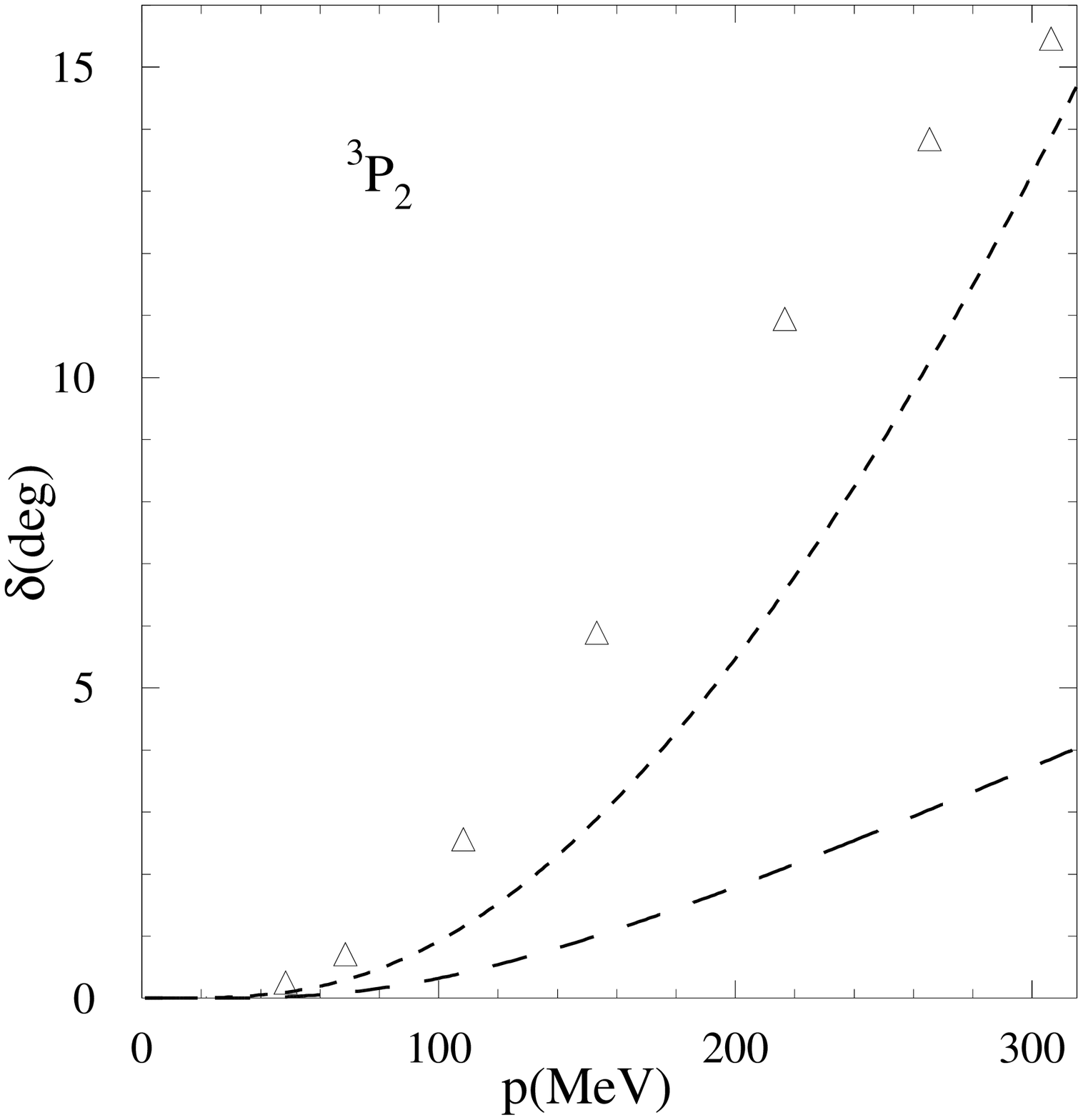}  }
{\tighten \caption[1]{P-wave phase shifts for NN scattering. The triangles are
values from Nijmegen's single energy analysis and have errors that are
invisible on the scale shown. At LO these phase shifts are zero. The
long dashed line is the NLO result and the short dashed line is the NNLO
result.   There are no free parameters at this order.} \label{fig_pwaves}}
\end{figure}
\begin{figure}[!t]
  \centerline{\epsfxsize=8.truecm \epsfbox{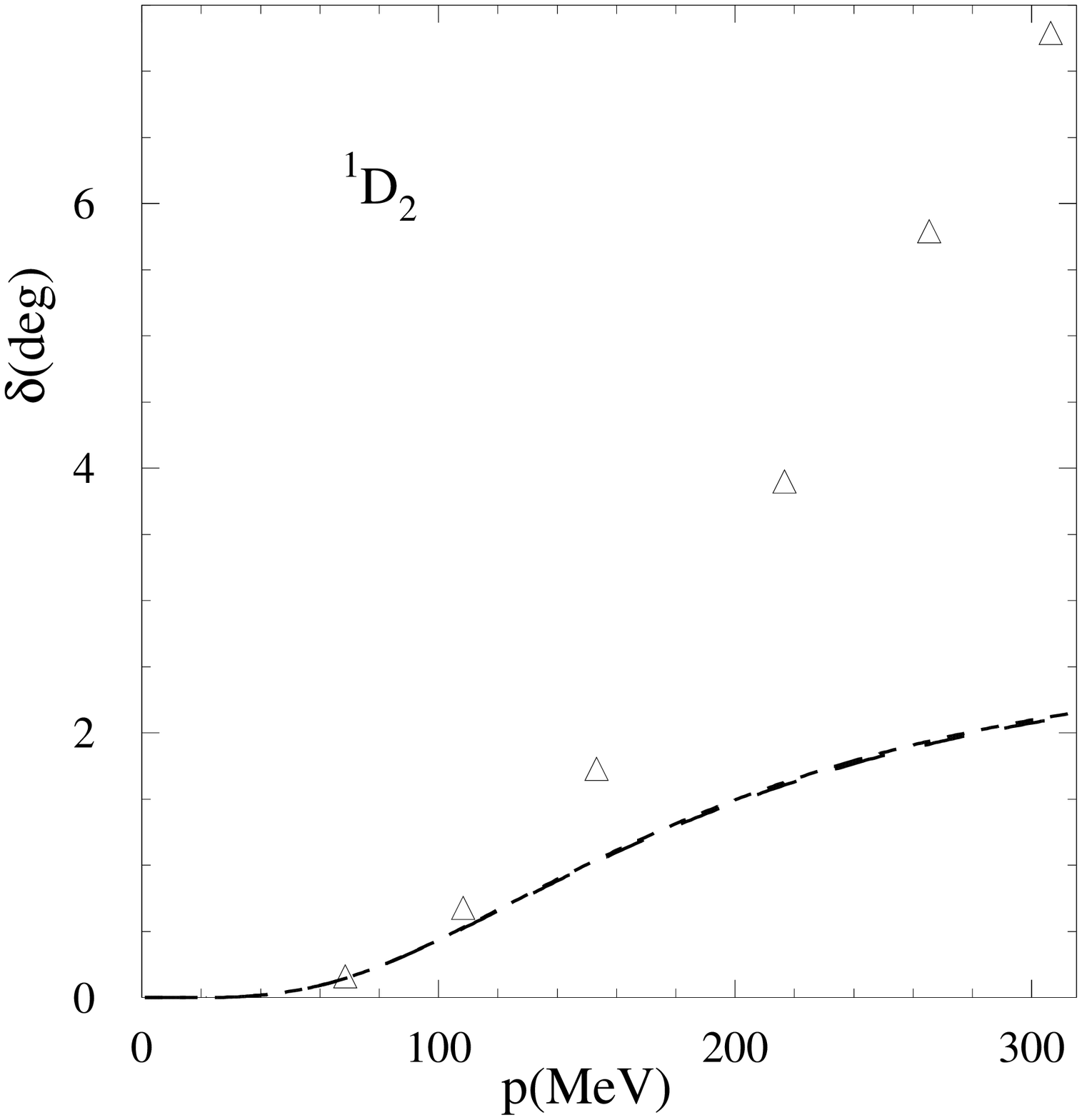}
       \epsfxsize=8.truecm \epsfbox{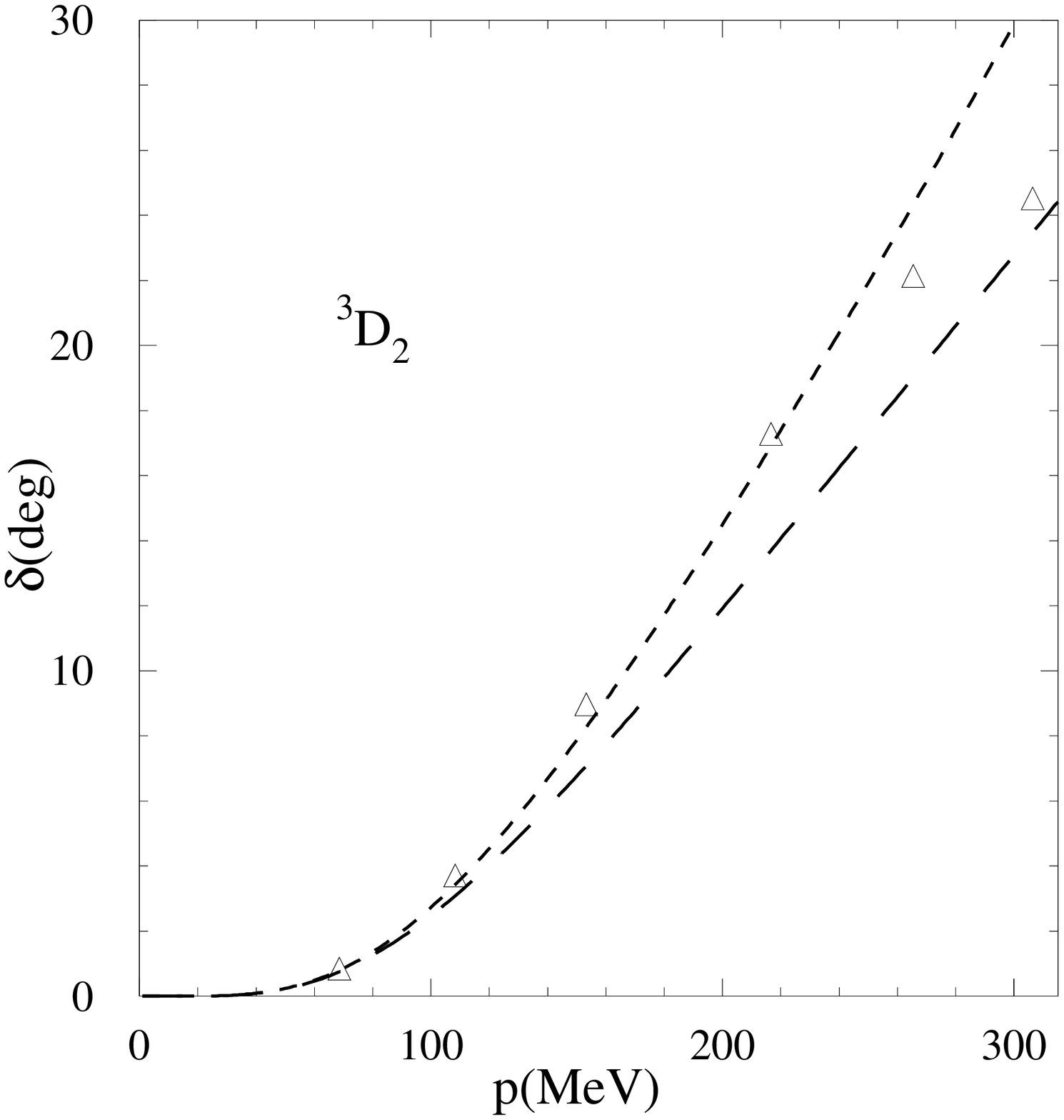}  } \vspace{.5cm}
  \centerline{\epsfxsize=8.truecm \epsfbox{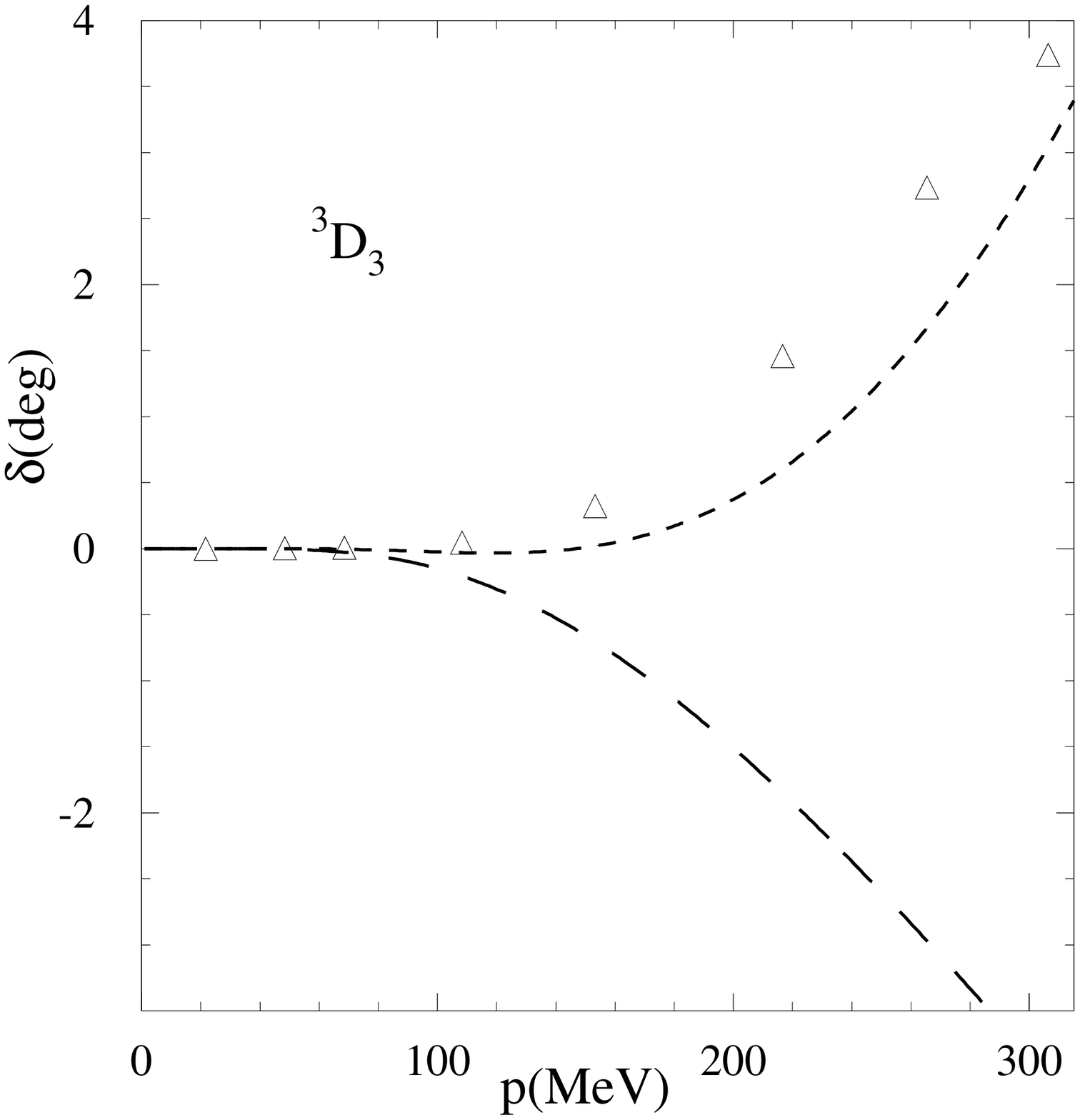} }
{\tighten \caption[1]{D-wave phase shifts for NN scattering. The triangles are
values from Nijmegen's single energy analysis and have errors that are
invisible on the scale shown. At LO these phase shifts are zero. The
long dashed line is the NLO result and the short dashed line is the NNLO
result.  There are no free parameters at this order.} \label{fig_dwaves} }
\end{figure}
Results for the P and D wave amplitudes are given in Figs.~\ref{fig_pwaves} and
\ref{fig_dwaves} respectively.  The potential box diagram gives a very small
contribution in the singlet channels, in contrast to the triplet channels.  For
momenta $p\lesssim 110\,{\rm MeV}$ the NNLO calculation gives reasonable agreement
in the $^1P_1$, $^3P_1$, $^1D_2$, and $^3D_{2,3}$ channels, but not in the
$^3P_{0,2}$ channels. For larger momentum, $p \sim 300\,{\rm MeV}$, the error in
the $^1P_1$, $^3P_{0,1}$, and $^1D_2$ channels is very large.  This is less
of a concern because the KSW power counting is not designed to work for $p \gg
m_\pi$, but it does indicate a need to modify the KSW power counting for momentum
greater than the pion mass.

To get a better idea of what is happening at large momenta it is useful to
look at the $m_\pi\to 0$ limit of the amplitudes:
\begin{eqnarray} \label{nonan}
   {M\over 4\pi} {\cal A}(^1P_0) &=& 0 \,, \\
   {M\over 4\pi} {\cal A}(^3P_0) &=& \frac{1}{\Lambda_{NN}} + \frac{1}
    {\Lambda_{NN}^2} \Big(\, i p+\frac{2\pi\,p}{5}\, \Big) \,, \nn \\
   {M\over 4\pi} {\cal A}(^3P_1) &=& \frac{-1}{2\Lambda_{NN}} + \frac{1}
    {\Lambda_{NN}^2} \Big(\, \frac{i p}{4}+\frac{\pi\,p}{10}\, \Big) \,, \nn\\
   {M\over 4\pi} {\cal A}(^3P_2) &=& \frac{1}{10\Lambda_{NN}} + \frac{1}
    {\Lambda_{NN}^2} \Big(\, \frac{i p}{60}+\frac{3\pi\,p}{50}\, \Big) \,,\nn\\
   {M\over 4\pi} {\cal A}(^1D_2) &=& 0 \,, \nn\\
   {M\over 4\pi} {\cal A}(^3D_2) &=& \frac{1}{2\Lambda_{NN}} + \frac{1}
    {\Lambda_{NN}^2} \Big(\, \frac{i p}{4}+\frac{3\pi\,p}{70}\, \Big)\,,\nn\\
   {M\over 4\pi} {\cal A}(^3D_3) &=& \frac{-1}{7\Lambda_{NN}} + \frac{1}
    {\Lambda_{NN}^2} \Big(\, \frac{i p}{28}+\frac{3\pi\,p}{49}\, \Big) \,.\nn
\end{eqnarray}
\begin{table}[t!]
\begin{center} \begin{tabular}{ccc|ccccccc}
 & $p=300\,{\rm MeV}$ &  & $^3P_0$ & $^3P_1$ & $^3P_2$ &  $^3D_2$ & $^3D_3$ \\
  \hline
 & $\delta^{(2)}$ & & $70^\circ$ & $18.4^\circ$ & $9.6^\circ$ & $7.1^\circ$
   & $6.6^\circ$  \\
 & $\lim_{m_\pi\to 0} \delta^{(2)}$ & & $75^\circ$ & $18.8^\circ$ & $11.3^\circ$
   & $8.0^\circ$ & $11.5^\circ$
\end{tabular} \end{center}
{\tighten \caption{Comparison of the NNLO part of the phase shift to its $m_\pi\to 0$ limit at $p=300\,{\rm MeV}$.} \label{compare} }
\end{table}
For the spin singlet channels there are no corrections which grow with $p$,
while the spin triplet channels have non-analytic corrections proportional to
$\pi p$. This short distance behavior is similar to what is seen in Section
III.  At $p = 300\,{\rm MeV}$ these particular non-analytic terms dominate all
other NNLO corrections as can be seen from Table~\ref{compare}.  In the $^3P_2$,
$^3D_{2,3}$ channels these corrections improve the agreement with data, while in
the $^3P_{0,1}$ channels they do not.

At lower momenta $p\sim 50\,{\rm MeV}$ the effective theory does a better job
of reproducing the phase shifts.  Therefore, it seems possible that in these
channels predictions for terms in the effective range expansions,
$p^{2L+1}\cot\delta = -1/a + r_0 p^2/2+ \ldots$, might work fairly well.
Equivalently one can match onto the theory with pions integrated out to make
predictions for the coefficients of four nucleon operators in the P and D
waves. Such an investigation is beyond the scope of this paper.

\section{Discussion}  \label{discuss}

In this section we summarize our results for the $S$, $P$, and $D$ wave phase
shifts.  We also discuss in greater detail the nature of the perturbative expansion
in the spin singlet and triplet channels.

Errors in each channel at $p = 50\,{\rm MeV}$ and $p=m_\pi$ are given in
Table~\ref{perrs}. For an expansion parameter of 1/2, we expect roughly 50\%
error at NLO ($Q^0$), and 25\% error at NNLO ($Q$). (For the
two S-wave phase shifts, which start at one lower order in the expansion, the
expected error at NLO and NNLO is 25\% and 12.5\%, respectively.) At $p =
m_\pi$, errors are significantly larger than expected in the $^3S_1$,
$^3P_{0,2}$, and $^3D_1$ channels. In the case of the $^3D_3$ channel the
percent error is exaggerated due to the smallness of the phase shift, so in this
channel the percent error is probably not a figure of merit for examining the
quality of the expansion. However, the LO correction in this channel has the
wrong sign so there is no sign of convergence of the perturbative expansion.  At
$p \sim m_\pi$, the performance of the effective theory is erratic, working some
but not all of the time. The overall agreement with data at $50\,{\rm MeV}$ is
better, but there are still channels ($^3P_{0,2},{}^3D_1$) in which the agreement
with data is worse than one expects.
\begin{table}[t!]
\begin{center} \begin{tabular}{ccc|ccccccccc}
 $p=50\,{\rm MeV}$ & $^1S_0$ & $^3S_1$ & $^3S_1-{}^3D_1$ & $^1P_1$ & $^3P_0$ &
 $^3P_1$ & $^3P_2$ & $^1D_2$ & $^3D_1$ & $^3D_2$ & $^3D_3$ \\ \hline
 NLO & 0.4\% & 0.2\% & 42\% & 4\% & 10\% & 23\% & 90\% & 3\% & 35\% & 9\% &
   -320\%$^*$ \\
 NNLO & 0.2\% & 0.1\% & 14\% & 5\% & 50\% & 0.2\% & 61\% & 4\% & 48\% & 5\% &
    88\%$^*$ \\ \hline\hline
 $p=137\,{\rm MeV}$ & $^1S_0$ & $^3S_1$ & $^3S_1-{}^3D_1$ & $^1P_1$ & $^3P_0$ &
 $^3P_1$ & $^3P_2$ & $^1D_2$ & $^3D_1$ & $^3D_2$ & $^3D_3$ \\ \hline
 NLO & 17\% & 0.4\% & 25\% & 3\% & 54\% & 32\% & 83\% & 34\% & 24\% & 19\% &
   -370\%$^*$ \\
 NNLO & 0.3\% & 36\% & 19\% & 13\% & 170\% & 15\% & 52\% & 33\% & 70\% & 8\% &
   -110\%$^*$
\end{tabular} \end{center}
{\tighten \caption{Percent errors in the phase shifts for $p=50\,{\rm MeV}$ and
$p=137\,{\rm MeV}$ at NLO and NNLO. ($^*$ Since the $^3D_3$ phase shift is
close to zero the percent errors are not very meaningful. The absolute errors
for this phase shift have the expected size.)} \label{perrs} }
\end{table}

\begin{figure}[!t]
   \centerline{\epsfysize=2.0 cm \epsfbox{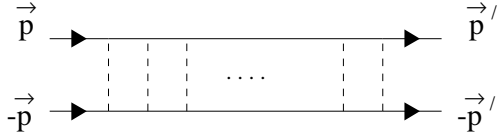}  }
 { \caption[1]{Potential pion ladder diagrams.} \label{fig_ladder} }
\end{figure}

The perturbative expansions in the spin triplet and singlet channels are
qualitatively different. All triplet channels have non-analytic corrections
that grow with p, while the singlet channels do not. This can be understood as
follows. First consider the spin singlet channel. In this channel, the
potential due to one pion exchange is the sum of a delta function and a Yukawa
potential,
\begin{eqnarray}
  {\vec q \cdot \vec \sigma_{\alpha\beta}\ \vec q \cdot \vec
 \sigma_{\gamma\delta}
  \over  \vec q\,^2 +m_\pi^2 }\quad \rightarrow \quad 1 \ \ -\ \  {m_\pi^2
 \over
    (\vec q\,^2+ m_\pi^2 )} \,. \label{ppip}
\end{eqnarray}
The effect of the delta function part of one pion exchange is indistinguishable
from the $C_0^{(^1S_0)}$ operator and therefore only contributes to S-wave
scattering. A well known theorem from quantum mechanics shows that at large
energy the Born approximation becomes more accurate for a Yukawa potential.
From the point of view of the field theory, this means that the ladder graphs
shown in Fig.~\ref{fig_ladder} with Yukawa exchange at the rungs are suppressed
by powers of the momentum.
Using dimensional analysis we see that adding a Yukawa rung gives a factor of
\begin{eqnarray}
 \frac{g_A^2 M}{2 f^2} \: \int {d^n k \over (2\pi)^n} { m_\pi^2 \over
   ({\vec k}\,^2+  2\vec k \cdot \vec p\, )\ ({\vec k}\,^2 + m_\pi^2)
 \cdots }
   \ \ \sim\ \ \frac{m_\pi^2}{p\:\Lambda_{NN}}
\end{eqnarray}
for $p \gg m_\pi$. The one loop pion box diagram in the spin singlet channel gives a
contribution which can be eliminated by a shift in $C_0^{(^1S_0)}$ and other terms
that are suppressed by powers of $m_\pi/p$ (see for e.g., Eq.~(\ref{2p1s0})). Once
the short distance effects of pions are absorbed into $C_0^{(^1S_0)}$, the
remaining piece of two pion exchange is never larger than the estimate given in
Eq.~(\ref{pilps}) and gets smaller as $p$ increases. This is good for the
convergence of the perturbative expansion because it means higher order potential
pion corrections in singlet channels will be well behaved.  In fact the tree level
pion exchange graph gives almost the same prediction for the $^1P_1$ and $^1D_2$
phase shifts as the leading order prediction in the Weinberg expansion that sums
potential pions to all orders.  Thus, the evidence for the behavior of the
spin singlet channels is independent of how the parameters in the $^1S_0$ channel 
are fit to the data.

The S, P and D wave phase shifts are calculated to NNLO within the Weinberg
expansion in Refs.~\cite{ork} and \cite{Ulf}.  These studies are complementary since
Ref.~\cite{ork} gives the uncertainty in their NNLO predictions by varying the
cutoff from $0.5\,{\rm GeV}$ to $1.0\,{\rm GeV}$, while Ref.~\cite{Ulf} explicitly
displays their LO, NLO and NNLO results (which are respectively order $p^0$, $p^2$,
and $p^3$ in the potential). In comparing our results with those of
Refs.~\cite{ork,Ulf} it must be noted that these calculations include many effects
which do not appear until higher order in the KSW expansion. For example, P-wave
contact interactions are included at NLO so the predictions in these channels have a
free parameter which is fit to the data. Soft pion effects\footnote{\tighten Soft
pion diagrams with nucleons and $\Delta$'s were calculated in
Refs.~\cite{kaiser,kaiser2}} also enter at this order.  These effects enter at
order $Q^2$ in the KSW power counting (${\rm N}^3{\rm LO}$). In the Weinberg
expansion the singlet and some triplet phase shifts cannot be fit until these
interactions are included \cite{ork,Ulf}.

At $p=306\,{\rm MeV}$ the LO result in Ref.~\cite{Ulf} is
$\delta(^1P_1)=-12^\circ$ and $\delta(^1D_2)=2.2^\circ$ which is very close to
tree level pion exchange which gives $\delta(^1P_1)=-12^\circ$ and
$\delta(^1D_2)=2.1^\circ$. Thus, as expected, the discrepancy between theory
and experiment seen in the $^1P_1$ and $^1D_2$ channels in
Figs.~\ref{fig_pwaves} and \ref{fig_dwaves} is not removed by summing potential
pion diagrams.     The LO predictions in the Weinberg expansion are shown in
Table~\ref{loW}.  In the $^1S_0$ channel the result in Table~\ref{loW} is only
slightly better than the LO result in Fig.~\ref{fig_fits}a.  We conclude that
there is little to be gained by summing potential pions in spin-singlet
channels.
\begin{table}[t!]
\begin{center} \begin{tabular}{c|ccccccccccc}
 phase shifts & $^1S_0$ & $^3S_1$ & $\bar\epsilon_1$ & $^1P_1$ & $^3P_0$ &
 $^3P_1$ & $^3P_2$ & $^1D_2$ & $^3D_1$ & $^3D_2$ & $^3D_3$ \\ \hline
 $p=153\,{\rm MeV}$  & $68^\circ$ & $66^\circ$ & $3.2^\circ$
   & $-8.6^\circ$ & $53^\circ$ & $-9.1^\circ$ & $2.0^\circ$ & $1.1^\circ$
   & $-6.4^\circ$ & $9.0^\circ$ & $-0.2^\circ$ \\
 $p=306\,{\rm MeV}$ & $57^\circ$ & $30^\circ$ & $8.4^\circ$ & $-12^\circ$
   & $73^\circ$ & $-22^\circ$ & $5.6^\circ$ & $2.2^\circ$ & $-28^\circ$
   & $31^\circ$ & $-1.2^\circ$ \\
\end{tabular} \end{center}
{\tighten
\caption{Predictions for the phase shifts when diagrams with
insertions of $C_0^{(s)}$ and potential pions are summed to all orders.  The
results shown are the leading order predictions from Ref.~\protect\cite{Ulf}. }
\label{loW} }
\end{table}

In the spin triplet channel, the potential from one pion exchange is much more
singular and has terms that go like $1/r^3$ for small $r$ (where $r$ is the
nucleon separation). In fact, without introducing an ultraviolet cutoff, it is
not possible to solve the Schr\"{o}dinger equation for such a singular potential.
In field theory this means that higher potential pion ladder graphs have
ultraviolet divergences of the form $p^{2 m}/\epsilon$, which must be cancelled
by a four nucleon operator with $2 m$ derivatives.  (Examples of two loop graphs
with $p^2/\epsilon$ divergences were computed in Ref.~\cite{ms1}.)  In the spin
triplet channel perturbative pions give corrections which go like
$(p/\Lambda_{NN})^k$, where $k$ is the number of loops.  Loop graphs with pions
in the spin triplet channel can therefore have finite corrections which grow with
$p$ and are non-analytic in $p^2$.  These short distance pion corrections cannot
be compensated by any short distance operator.  For the $^3S_1$, $^3P_0$, and
$^3D_1$ channels the non-analytic corrections are large and ruin the agreement
with the data.  In the $^3P_2$ and $^3D_3$ channels the quality of the expansion
is poor because the non-analytic correction makes the NLO and NNLO corrections
comparable in size.

In the $^3S_1$ channel calculations within the Weinberg approach\cite{ork,Ulf} have
small cutoff dependence and agree much better with the Nijmegen partial wave
analysis than Fig.~\ref{fig_3s1}.  The $^3P_1$ and $^3D_1$ channels are also in good
agreement.  In these channels the summation of potential pions improves the
agreement with data.  However, in other $P$ and $D$ wave channels the summation of
potential pions is not as helpful. At LO Ref.~\cite{Ulf} finds large disagreement in
the $^3P_0$ and $^3P_2$ channels as can be seen from Table~\ref{loW}. These
predictions are similar to what is given by tree level pion exchange.  At NNLO the
$^3P_0$ phase shift is in reasonable agreement with the data with small cutoff
dependence\cite{ork}. In the $^3P_2$ channel there is larger cutoff
dependence\cite{ork} and no sign of convergence of the perturbative expansion at
$p\sim 300\,{\rm MeV}$\cite{Ulf}.  In the $^3P_{0,2}$ channels agreement with data
is only achieved at the order that a free parameter appears. Soft pion graphs and
four nucleon operators appear to be more important than summing potential pions. In
the $^3D_{2,3}$ channels at $p\sim 300\,{\rm MeV}$ our NNLO prediction is of similar
quality to the NNLO prediction in Ref.~\cite{Ulf}, so the summation of potential
pions does not seem to be necessary.

The large NNLO corrections in the $^3S_1$, $^3D_1$, and $^3P_{0,2}$ channels
cast considerable doubt on the effectiveness of the KSW power counting for
pions.  The $\sim 10\%$ accuracy of NLO results \cite{oNLO} remains somewhat
mysterious. For momenta $p< m_\pi$ the pion can be integrated out. This low
energy theory has been shown to be effective in calculations at NNLO
\cite{nopi,nopi2,resd}.

Large perturbative corrections from two pion exchange suggest that a nonperturbative
treatment of pions is necessary for nuclear two body problems in some spin triplet
channels.  This is achieved in Weinberg's power counting \cite{weinberg} because the
potential pion exchange diagrams are summed at leading order.  However, the graphs
which are resumed by solving the Schr\"{o}dinger equation have logarithmic
divergences of the form $m_\pi^{2m}~{\rm ln}(\Lambda)$ or $p^{2m}~{\rm ln}(\Lambda)$
(this has been shown explicitly for the case $m=1$ in Refs.~\cite{ksw2} and
\cite{ms2}). The short distance counterterms necessary to cancel the ${\rm
ln}(\Lambda)$ dependence of these graphs are not included until higher order (for
$m=1$ this would be $C_2 p^2$ and $D_2 m_\pi^2$).  The residual cutoff dependence is
of the same size as higher corrections.  However, it is not a priori clear why the
contribution of the graphs included in the summation is larger than the omitted
counterterms (see for instance Ref.~\cite{lm}).

We have seen that in spin triplet channels there are perturbative corrections
which survive in the $m_\pi \rightarrow 0$ limit, and are enhanced by large
numerical factors (the $\pi$'s in Eq.~(\ref{nonan})). Empirically these terms
tend to dominate the NNLO correction. Furthermore, these corrections are
non-analytic in $p^2$. Since no unknown counterterm can contribute to their
coefficient they can be calculated unambiguously. It would be interesting to see
if there are similar large calculable corrections at higher orders. If so, then
the advantage of the Weinberg approach relative to the method used here is that
it sums these important contributions along with smaller scheme dependent
corrections.  In three body problems \cite{3bdy}, a power counting similar to KSW
gives accurate results at very low energies. In these computations, the
perturbative treatment of pions and higher derivative operators is crucial
because it renders the calculations more tractable (see Ref.~\cite{gp}) than
conventional potential model approaches. For this reason, an approach to
two body forces which sums genuinely large calculable corrections from pion
exchange analytically or semi-analytically would be worth pursuing.

In this paper we extended calculations of the $^1S_0$, $^3S_1$, and $^3D_1$ phase
shifts to NNLO in the KSW power counting, including a complete calculation of
radiation pion contributions. At this order the predictions for the $P$ wave and
remaining $D$ wave channels were also examined.  In spin singlet channels a
perturbative treatment of potential pions is justified.  The large disagreement for
the $^3S_1$ phase shift provides an unambiguous indication that the KSW expansion
for pions needs to be modified.  This is supported by the failure in the $^3P_0$
channel and the lack of convergence at $p\sim m_\pi$ in the $^3D_1$, $^3P_2$, and
$^3D_3$ channels.

S.F. was supported in part by NSERC and wishes to thank the Caltech theory group
for their hospitality.  T.M and I.W.S. were supported in part by the Department
of Energy under grant numbers DE-FG03-92-ER 40701 and DOE-FG03-97ER40546.


\newpage \phantom{stuff}

\appendix
\section{Partial Wave Projection Technique} \label{AppProj}

In this Appendix we discuss a method for obtaining the contribution of a
Feynman diagram to a particular partial wave amplitude. We use a trace
formalism which allows us to project out the partial wave amplitude before
doing loop integrations.  This approach has the advantage of being well adapted
to situations in which spin (and isospin) traces should be performed in
$n=3-2\epsilon$ dimensions.

Consider the process $N(\vec k/2 -\vec p\,) N(\vec k/2 +\vec p\,) \to N(\vec k/2
+\vec p\,'\,) N(\vec k/2 -\vec p\,'\,)$.  We begin by defining two nucleon
states\cite{nopi}
\begin{eqnarray} \label{NNstate}
 \left| NN(s; \vec k, p) \right\rangle= \frac{p}{\sqrt{4\pi}} \frac1{(2\pi)^3}
  \int d\Omega_{\vec p} \left[ N^T(\vec k/2+\vec p) P^{(s)}
   N(\vec k/2-\vec p) \right]^{\dagger} \: |\, 0\, \rangle \,,
\end{eqnarray}
where $s={}^{2S+1}L_J$, $p=|\vec p\,|$, and the matrix $P^{({}^{2S+1}L_J)}$
projects onto the desired partial wave.  The normalization of the states in
Eq~(\ref{NNstate}) is chosen so that averaging over polarizations
\begin{eqnarray}
 \sum_{\rm pol.\ avg} \left\langle NN(\,s'; \vec k', p'\,) \:\right.  \left|\:
  NN (\,s; \vec k, p )  \right\rangle = \delta^3(\vec k'-\vec k)\:
  \delta(p'-p)\: \delta^{s's} \,,
\end{eqnarray}
with the projection matrices satisfying
\begin{eqnarray}
 \sum_{\rm pol.\ avg}\:  {\rm Tr}[ P^{(s)}  P^{(s)\dagger}\,]= \frac12 \,.
\end{eqnarray}
Here Tr denotes a trace over spin and isospin. Evaluating the traces in $n$
dimensions gives the following normalization to the projection matrices for the
S, P, and D waves:
\begin{eqnarray} \label{proj2}
 && P^{({}^1\!S_0)} = { (i\sigma_2) \, (i\tau_2 \vec\tau\cdot \vec
 \epsilon_{\mbox{\tiny I}} ) \over 2\sqrt{2} }\,,
 \qquad\qquad\qquad\qquad\qquad\quad
  P^{({}^3\!S_1)} = { (i\sigma_2\, \vec\sigma\cdot \vec\epsilon  ) \, (i\tau_2)
  \over 2\sqrt{2} }\,,  \\
 && P^{({}^1P_1)} =  {\sqrt{n}\ {\hat p}\cdot{\vec\epsilon}\ (i\sigma_2)\,
     (i\tau_2)\over 2\sqrt{2} }\,, \qquad\qquad\qquad\qquad\quad
  P^{({}^3P_0)} = { (i\sigma_2\, \vec\sigma \cdot {\hat p}\,) \, (i\tau_2\,
   \vec\tau \cdot {\vec \epsilon_{\mbox{\tiny I}} }\,) \over 2\sqrt{2} }\,,
   \nn \\
 && P^{({}^3P_1)} = {\sqrt{n}\: \epsilon^{ijk}\:\epsilon^i\: {\hat p}\,^j\:
   (i\sigma_2\, \sigma^k)\, (i\tau_2\, \vec\tau\cdot {\vec\epsilon_{
   \mbox{\tiny I}}}\,)\over 4} \,, \qquad\qquad\!\!\!
  P^{({}^3P_2)} = {\sqrt{n}\ \epsilon^{ij}\: {\hat p}\,^i\: (i\sigma_2\,
   \sigma^j)\, (i\tau_2\,\tau\cdot {\vec\epsilon_{\mbox{\tiny I}}}\,) \over
   2\sqrt{2} }\,,\nn \\
 && P^{({}^1D_2)} = { \sqrt{n(n+2)}\: \epsilon^{ij}\, (i\sigma_2)\, (i\tau_2\,
   \vec\tau \cdot \vec \epsilon_{\mbox{\tiny I}} ) \over 4 }\ {\hat p}\,^i\:
   {\hat p}\,^j\,, \qquad\quad\!\!\!\!\!\!
   P^{({}^3D_1)} =\frac{n (i\sigma_2\,\sigma^i \epsilon^j)\,(i\tau_2)}
  {2\sqrt{2(n-1)}} \: \Big( {\hat p}\,^i {\hat p}\,^j -\frac{\delta^{ij}}{n}
  \Big) \,, \nn\\
&& P^{({}^3\!D_2)} \! = \! {\sqrt{n+2}\:\epsilon^{lkj}\: \epsilon^{il}\,
   (i\sigma_2\,\sigma^k)(i\tau_2) \over 2\sqrt{2} } \Big( {\hat p}\,^i {\hat p}
   \,^j\! -\!\frac{\delta^{ij}}{n} \Big) \,,\quad\!\!
   P^{({}^3\!D_3)} \!=\!  { \sqrt{n(n+2)}\, b^{ijk}
   (i\sigma_2 \sigma^k ) (i\tau_2) \over 4} \: {\hat p}\,^i {\hat p}
   \,^j  \nn \,,
\end{eqnarray}
where $\vec\epsilon$, $\epsilon^{ij}$, and $b^{ijk}$ are the $J=1,2,3$
polarization tensors and ${\hat p}=\vec p/|\vec p|$. The $^3S_1$, $^1P_1$ and
$^3D_{1,2,3}$ waves are isosinglets, while the $^1S_0$, $^3P_{0,1,2}$ and
$^1D_2$ states are isovectors labelled by $\vec \epsilon_{\mbox{\tiny I}}$.
Averaging over the polarization states in $n$ dimensions gives
\begin{eqnarray} \label{polsums}
 &&  \epsilon^i\,\epsilon^{*j} \to \frac{\delta^{ij}}{n} \,,\qquad \quad
   \epsilon^{ij}\,\epsilon^{*kl} \to \frac{1}{(n+2)(n-1)} \bigg[
     {\delta^{ik}\delta^{jl}+\delta^{il}\delta^{jk}}-\frac{2\delta^{ij}
     \delta^{kl}}{n} \bigg] \,,\\[5pt]
 && b^{ijk} b^{*lmq} \to \frac1{n(n-1)(n+4)} \bigg\{\frac{-2}{n+2}
   \Big[\, \delta^{mq}(\delta^{ij}\delta^{kl}+\delta^{ik}\delta^{jl}+
   \delta^{il}\delta^{jk}) +( m\leftrightarrow l)+(q\leftrightarrow l)\Big] \nn\\
 && \qquad\qquad\qquad + \Big[ (\delta^{il}\delta^{jm}\delta^{kq} +\delta^{il}
   \delta^{jq}\delta^{km}) +(i\to j\to k\to i)+(i\to k\to j\to i) \Big]
   \bigg\}   \,.\nn
\end{eqnarray}
To evaluate the matrix element of an operator ${\cal O}$, we write ${\cal
O}=N^*_a N^*_b {\cal O}_{ab\,;\,cd} N_c N_d$, so the scattering amplitude is
\begin {eqnarray} \label{mef}
i{\cal A}\equiv \sum_{\rm pol.\ avg} \left\langle NN(\,s'; \vec k', p'\,)
\:\right| {\cal O}
  \left|\: NN (\,s; \vec k, p )  \right\rangle = 4\ \int_{-1}^1
  \frac{d\cos\theta}{2}\ \: P_{ab}\: {\cal O}_{ab\,;\,cd}\: P^\dagger_{cd}\: \,,
\end{eqnarray}
where $\vec p\,' \cdot \vec p = p^2 \cos\theta$ and the indices ($a$,$b$,$c$,$d$)
are for both spin and isospin.

Examples of the use of Eq.~(\ref{mef}) are:
\begin{eqnarray} \label{mefeg}
\nn \\[-10pt]
&& \begin{picture}(20,10)(1,1)
      \put(1,3){\line(1,1){10}} \put(1,3){\line(1,-1){10}}
      \put(1,3){\line(-1,1){10}} \put(1,3){\line(-1,-1){10}}
      \put(-7,17){\mbox{\footnotesize $C_0^{(^1S_0)}$}}
      \put(-24,0){\mbox{\scriptsize ${}^1S_0$}}
      \put(15,0){\mbox{\scriptsize ${}^1S_0$}}
  \end{picture}
  \quad =\Big( \frac12 \int d\!\cos\theta\Big)\  4\ (-iC_0) {\rm Tr}\Big[
   P^{(^1S_0)} P_i^{(^1S_0)}\,^\dagger \Big]\ {\rm Tr}\Big[ P_i^{(^1S_0)}
   P^{(^1S_0)}\,^\dagger \Big]= -iC_0 \,,
\end{eqnarray}
where $P^{(^1S_0)}_i$ is given in Eq.~(\ref{proj}) and we have averaged over
the isospin polarizations, and
\begin{eqnarray} \label{mefeg2}
 &&  \begin{picture}(25,20)(1,1)
   \put(-10,-6){\line(1,0){20}} \put(-10,14){\line(1,0){20}}
     \multiput(0,-5)(0,7){3}{\line(0,1){4}}
      \put(-24,0){\mbox{\scriptsize ${}^1S_0$}}
      \put(15,0){\mbox{\scriptsize ${}^1S_0$}}
  \end{picture}\ \ =  \Big( \frac12 \int d\!\cos\theta\Big)\  2\
  \Big(i\frac{g_A^2}{2f^2} \Big) \frac{
  {\rm Tr} \Big[ P^{(^1S_0)}\: \vec\sigma \! \cdot (\vec p\,'-\vec p)\:\tau^k
 P^{(^1S_0)} \,^\dagger \: \vec\sigma\,^T \! \cdot (\vec p\,'-\vec p)\:(\tau^k)^{T}
  \Big]} {(\vec p\,'-\vec p)\,^2+m_\pi^2} \,, \nn \\[-5pt]
\end{eqnarray}
where in evaluating this trace it is useful to recall that $(i\sigma_2)\:
\vec\sigma\,^T \: (i\sigma_2) = \vec\sigma$. The factors of $4$ and $2$ in
Eqs.~(\ref{mefeg}) and (\ref{mefeg2}) are symmetry factors for the graphs.
Projecting the tree level one pion exchange diagram onto the various P and D
waves gives the order $Q^0$ amplitude in these channels:
\begin{eqnarray} \label{pd1pi}
  i\,{\cal A}_0({}^1P_1) &=& i \frac{g_A^2}{2 f^2} \bigg[ \frac{3m_\pi^2}{2p^2}
   -\Big(\frac{3m_\pi^4}{8p^4}+\frac{3m_\pi^2}{4 p^2} \Big)
  \ln\Big(1+\frac{4p^2}{m_\pi^2}\Big) \bigg] \,, \\
  i\,{\cal A}_0({}^3P_0) &=& i \frac{g_A^2}{2 f^2} \bigg[ 1-\frac{m_\pi^2}{4 p^2}
    \ln\Big(1+\frac{4p^2}{m_\pi^2}\Big) \bigg] \,, \nn \\
  i\,{\cal A}_0({}^3P_1) &=& -i \frac{g_A^2}{2 f^2} \bigg[ \frac12 -\frac{m_\pi^2}
    {4p^2}+\frac{m_\pi^4}{16 p^4} \ln\Big(1+\frac{4p^2}{m_\pi^2}\Big) \bigg] \,,
    \nn \\
  i\,{\cal A}_0({}^3P_2) &=& i \frac{g_A^2}{2 f^2} \bigg[ \frac{1}{10}+
   \frac{3m_\pi^2}{20p^2}-\Big(\frac{3 m_\pi^4}{80 p^4}+\frac{m_\pi^2}{10p^2}\Big)
    \ln\Big(1+\frac{4p^2}{m_\pi^2}\Big) \bigg] \,, \nn \\
  i\,{\cal A}_0({}^1D_2) &=& -i \frac{g_A^2}{2 f^2} \bigg[ \frac{3m_\pi^2}{4p^2}+
    \frac{3m_\pi^4}{8p^4}-\Big(\frac{3 m_\pi^6}{32p^6}+\frac{3m_\pi^4}{8p^4}
   +\frac{m_\pi^2}{4p^2}\Big) \ln\Big(1+\frac{4p^2}{m_\pi^2}\Big) \bigg] \,,\nn\\
  i\,{\cal A}_0({}^3D_1) &=& i \frac{g_A^2}{2 f^2} \bigg[ -\frac12-\frac{3m_\pi^2}
   {4p^2}+\Big(\frac{3 m_\pi^4}{16p^4}+\frac{m_\pi^2}
   {2p^2}\Big) \ln\Big(1+\frac{4p^2}{m_\pi^2}\Big) \bigg] \,, \nn \\
  i\,{\cal A}_0({}^3D_2) &=& i \frac{g_A^2}{2 f^2} \bigg[ \frac12-\frac{3m_\pi^2}
   {4p^2}-\frac{3m_\pi^4}{4p^4}+\Big(\frac{3 m_\pi^6}{16p^6}+\frac{9m_\pi^4}
   {16p^4}\Big) \ln\Big(1+\frac{4p^2}{m_\pi^2}\Big) \bigg] \,, \nn \\
  i\,{\cal A}_0({}^3D_3) &=&  -i\frac{g_A^2}{2 f^2} \bigg[ \frac17 +
    \frac{3m_\pi^2}{4p^2} + \frac{15m_\pi^4}{56p^4} -\Big( \frac{9m_\pi^2}
    {28p^2} + \frac{9m_\pi^4}{28p^4}+\frac{15m_\pi^6}{224p^6} \Big)
    \ln\Big(1+\frac{4p^2}{m_\pi^2}\Big)  \bigg] \nn \,.
\end{eqnarray}
These expressions agree with Ref.~\cite{kaiser}.


\section{Evaluation of order $Q$ loop diagrams}  \label{AppQ}

In this Appendix explicit expressions are given for the individual graphs in
Fig.~\ref{figQ} in the $^1S_0$ and $^3S_1$ channels and the graphs in
Fig.~\ref{figQD} for the $^3D_1$ channel.  Details on the evaluation of the
three non-trivial two pion exchange diagrams (Fig.~\ref{figQ}i,k,m) are also
presented.

Our calculation is performed using the Power Divergence Subtraction (PDS)
\cite{ksw1,ksw2} renormalization scheme in $d=n+1$ dimensions.  A factor of
$(\mu/2)^{3-n}$ is included with each loop and we work in the center of momentum
frame, $N(\vec p\,) N(-\vec p\,) \to N(\vec p\,'\,) N(-\vec p\,'\,)$. A
detailed description of the method used to implement the PDS scheme can be
found in Ref.~\cite{ms2}. Our results are slightly different than
Ref.\cite{ms1} because all spin and isospin traces are performed in $n$
dimensions rather than 3 dimensions.  For a four-nucleon operator with coupling
$C$, there are subtractions for ultraviolet divergences in $n=3$, $\delta^{\rm
uv}C$, and we define the renormalized coupling $C(\mu)$ by:
\begin{eqnarray}
   C^{\rm bare} = C^{\rm finite} - \delta^{\rm uv}C \,, \qquad C^{\rm finite} =
     C(\mu) - \sum_{m=1}^\infty \delta^mC(\mu) \,.
\end{eqnarray}
Here $\delta^m C(\mu)$ is the \emph{finite} $m$-loop PDS counterterm, which is
defined by canceling overall poles in $n=2$ (linear divergences) and then
continuing back to $n=3$. This procedure correctly accounts for the unusual
scaling of the four nucleon operators due to the presence of the non-trivial
fixed point. $C(\mu)$ may also cancel $\ln(\mu)$ dependence in the amplitude.
The beta functions in Eqs.~(\ref{C0RGE}) and (\ref{C0RGE2}) are computed using
\begin{eqnarray}
 \beta = \mu { \partial \over \partial\mu} C(\mu) =
   \sum_{m=1}^\infty \: { \partial \over \partial\mu}\: \delta^m C(\mu) \,.
\end{eqnarray}
Renormalized PDS diagrams are defined by adding graphs with counterterm vertices
to the original diagram.

\subsection{Basic Strategy for evaluating non-relativistic loop integrals}
\label{stradegy}

Our basic strategy for evaluating massive multiloop potential diagrams
analytically consists of the following three steps:
\begin{enumerate}

\item  Evaluate the spin and isospin traces, then do the energy integrals using
contour integration. This leaves integrations over loop three-momenta which
will be evaluated using dimensional regularization in $n=3-2\epsilon$
dimensions. When nucleon poles are taken in doing the contour integrals
in the $n+1$ dimensional non-relativistic theory, the remaining loop integrals
have the same form as $n$ dimensional loop integrals in a Euclidean
relativistic theory. The corresponding diagram in the $n$ dimensional Euclidean
theory can be found simply by shrinking to a point the nucleon propagators whose
pole is taken. This gives a graph with a ``reduced topology''. Two examples
of this are given in Fig.~\ref{reduce}.
\begin{figure}[!t]
  \centerline{\epsfysize=2.truecm \epsfbox{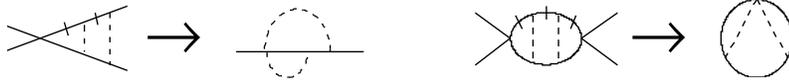}  }
{\tighten \caption[1]{Reduction in the topology of non-relativistic loop graphs
from performing the energy contour integrals and picking poles from the marked
nucleon propagators.} \label{reduce} }
\end{figure}
In the first example the energy integrals are performed using the poles in the
marked nucleon lines and the two loop graph becomes a two-point function.  Only
one momentum is relevant to the evaluation of this diagram because in the
original graph the loops only depend on the relative momentum between the two
outgoing lines.  In the second example choosing nucleon poles as indicated the
three loop graph becomes a vacuum bubble.  In the original diagram the loops
only see the total incoming energy. This energy will appear in mass terms in
the reduced diagram.

\item Eliminate factors of momenta in the numerator.  We begin by canceling terms
in the numerator against terms in the denominator (partial fractioning).
Numerators which can not be reduced by partial fractioning are labeled
irreducible.  These numerators are dealt with using the integration by parts
technique \cite{intbyparts}, using the tensor decomposition technique
\cite{tensdec}, and/or by using relations due to Tarasov\cite{Tarasov}.
Tarasov's method is to derive relations between integrals in $n$ dimensions
with irreducible numerators and integrals in $n+2, n+4, \ldots$ dimensions with
trivial numerators. These integrals are then reduced to $n$ dimensional
integrals with trivial numerators. (This method was automated for two loop
graphs in Ref.~\cite{mertig} using a Mathematica program called Tarcer). A
review of these techniques is given in Ref.~\cite{looprev}.

\item Evaluate the remaining scalar integrals. This can be done directly using
Feynman parameters, however it is often more useful to switch to
position space using
\begin{eqnarray}
 \frac1{{\vec k}^2+m^2} = \int d^nR \: e^{-i \vec k \cdot \vec R} \:\
 G(\vec R,m) \,,
\end{eqnarray}
where $G$ is the position space Green's function. An $m$-loop momentum space
integral with $k$ propagators becomes a $(k-m)$-loop integral in position
space. In $n$ dimensions the Green's function is
 \begin{eqnarray}
  G(\vec R, m) = \int\!\! {d^nk \over (2\pi)^n} {e^{i \vec k \cdot \vec R} \over
  {\vec k}\,^2 + m^2 } = \frac1{(2\pi)^{n/2}} \Big( \frac{M}{R} \Big)^{n/2-1}
   K_{1-n/2}(m R) \,,
 \end{eqnarray}
where $K$ is a modified Bessel function.  For odd $n$ the Bessel function
becomes an exponential; for $n=3$, $G(\vec R, m)= e^{-m R}/(4\pi R)$.  If the
reduced topology is that of a zero or two point function there are no
non-trivial angular integrations.  Since these are exponential integrals the
finite part of graphs are easy to evaluate. To evaluate ultraviolet divergent
integrals we follow Ref.~\cite{braaten} and split the $d^nR$ spatial
integration region into two parts $\int_0^\infty dR= \int_0^L dR +
\int_L^\infty dR$.  Ultraviolet divergences occur for $R\to 0$ so the
$\int_L^\infty dR$ integral can be done with $n=3$, discarding terms that
vanish as $L \to 0$. For the $\int_0^L dR$ integral we expand the Bessel
functions about $R=0$ using
\begin{eqnarray}
   K_\nu(z) = \frac{\Gamma(\nu)\Gamma(1-\nu)}{2} [I_{-\nu}(z)\!-\!I_{\nu}(z)]\,,
  \qquad I_\nu(z) = \sum_{k=0}^\infty \frac{(z/2)^{2k+\nu}}{k!\,\Gamma(\nu+k+1)
  }\,,
\end{eqnarray}
and then do the integration.  Ultraviolet divergences are expressed as
$1/\epsilon$ poles just as if the integration had been carried out in momentum
space.  When the integrals from $0$ to $L$ and from $L$ to $\infty$ are added
the $L$ dependent terms cancel.  For $n=3$, the scalar two-point and vacuum diagrams
with arbitrary masses have been evaluated to two and three loops respectively
in Ref.~\cite{Rajan}.

\end{enumerate}

\subsection{The order $Q$ potential diagrams}

At order $Q$  the potential diagrams that contribute to S-wave NN scattering
are shown in Fig.~\ref{figQ}.  The evaluation of the graphs in
Fig.~\ref{figQ}a-h,j,l,n is the same in the $^1S_0$ and $^3S_1$ channels, while
Fig.~\ref{figQ}i,k,m differ.  In the $^1S_0$ channel the order $Q$ diagrams
have also been evaluated in Ref.~\cite{rupak1}, however our results are
slightly different since all traces are performed in $n$ dimensions.

The graphs in Fig.~\ref{figQ}a-f are simple to evaluate:
\begin{eqnarray}
a)+d) &=&  -i\, [{\cal A}_{-1}]^2\, \bigg(\frac{C_4}{(C_0)^2} - \frac{(C_2)^2}
  {(C_0)^3} \bigg)\, p^4 + i\, [{\cal A}_{-1}]^3\, \frac{(C_2)^2}{(C_0)^4}\, p^4
  \,, \nn \\
b)+e) &=& -i\, [{\cal A}_{-1}]^2\, \bigg( \frac{E_4}{(C_0)^2}-\frac{2\,C_2
  D_2}{(C_0)^3} \bigg)\, p^2 m_\pi^2 + 2\, i\, [{\cal A}_{-1}]^3\, \frac{C_2
  D_2}{(C_0)^4}\, p^2 m_\pi^2 \,, \\
c)+f) &=&  -i\, [{\cal A}_{-1}]^2\, \bigg( \frac{D_4}{(C_0)^2} - \frac{(D_2)^2}
 {(C_0)^3} \bigg)\, m_\pi^4 + i\, [{\cal A}_{-1}]^3\, \frac{(D_2)^2}{(C_0)^4}\,
 m_\pi^4  \,. \nn
\end{eqnarray}
The diagrams in Fig.~\ref{figQ}g,h are also straightforward. Renormalized
diagrams are calculated by adding diagrams with the appropriate PDS
counterterms. The two basic renormalized diagrams needed to evaluate the
diagrams in Fig.~\ref{figQ}g,h are:
\begin{eqnarray} \label{C2pig}
 2\qquad\ \ \begin{picture}(25,20)(1,1)
   \put(-18,4){\line(-1,1){10}} \put(-18,4){\line(-1,-1){10}}
   \put(-18.4,0){\mbox{{$\Diamond$}}}
   \put(-10,4){\line(1,1){15}} \put(-10,4){\line(1,-1){15}}
     \multiput(0,-5)(0,7){3}{\line(0,1){4}}
  \end{picture}\!\!\!\!\!\!\! &=& 2 i\, (C_2 p^2+D_2 m_\pi^2)\,
  \frac{M g_A^2}{8\pi f^2}\,
     \bigg[ i p-\frac{i m_\pi^2}{2 p} \ln\Big(1-\frac{2 i p}{m_\pi}\Big) \bigg]
     + i\, \frac{M g_A^2}{8\pi f^2}\, C_2 m_\pi^3 \,,  \\
 2\qquad\ \ \begin{picture}(25,20)(1,1)
   \put(-18,4){\line(-1,1){10}} \put(-18,4){\line(-1,-1){10}}
   \put(-18.4,0){\mbox{{$\Diamond$}}}
   \put(0,4){\oval(20,18)} \put(14,4){\circle*{8}}
   \put(18.5,4){\line(1,1){10}} \put(18.5,4){\line(1,-1){10}}
     \multiput(0,-5)(0,7){3}{\line(0,1){4}}
  \end{picture} \ \ &=& 2 i\, \frac{M{\cal A}_{-1}}{4\pi}\,
    (C_2 p^2+D_2 m_\pi^2)\,
     \frac{M g_A^2}{8\pi f^2}\, \bigg[ -p^2 -\mu^2-\frac{m_\pi^2}{2}
     \ln\Big(\frac{\mu^2}{m_\pi^2}\Big)\nn \\
  &&  + m_\pi^2 \ln\Big(1-\frac{2 i p}{m_\pi}\Big) \bigg]+ i\,
    \frac{M{\cal A}_{-1}}{4\pi}\, (ip+\mu) \, \frac{M g_A^2}{8\pi f^2}\, C_2
    m_\pi^3 \,, \label{C2pig2}
\end{eqnarray}
where the sum of the $C_2$ and $D_2$ operators is represented by a diamond. The
diagram in Eq.~(\ref{C2pig2}) is ultraviolet divergent and in defining the
renormalized graph we have introduced two counterterms to cancel the
$1/\epsilon$ poles ($n=3-2\epsilon$):
\begin{eqnarray}
  \delta^{2,uv}E_4 &=& - \frac{C_0^{\rm finite} C_2^{\rm finite}}{2} \,\frac{M}{4\pi} \,
    \frac{M g_A^2}{8\pi f^2}\, \Big( \frac{1}{\epsilon} -2\gamma+2\ln\pi \Big) \,, \nn\\
  \delta^{2,uv}D_4 &=& - \frac{C_0^{\rm finite} D_2^{\rm finite}}{2} \,\frac{M}{4\pi} \,
    \frac{M g_A^2}{8\pi f^2}\, \Big( \frac{1}{\epsilon} -2\gamma+2\ln\pi \Big) \,.
\end{eqnarray}
The PDS renormalization scheme is being used, so there are also finite
subtractions that correspond to poles in three dimensions. The graph in
Eq.~(\ref{C2pig}) does not require a PDS counterterm because we are evaluating
spin and isospin traces in $n$ dimensions, and the isospin traces gives a
factor of $n-2$ which cancels the $1/(n-2)$ pole in the loop integration. The
graph in Eq.~(\ref{C2pig}) has $C_2$ and $D_2$ PDS counterterms which produce
the factor of $\mu^2$, while the factor of $\mu$ is from the first graph with a
$\delta^1C_2$ counterterm in place of the diamond. The remaining diagrams in
Fig.~\ref{figQ}g,h follow by dressing the results in Eqs.~(\ref{C2pig}) and
(\ref{C2pig2}) with $C_0$ bubbles and adding the appropriate counterterm
diagrams. The final result for the $^1S_0$ and $^3S_1$ channel is
\begin{eqnarray}
 g)+h) &=& 2i\, [{\cal A}_{-1}]^2\, \frac{M g_A^2}{8\pi f^2}\, \frac{(C_2 p^2 +
   D_2 m_\pi^2)} {(C_0)^2} \Bigg\{ \gamma - \frac{i m_\pi^2}{2 p} \ln\Big(1-
   \frac{2 i p}{m_\pi}\Big) + \frac{M{\cal A}_{-1}}{4\pi} \bigg[ \gamma^2
   -\mu^2 \nn\\
 && -\frac{m_\pi^2}{2} \ln\Big(\frac{\mu^2}{m_\pi^2}\Big)+m_\pi^2
    \ln\Big(1-\frac{2 i p}{m_\pi}\Big) \bigg] \Bigg\} + i \, [{\cal A}_{-1}]^2\, \frac{M g_A^2}
    {8\pi f^2}\, \frac{C_2 m_\pi^3}{(C_0)^2} \,.
\end{eqnarray}

Next consider the graphs in Fig.~\ref{figQ} with two potential pions. Diagrams
j), l) and n) can be obtained using the expressions for the NLO one pion
exchange diagrams:
\begin{eqnarray}
2\qquad\ \  \begin{picture}(25,20)(1,1) \put(-14.5,4){\circle*{8}}
   \put(-19.2,4){\line(-1,1){10}} \put(-19.2,4){\line(-1,-1){10}}
   \put(-10,4){\line(1,1){15}} \put(-10,4){\line(1,-1){15}}
     \multiput(0,-5)(0,7){3}{\line(0,1){4}}
  \end{picture}\!\!\!\!\!\! &=& -2i\,{\cal A}_{-1}\, \frac{M g_A^2}{8\pi f^2}\, \bigg[ ip
    - \frac{i m_\pi^2}{2 p} \ln\Big(1-\frac{2 i p}{m_\pi}\Big) \bigg] \,, \\
\begin{picture}(25,20)(1,1)
   \put(-18,4){\line(-1,1){10}} \put(-18,4){\line(-1,-1){10}}
   \put(0,4){\oval(20,18)} \put(14,4){\circle*{8}} \put(-14,4){\circle*{8}}
   \put(18.5,4){\line(1,1){10}} \put(18.5,4){\line(1,-1){10}}
     \multiput(0,-5)(0,7){3}{\line(0,1){4}}
  \end{picture} \ \ &=& -i\, \frac{M[{\cal A}_{-1}]^2}{4\pi} \, \frac{M g_A^2}{8\pi f^2}\,
     \bigg[ -p^2 -\mu^2 -\frac{m_\pi^2}{2} \ln\Big(\frac{\mu^2}{m_\pi^2}\Big)
     +m_\pi^2 \ln\Big( 1- \frac{2 i p}{m_\pi}\Big) \bigg] \,, \label{qqq}
\end{eqnarray}
giving the following expressions valid for the $^1S_0$ and $^3S_1$ channels:
\begin{eqnarray} \label{qqqr}
  \begin{picture}(25,20)(1,1) \put(-14.5,4){\circle*{8}}
   \put(-19.5,4){\line(-1,1){15}} \put(-19.5,4){\line(-1,-1){15}}
   \put(-10,4){\line(1,1){15}} \put(-10,4){\line(1,-1){15}}
     \multiput(0,-5)(0,7){3}{\line(0,1){4}}   \multiput(-30,-5)(0,7){3}{\line(0,1){4}}
  \end{picture}\!\!\!\!\!\! &=& i\,[{\cal A}_{-1}]\, \Big( \frac{M g_A^2}
   {8\pi f^2}\Big)^2\, \bigg[ ip - \frac{i m_\pi^2}{2 p} \ln\Big(1-\frac{2 i p}{m_\pi}\Big)
    \bigg]^2 \,, \nn \\[5pt]
2\qquad\ \ \  \begin{picture}(25,20)(1,1)
   \put(-18,4){\line(-1,1){10}} \put(-18,4){\line(-1,-1){10}}
   \put(0,4){\oval(20,18)}  \put(14,4){\circle*{8}} \put(-14,4){\circle*{8}}
   \put(18.5,4){\line(1,1){15}} \put(18.5,4){\line(1,-1){15}}
     \multiput(0,-5)(0,7){3}{\line(0,1){4}}  \multiput(28,-5)(0,7){3}{\line(0,1){4}}
  \end{picture} \ \  &=& 2i \frac{M[{\cal A}_{-1}]^2}{4\pi}\, \Big(\frac{M g_A^2}
   {8\pi f^2}\Big)^2\, \bigg[ ip - \frac{i m_\pi^2}{2 p} \ln\Big(1-\frac{2 i p}{m_\pi}\Big)
    \bigg]  \\
  && \qquad \times  \bigg[ -p^2 -\mu^2 -\frac{m_\pi^2}{2} \ln\Big(\frac{\mu^2}
  {m_\pi^2}\Big)   +m_\pi^2 \ln\Big( 1- \frac{2 i p}{m_\pi}\Big) \bigg] \,,\nn \\
\begin{picture}(25,20)(1,1)
   \put(-18,4){\line(-1,1){10}} \put(-18,4){\line(-1,-1){10}}
   \put(0,4){\oval(20,18)}  \put(14,4){\circle*{8}} \put(-14,4){\circle*{8}}
   \put(28,4){\oval(20,18)}  \put(42.5,4){\circle*{8}}
     \multiput(0,-5)(0,7){3}{\line(0,1){4}}  \multiput(28,-5)(0,7){3}{\line(0,1){4}}
  \put(48.,4){\line(1,1){10}} \put(48.,4){\line(1,-1){10}}
   \end{picture} \qquad  \ \  &=& i\, \frac{M^2[{\cal A}_{-1}]^3}{(4\pi)^2}\,
   \Big(\frac{M g_A^2}{8\pi f^2}\Big)^2\, \bigg[-p^2\!-\!\mu^2\!-\!\frac{m_\pi^2}{2}
   \ln\Big(\frac{\mu^2}{m_\pi^2}\Big) +m_\pi^2 \ln\Big( 1- \frac{2 i p}{m_\pi}\Big)
   \bigg]^2 \,. \nn
\end{eqnarray}
The last diagram required a new ultraviolet counterterm
\begin{eqnarray}
  \delta^{2,uv}D_4 &=& -\frac1{16} \Big[C_0^{\rm finite}\Big]^3 \Big( \frac{M}
  {4\pi} \Big)^4
   \Big(\frac{g_A^2}{2f^2}\Big)^2 \, \Big( \frac{1}{\epsilon}
   -2\gamma+2\ln\pi \Big)^2 \,,
\end{eqnarray}
while the other poles in the graphs in Eq.~(\ref{qqqr}) are cancelled by diagrams
with the $D_2$ counterterm defined in renormalizing the graph in
Eq.~(\ref{qqq}).

To evaluate the diagrams in Fig.~\ref{figQ}i,k,m we follow the three steps
discussed in section~\ref{stradegy}.  In the $^1S_0$ channel
step 2 may be accomplished by canceling terms in the numerator against those
in the denominator. For example, after doing the contour integrals the
integrand of the one-loop box diagram is
\begin{eqnarray} \label{boxr}
 &&\int\! {d^nk \over (2\pi)^n}  { {\vec k}\,^2 \: (\vec k-\vec q\,)^2 \over
    [\,{\vec k}\,^2+2 \vec k\cdot \vec p\,][\,{\vec k}\,^2 +m_\pi^2\,]
    [\, (\vec k-\vec q\,)^2 + m_\pi^2 \,]} \\
 && = \int\! {d^nk \over (2\pi)^n}
   \frac{1}{[{\vec k}\,^2+2 \vec k\cdot \vec p\,]} \Bigg\{ 1 - \frac{m_\pi^2}{[{\vec k}\,^2
  +m_\pi^2]} - \frac{m_\pi^2}{[(\vec k-\vec q\,)^2 + m_\pi^2] } + \frac{m_\pi^4}
  { [{\vec k}\,^2 +m_\pi^2][ (\vec k-\vec q\,)^2 + m_\pi^2] } \Bigg\} \nn \,,
\end{eqnarray}
where $\vec q=\vec p\,'-\vec p$. The integral over $\vec k$ can be evaluated
using Feynman parameters. The term with three propagators requires the most
effort and gives an answer involving di-logarithms, $\ply$. Integrating over
$\cos\theta = \vec p \cdot \vec p\,'/p^2$ to project out the $^1S_0$ partial wave
gives
\begin{eqnarray}  \label{I90}
  && \int_{-1}^1 d\cos\theta\: \int\! {d^3k \over (2\pi)^3} {1 \over
  [{\vec k}\,^2+2 \vec k\cdot \vec p\,][{\vec k}\,^2 +m_\pi^2]
  [ (\vec k-\vec q\,)^2 + m_\pi^2] } \\
&& = \frac{1}{8\pi p^3} \Bigg\{ \frac{i}{4} \ln^2\Big( 1+ \frac{4p^2}{m_\pi^2} \Big)
  + {\rm Im}\: \ply \Big( \frac{2 p^2 -i\, p\, m_\pi}{m_\pi^2+4 p^2} \Big) +
     {\rm Im}\: \ply \Big(\frac{-2 p^2+i\, p\, m_\pi}{m_\pi^2} \Big) \: \Bigg\} \,.
  \nn
\end{eqnarray}
Manipulations similar to those in Eq.~(\ref{boxr}) allow us to eliminate the
numerators in Fig.~\ref{figQ}k and Fig.~\ref{figQ}m.  For these diagrams all
the remaining scalar integrals were evaluated by Rajante, in Ref.~\cite{Rajan}.
A $D_2$ counterterm is introduced to cancel an $m_\pi^2/\epsilon$ divergence in
Fig.~\ref{figQ}m,
\begin{eqnarray}
  \delta^{uv}D_2 = -C_0^{\rm finite} \Big( \frac{Mg_A^2}{8\pi f^2} \Big)^2 \Big(
     \frac{1}{2\epsilon} -\gamma_E+\ln\pi+2-2\ln 2 \Big) \,.
\end{eqnarray}
The final result for Fig.~\ref{figQ}i,k,m in the $^1S_0$ channel is then:
\begin{eqnarray}  \label{2p1s0}
\begin{picture}(25,20)(1,1)
   \put(-15,-6){\line(1,0){30}} \put(-15,14){\line(1,0){30}}
     \multiput(-8,-6)(0,7){3}{\line(0,1){4}}  \multiput(8,-6)(0,7){3}{\line(0,1){4}}
 \end{picture}
 &=&
  \frac{i M}{4\pi} \Big( \frac{g_A^2}{2f^2} \Big)^2 \Bigg\{ i p -\frac{i m_\pi^2}{p}
  \ln\Big(1-\frac{2 i p}{m_\pi} \Big) + \frac{m_\pi^4}{4 p^3} \bigg[ \frac{i}{4}
   \ln^2\Big( 1+ \frac{4p^2}{m_\pi^2} \Big)  \nn\\
  && \qquad\qquad\quad + {\rm Im}\: \ply \Big( \frac{2 p^2 -i\, p\, m_\pi}
 {m_\pi^2+4 p^2} \Big) + {\rm Im}\: \ply \Big(\frac{-2 p^2+i\, p\, m_\pi}{m_\pi^2}
  \Big) \: \bigg] \Bigg\} \,, \nn\\[5pt]
2\qquad\ \ \begin{picture}(25,20)(1,1)  \put(-14,4){\circle*{8}}
   \put(-18,4){\line(-1,1){10}} \put(-18,4){\line(-1,-1){10}}
   \put(-10,4){\line(1,1){25}} \put(-10,4){\line(1,-1){25}}
     \multiput(0,-5)(0,7){3}{\line(0,1){4}}  \multiput(10,-16)(0,7){6}{\line(0,1){4}}
  \end{picture}
 &=& -2 i\, {\cal A}_{-1}\, \Big(  \frac{Mg_A^2}{8\pi f^2} \Big)^2 \Bigg\{ p^2 +
  \frac{m_\pi^2}{2} \Big[ \ln\Big(\frac{\mu^2}{m_\pi^2} \Big) -3 +2\ln 2 \Big]- \frac32
  m_\pi^2 \ln\Big(1-\frac{2 i p}{m_\pi} \Big) \nn\\
 &&  + \frac{m_\pi^4}{4 p^2} \bigg[ \frac32
  \ln^2\Big(1-\frac{2 i p}{m_\pi} \Big) + 2 \ply \Big( \frac{-m_\pi+2 i p}{m_\pi} \Big)
  + \ply \Big( \frac{m_\pi+2 i p}{-m_\pi+ 2 i p} \Big) +\frac{\pi^2}{4} \bigg] \Bigg\} \,,
  \nn\\[5pt] \phantom{xxxx}
\begin{picture}(25,20)(1,1) \put(-23.5,4){\circle*{8}}
   \put(-27,4){\line(-1,1){10}} \put(-27,4){\line(-1,-1){10}}
   \put(0,4){\oval(40,25)} \put(23.5,4){\circle*{8}}
   \put(27,4){\line(1,1){10}} \put(27,4){\line(1,-1){10}}
     \multiput(-7,-9)(0,7){4}{\line(0,1){4}}
     \multiput(7,-9)(0,7){4}{\line(0,1){4}}
  \end{picture} \quad
&=& [{\cal A}_{-1}]^2 \frac{M}{4\pi}  \Big(  \frac{Mg_A^2}{8\pi f^2} \Big)^2 \Bigg\{
  p^3 + i \mu^3 + p\, m_\pi^2 \Big[ \ln\Big(\frac{\mu^2}{m_\pi^2} \Big) -3 + 2\ln 2
  -2\ln\Big(1-\frac{2 i p}{m_\pi} \Big)  \Big] \nn\\
 && - \frac{m_\pi^4}{p} \bigg[ \ply\Big(\frac{m_\pi}
  {-m_\pi + 2 i p} \Big) + \frac{\pi^2}{12} \bigg]  \Bigg\}  \,.
\end{eqnarray}
Only the three loop graph requires a PDS counterterm because the isospin trace
with two pions gives a factor of $(n-2)^2$ while each loop gives at most a
$1/(n-2)$ pole. Our analytic expression for the box diagram agrees numerically with
the result in Ref.~\cite{kaiser}.

The evaluation of Fig.\ref{figQ}i,k,m in the $^3S_1$ channel is more difficult
because of the more complicated numerators.  For the box graph we can again
perform step 2 of the previous section by partial fractioning,
\begin{eqnarray}
&&\int\! {d^nk \over (2\pi)^n}\: { 4[ (\vec k-\vec q)\cdot \vec k]^2 +(n-4)
 {\vec k}\,^2 \: (\vec k-\vec q\,)^2 \over [{\vec k}\,^2+2 \vec k\cdot \vec p\,]
 [{\vec k}\,^2+m_\pi^2] [ (\vec k-\vec q\,)^2 + m_\pi^2 ]}  = \int\! {d^3k
 \over (2\pi)^3}\: \frac{1}{[{\vec k}\,^2+2 \vec k\cdot \vec p\,]}\: \Bigg\{ 1
 \nn \\[5pt]
&& \qquad +\frac{{\vec k}\,^2 - 2 m_\pi^2-2 {\vec q}\,^2}{[(\vec k-\vec q\,)^2
 +m_\pi^2] } + \frac{({\vec k}-\vec q)^2-2 m_\pi^2-2 {\vec q}\,^2}{[\vec k\,^2
 + m_\pi^2] } + \frac{3 m_\pi^4 + 4 m_\pi^2 {\vec q}\,^2 + {\vec q}\,^4}
  { [{\vec k}\,^2 +m_\pi^2][ (\vec k-\vec q\,)^2 + m_\pi^2] } \Bigg\}  \,.
\end{eqnarray}
Since this graph is finite we have set $n=3$.  The terms with three propagators
require Eq.~(\ref{I90}) and the following two integrals
\begin{eqnarray} \label{I912}
&& \int_{-1}^1 d\!\cos\theta\: (1-\cos\theta)  \int\! {d^3k \over (2\pi)^3}
 {1 \over  [{\vec k}\,^2+2 \vec k\cdot \vec p\,][{\vec k}\,^2 +m_\pi^2]
  [ (\vec k-\vec q\,)^2 + m_\pi^2] } \nn \\
  && = \frac{1}{8\pi p^3} \Bigg\{ \frac{(m_\pi^2 + 2p^2)}{p^2} \tan^{-1}\Big(
  \frac{m_\pi p}{m_\pi^2+2 p^2} \Big) - \frac{m_\pi^3}{2p^3} \ln\Big(1+
  \frac{p^2}{m_\pi^2} \Big) \nn\\
 &&\quad  - \Big(\frac{m_\pi^2}{p^2} +\frac{m_\pi^4}{4 p^4} \Big)
  \bigg[ {\rm Im}\: \ply \Big( \frac{2 p^2 -i\, p\, m_\pi}{m_\pi^2+4 p^2} \Big)
  + {\rm Im}\: \ply \Big(\frac{-2 p^2+i\, p\, m_\pi}{m_\pi^2} \Big) \bigg] \:
  \nn\\
 && \quad-i +  i \Big( 1 + \frac{m_\pi^2}{2p^2} \Big) \ln\Big( 1 + \frac{4p^2}
  {m_\pi^2}\Big) - \frac{i}{4} \Big( \frac{m_\pi^2}{p^2}+\frac{m_\pi^4}{4 p^4}
  \Big) \ln^2\Big(1+ \frac{4p^2}{m_\pi^2} \Big)  \Bigg\} \,, \nn \\[5pt]
 && \int_{-1}^1 d\!\cos\theta\: (1-\cos\theta)^2  \int\! {d^3k \over
 (2\pi)^3}{1 \over  [{\vec k}\,^2+2 \vec k\cdot \vec p\,][{\vec k}\,^2 +m_\pi^2]
  [ (\vec k-\vec q\,)^2 + m_\pi^2] } \nn \\
 && = \frac{1}{6\pi p^3} \Bigg\{ -\frac{3m_\pi^3}{8p^3} + \Big(
  \frac{15 m_\pi^5}{16p^5}+ \frac{9m_\pi^7}{64 p^7}\Big) \ln\Big(1+\frac{p^2}
  {m_\pi^2}\Big)-\frac34 \Big( 1+\frac{m_\pi^2}{2p^2} \Big) \Big( -2+
  \frac{3m_\pi^2}{p^2}+\frac{3 m_\pi^4}{4 p^4} \Big) \nn\\ && \quad \times
  \tan^{-1}\Big(\frac{m_\pi p}{m_\pi^2+2 p^2} \Big)
 +\frac98 \Big( \frac{m_\pi^2}{p^2}+\frac{m_\pi^4}{4 p^4} \Big)^2
  \bigg[ {\rm Im}\: \ply \Big( \frac{2 p^2 -i\, p\, m_\pi}{m_\pi^2+4 p^2} \Big)
  +{\rm Im}\: \ply \Big(\frac{-2 p^2+i\, p\, m_\pi}{m_\pi^2} \Big) \bigg] \:
  \nn\\
 && \quad + \frac{3i}{8} \bigg[ -1 +\frac{3 m_\pi^2}{p^2} +\frac{3 m_\pi^4}
  {4 p^4}- \Big( 1+\frac{m_\pi^2}{2 p^2}\Big)  \Big( -2+\frac{3m_\pi^2}{p^2}
  +\frac{3 m_\pi^4}{p^4} \Big) \ln\Big(1+ \frac{4p^2}{m_\pi^2} \Big) \nn\\
 && \quad + \frac34  \Big( \frac{m_\pi^2}{p^2}+\frac{m_\pi^4}{4 p^4} \Big)^2
  \ln^2\Big( 1+ \frac{4p^2}{m_\pi^2} \Big) \bigg]  \Bigg\} \,.
\end{eqnarray}
Using these results we find that the renormalized box graph in the $^3S_1$
channel is
\begin{eqnarray} \label{boxt}
\begin{picture}(25,20)(1,1)
   \put(-15,-6){\line(1,0){30}} \put(-15,14){\line(1,0){30}}
     \multiput(-8,-6)(0,7){3}{\line(0,1){4}}  \multiput(8,-6)(0,7){3}{\line(0,1){4}}
 \end{picture}
 &=& 3\,
  \frac{i M}{4\pi} \Big( \frac{g_A^2}{2f^2} \Big)^2 \Bigg\{-2 m_\pi - \frac{m_\pi^3}
  {2 p^2} +\frac{4\mu}{3}+\Big( \frac{3 m_\pi^6}{8 p^5} + \frac{m_\pi^4}{4 p^3} -
  \frac{2 m_\pi^2}{p} - 2 p \Big) \tan^{-1}\Big(\frac{p}{m_\pi} \Big) \nn\\
 && -\Big( \frac{3 m_\pi^6}{8 p^5} + \frac{m_\pi^4}{4 p^3} \Big) \tan^{-1}\Big(
  \frac{2p}{m_\pi} \Big) + \Big( \frac{3 m_\pi^7}{16 p^6} +\frac{m_\pi^5}{4 p^4} \Big)
  \ln\Big(1+\frac{p^2}{m_\pi^2} \Big)  \nn\\
 && - \Big( \frac{3 m_\pi^8}{32 p^7} +\frac{m_\pi^6}{4 p^5}
  + \frac{m_\pi^4}{4 p^3} \Big) \bigg[ {\rm Im}\: \ply \Big( \frac{2 p^2 +i\, p\, m_\pi}
 {m_\pi^2+4 p^2} \Big) + {\rm Im}\: \ply \Big(\frac{-2 p^2-i\, p\, m_\pi}{m_\pi^2}
  \Big) \: \bigg] \nn\\
 && +\frac{3 i m_\pi^4}{8 p^3} - \frac{i m_\pi^2}{2 p} +\frac{i p}{2} - i \Big(
  \frac{3 m_\pi^6}{16 p^5}+\frac{m_\pi^4}{8 p^3} \Big) \ln\Big(1+\frac{4 p^2}{m_\pi^2}
  \Big) +i \Big( \frac{3 m_\pi^8}{128 p^7} + \frac{m_\pi^6}{16 p^5} \nn\\
 && + \frac{m_\pi^4} {16 p^3} \Big)  \ln^2\Big(1+\frac{4 p^2}{m_\pi^2} \Big)
   \Bigg\} \,.
\end{eqnarray}
The $\mu$ dependence comes from adding a $\delta^1C_0$ counterterm at tree
level to cancel a $1/(n-2)$ pole.  For $\mu=0$, Eq.~(\ref{boxt}) agrees
numerically with the result in Ref.~\cite{kaiser}.

For the $^3S_1$ channel, the two loop graph in Fig.~\ref{figQ}k requires
evaluating
\begin{eqnarray}
   &&\int\!\! {d^nk \over (2\pi)^n}\int\!\! {d^n\ell \over (2\pi)^n}\: { 4[ (\vec k-\vec \ell)
  \cdot \vec k\:]^2 +(n-4)\; {\vec k}\,^2 \: (\vec k-\vec \ell\,)^2 \over  [{\vec \ell}\,^2
  +2 \vec \ell\cdot \vec p\,'\,] [{\vec k}\,^2+2 \vec k\cdot \vec p\,'\,][{\vec k}\,^2 +
  m_\pi^2] [ (\vec \ell-\vec k\,)^2 + m_\pi^2 ]}  \,.
\end{eqnarray}
We begin by eliminating the loop momenta from the numerator. This may be done
using the computer program\footnote{\tighten For this loop integral partial
fractioning is insufficient to eliminate the numerator.  This can only occur when
the reduced graph has a four-point vertex. This is why partial
fractioning was enough to eliminate the numerator for the box diagram.} in
Ref.~\cite{mertig}, that implements a set of reduction formulae due to Tarasov
\cite{Tarasov}.  The remaining scalar integrals can then be found in
Ref.~\cite{Rajan}. We have checked by hand that this program gives the same
final result as using tensor decomposition along with integration by parts and
partial fractioning.  The following counterterms are needed to cancel $1/(n-3)$
poles:
\begin{eqnarray} \label{2l3s1ct}
  \delta^{uv}C_2 &=& -6\,C_0^{\rm finite} \Big( \frac{Mg_A^2}{8\pi f^2} \Big)^2
  \Big(\frac{1}{2\epsilon} -\gamma_E+\ln\pi+2-2\ln 2 \Big) \,, \nn\\
  \delta^{uv}D_2 &=& -6\,C_0^{\rm finite} \Big( \frac{Mg_A^2}{8\pi f^2} \Big)^2
  \Big(\frac{1}{2\epsilon} -\gamma_E+\ln\pi+2-2\ln 2 \Big) \,.
\end{eqnarray}
We find that in the $^3S_1$ channel the PDS renormalized diagram is
\begin{eqnarray}
\nn \\[-10pt] 2\qquad\ \
 \begin{picture}(25,20)(1,1) \put(-14.5,4){\circle*{8}}
   \put(-18,4){\line(-1,1){10}} \put(-18,4){\line(-1,-1){10}}
   \put(-10,4){\line(1,1){25}} \put(-10,4){\line(1,-1){25}}
     \multiput(0,-5)(0,7){3}{\line(0,1){4}}  \multiput(10,-16)(0,7){6}{\line(0,1){4}}
 \end{picture}
 &=& 3 i\, {\cal A}_{-1}\, \Big(  \frac{Mg_A^2}{8\pi f^2} \Big)^2 \Bigg\{
  \frac{13 m_\pi^2}{6}\!-\!4 i m_\pi p\!-\! \frac{3 m_\pi^4}{2 p^2} \!-\! \frac{3 i m_\pi^3}
  {2 p} \!+\! \frac{3 i m_\pi^5}{4 p^3} \!+\! \frac{8 i \mu p}{3} \!+\! \frac{4 \mu^2}{3} \\
 &-& \Big( \frac{3 m_\pi^6}{4p^4}+\frac{m_\pi^4}{2 p^2} \Big) \ln 2 + \Big(
  \frac{3i m_\pi^7}{4 p^5} - \frac{3 m_\pi^6}{4 p^4} + \frac{i m_\pi^5}{p^3} -
  \frac{m_\pi^4}{2 p^2} + 4 m_\pi^2 + 4 p^2 \Big) \ln(1 - \frac{i p}{m_\pi} \Big) \nn\\
&-& \Big( \frac{3 m_\pi^8}
  {16 p^6} +\frac{m_\pi^6}{2 p^4} + \frac{m_\pi^4}{2 p^2} \Big) \bigg[ \frac32
  \ln^2\Big( 1 - \frac{2 ip}{m_\pi} \Big) + 2 \ply\Big(-1 + \frac{2 i p}{m_\pi} \Big) +
    \ply \Big( \frac{m_\pi+2 i p}{-m_\pi+ 2 i p} \Big) \nn\\
 &+& \frac{\pi^2}{4} \bigg] - 2 (p^2+m_\pi^2)  \ln\Big(\frac{\mu^2}{m_\pi^2} \Big) +
  \Big( \frac{-3im_\pi^7}{8 p^5} + \frac{9 m_\pi^6}{8 p^4} -\frac{i m_\pi^5}{2 p^3} +
  \frac{3 m_\pi^4}{4 p^2} \Big) \ln(1 - \frac{2i p}{m_\pi} \Big) \Bigg\} \,. \nn
\end{eqnarray}
The term proportional to $\mu^2$ is from a $\delta^2C_0(\mu)$ counterterm,
while the term proportional to $\mu$ is from a one-loop nucleon bubble with
$\delta^1C_0(\mu)$ and $C_0(\mu)$ vertices.

Now we turn to the three loop diagram in Fig.~\ref{figQ}m in the $^3S_1$
channel. After performing the traces and energy integration we are left with
the integral:
\begin{eqnarray}
&&\int\!\! {d^nq \over (2\pi)^n} \int\!\! {d^nk \over (2\pi)^n} \int\!\!
{d^n\ell \over (2\pi)^n}\: { 4[\,(\vec q-\vec \ell\,) \cdot (\vec q -\vec k\,)
  \,]^2 +(n-4)\, ({\vec q-\vec\ell \, })^2 \: (\vec q-\vec k\,)^2 \over
  [{\vec \ell}\,^2 -p^2][{\vec k}\,^2-p^2] [{\vec q}\,^2-p^2]
  [(\vec q-{\vec k})^2 + m_\pi^2 ][(\vec q-\vec \ell)^2 + m_\pi^2 ]}
  \,. \nn \\[5pt]
\end{eqnarray}
To eliminate the first term in this numerator we have implemented by hand the
procedure given in Ref.~\cite{Tarasov}.  The remaining $3-2\epsilon$ and
$5-2\epsilon$ dimensional scalar integrals are evaluated in position space as
described in Step 3 of the previous section.  The non-analytic ultraviolet
divergences in the result ($p^3/\epsilon$, $m_\pi^2 p/\epsilon$) are cancelled
by inserting the counterterms in Eq.~(\ref{2l3s1ct}) at one-loop as described
in Ref.~\cite{ms1}. The final result for the $^3S_1$ channel in the PDS scheme
is
\begin{eqnarray}
\phantom{xxxx}
\begin{picture}(25,20)(1,1) \put(-23.5,4){\circle*{8}}
   \put(-27,4){\line(-1,1){10}} \put(-27,4){\line(-1,-1){10}}
   \put(0,4){\oval(40,25)} \put(23.5,4){\circle*{8}}
   \put(27,4){\line(1,1){10}} \put(27,4){\line(1,-1){10}}
     \multiput(-7,-9)(0,7){4}{\line(0,1){4}}
     \multiput(7,-9)(0,7){4}{\line(0,1){4}}
  \end{picture} \quad
&=&-3 [{\cal A}_{-1}]^2 \frac{M}{4\pi} \Big(\frac{Mg_A^2}{8\pi f^2} \Big)^2
  \Bigg\{\!-\!\frac{7i m_\pi^3}{3} +\frac{5 m_\pi^2 p}{3} -4i m_\pi p^2 +
  \frac{p^3}{2} - \frac{9 m_\pi^4}{8 p}+\frac{i m_\pi^5}{p^2} \nn \\
 &+& \frac{3 m_\pi^6}{8 p^3} +\frac{4 i \mu p^2}{3} +\frac{4 \mu^2 p}{3}
  -\frac{4 i \mu^3}{9} + \Big( \frac{3 i m_\pi^7}{4p^4}-\frac{3 m_\pi^6}{4 p^3}
  +\frac{i m_\pi^5}{p^2}-\frac{m_\pi^4}{2 p} \Big) \ln 2 \nn\\
 &-& 2 (p^3+m_\pi^2 p) \ln\Big(\frac{\mu^2}{m_\pi^2} \Big)-\Big(
  \frac{3im_\pi^7}{4 p^4} - \frac{3 m_\pi^6}{4 p^3} +\frac{i m_\pi^5}{p^2}
  -\frac{m_\pi^4}{2 p} \Big) \ln(1 - \frac{2i p}{m_\pi} \Big) \nn\\
 &+& \Big( \frac{3i m_\pi^7}{4 p^4} - \frac{3 m_\pi^6}{4 p^3} +
  \frac{i m_\pi^5}{p^2}- \frac{m_\pi^4}{2 p} + 4 m_\pi^2 p + 4 p^3 \Big)
  \ln(1 - \frac{i p}{m_\pi} \Big)\nn \\
 &+&  \Big( \frac{m_\pi^4}{p}+\frac{m_\pi^6}{p^3}+\frac{3 m_\pi^8}{8 p^5} \Big)
  \bigg[ \ply\Big(\frac{m_\pi}{-m_\pi + 2 i p} \Big) + \frac{\pi^2}{12} \bigg]
  \Bigg\}  \,.
\end{eqnarray}
The terms with powers of $\mu$ are from a combination of tree, one, and two loop
PDS counterterm diagrams.

In  the $^3D_1$ channel the order $Q$ diagrams are shown in Fig.~\ref{figQD}.
The method used to evaluate the box diagram is the same as the $^3S_1$
channel, and the only difficult scalar integrals that appear are those in
Eqs.~(\ref{I90}) and (\ref{I912}).  We find
\begin{eqnarray}  \label{ddb}
\begin{picture}(25,20)(1,1)
   \put(-15,-6){\line(1,0){30}} \put(-15,14){\line(1,0){30}}
     \multiput(-8,-6)(0,7){3}{\line(0,1){4}}  \multiput(8,-6)(0,7){3}{\line(0,1){4}}
 \end{picture}
 &=& \frac32 \, \frac{i M}{4\pi} \Big( \frac{g_A^2}{2f^2} \Big)^2 \Bigg\{
  -\frac{2m_\pi}{7} + \frac{54m_\pi^5}{35 p^4}-\frac{19 m_\pi^3}{70 p^2}
  -\Big( \frac{9 m_\pi^6}{8 p^5} + \frac{7m_\pi^4}{4 p^3} \Big) \tan^{-1}\Big(
  \frac{2p}{m_\pi} \Big)\nn\\
 && +\Big(\frac{9m_\pi^6}{8 p^5} +\frac{7m_\pi^4}{4 p^3}+\frac{4m_\pi^2}{5 p}
  -\frac{2 p}{7} \Big) \tan^{-1}\Big(\frac{p}{m_\pi}\Big)  -  \Big( \frac{549
  m_\pi^7}{560p^6}+\frac{3 m_\pi^5}{4 p^4} \Big) \ln\Big(1+\frac{p^2}{m_\pi^2}
  \Big)  \nn\\
 && - \Big( \frac{9 m_\pi^8}{32 p^7} +\frac{m_\pi^6}{p^5}+\frac{m_\pi^4}{p^3}
 \Big) \bigg[ {\rm Im}\: \ply \Big( \frac{2 p^2 +i\, p\, m_\pi}
 {m_\pi^2+4 p^2} \Big) + {\rm Im}\: \ply \Big(\frac{-2 p^2-i\, p\, m_\pi}{m_\pi^2}
  \Big) \: \bigg] \nn\\
 && +\frac{9 i m_\pi^4}{8 p^3} - \frac{i m_\pi^2}{2 p} +\frac{i p}{2} - i \Big(
  \frac{9 m_\pi^6}{16 p^5}+\frac{7m_\pi^4}{8 p^3} \Big) \ln\Big(1+\frac{4 p^2}
  {m_\pi^2} \Big) +i \Big( \frac{9 m_\pi^8}{128 p^7} + \frac{m_\pi^6}{4 p^5} \nn\\
 && + \frac{m_\pi^4}{4 p^3} \Big)  \ln^2\Big(1+\frac{4 p^2}{m_\pi^2} \Big)
   \Bigg\} \,.
\end{eqnarray}
This expression agrees numerically with the result in Ref.~\cite{kaiser}.
Fig.~\ref{figQD}c can be evaluated using the result for the NLO one pion
exchange $^3S_1-{}^3D_1$ diagram
\begin{eqnarray}
 \qquad \begin{picture}(25,20)(1,1) \put(-14.5,4){\circle*{8}}
   \put(-19.2,4){\line(-1,1){10}} \put(-19.2,4){\line(-1,-1){10}}
   \put(-10,4){\line(1,1){15}} \put(-10,4){\line(1,-1){15}}
     \multiput(0,-5)(0,7){3}{\line(0,1){4}}
  \end{picture}\!\!\!\!\!\!\!\!\! &=& -\sqrt{2}i\,{\cal A}_{-1}\, \frac{M g_A^2}
  {8\pi f^2}\, \bigg[\!-\!\frac{3m_\pi^3}{4 p^2}\!-\!\frac{3i m_\pi^2}{4 p}\!+\!
  \frac{i p}{2} + i \Big( \frac{m_\pi^2}{2p}+\frac{3m_\pi^4}{8p^3} \Big)
  \ln\Big(1-\frac{2 i p}{m_\pi}\Big) \bigg] \,,
\end{eqnarray}
giving the following result for the $^3D_1-{}^3D_1$ transition:
\begin{eqnarray} \label{ddc}
 \qquad \begin{picture}(25,20)(1,1) \put(-14.5,4){\circle*{8}}
   \put(-19.5,4){\line(-1,1){15}} \put(-19.5,4){\line(-1,-1){15}}
   \put(-10,4){\line(1,1){15}} \put(-10,4){\line(1,-1){15}}
     \multiput(0,-5)(0,7){3}{\line(0,1){4}}
     \multiput(-30,-5)(0,7){3}{\line(0,1){4}}
  \end{picture}\!\!\!\!\!\! &=& 2i\,[{\cal A}_{-1}]\, \Big( \frac{M g_A^2}
   {8\pi f^2}\Big)^2\, \bigg[ \!-\!\frac{3m_\pi^3}{4 p^2}\!-\!\frac{3i m_\pi^2}
   {4 p}\!+\!\frac{i p}{2} + i \Big( \frac{m_\pi^2}{2p}+\frac{3m_\pi^4}{8p^3}
   \Big) \ln\Big(1-\frac{2 i p}{m_\pi}\Big) \bigg]^2 \,.
\end{eqnarray}


\section{Order $Q$ radiation pion contributions} \label{AppRad}

For interactions involving two nucleons it is useful to divide pions into three
classes: potential, radiation, and soft. This division is analogous to the
potential, soft, and ultrasoft regimes \cite{beneke} devised for calculating
non-relativistic diagrams with massless photons (NRQED) or gluons (NRQCD)
\cite{NRQCD0}.  To see how the different types of pion arise consider evaluating
the energy integrals for non-relativistic loop diagrams using contour
integration. When only residues of nucleon poles are taken, the pions in the
graph are potential pions. When the residue of a pion pole is taken, the pion is
either radiation or soft.  A soft pion has a momentum which is similar in size to
the momentum of the nucleons with which it is interacting.  A radiation pion
exchanges energy with nucleons but does not transfer three momentum.  Instead,
its momentum exchange is governed by a multipole expansion in powers of
$v_r=\sqrt{m_\pi/M}$.  Radiation pions are the only type which occur as external
particles.

Loops with only potential or soft pions give functions of $p/m_\pi$ where $p$ is
a nucleon momentum.  These graphs have a natural power counting in powers of
$Q\sim p \sim m_\pi$.  By natural power counting we mean that the graph scales
homogeneously with $Q$.  On the other hand, graphs with radiation pions give
functions of $p/Q_r$ where $Q_r=\sqrt{Mm_\pi}$ is the momentum threshold for pion
production.  These graphs have a natural power counting in powers of $Q_r$ at the
scale $p\sim Q_r$ \cite{ms2}.  This can be seen at the level of the Lagrangian.
In order to avoid double counting it also necessary to take $p \sim Q_r$ when
calculating soft contributions\footnote{\tighten At $p \sim Q_r$ the potential
and soft pion propagators should be expanded in $m_\pi/Q_r$. At $p \sim m_\pi$
there may then be factors of $m_\pi/p$ that must be resummed. See Ref.\cite{ms2}
for an explicit example.}.  For nucleons with $p\sim Q_r$ the three classes of
pion are characterized by different energy $(q_0)$ and momentum $({\vec q\,})$:
\begin{eqnarray}
 &{\rm potential} \qquad\quad &q_0 \sim {\vec q\,}^2/M \sim m_\pi \nn \\
 &{\rm radiation} \qquad\quad &q_0 \sim |{\vec q\,}| \sim m_\pi  \\
 &{\rm soft} \qquad\quad & q_0 \sim |{\vec q\,}| \sim Q_r =\sqrt{M m_\pi} .\nn
\end{eqnarray}

To implement the KSW expansion, which assumes $p \sim m_\pi$, we must expand the
result of a $Q_r^n$ radiation pion graph in powers of $Q$.  It turns out that the
leading $Q$ contribution of a radiation pion graph is not determined by the
substitution $Q_r \rightarrow Q^{1/2}$. Instead we will show that some radiation
pion graphs are enhanced by a factor of $1/Q$ so that an order $Q_r^k$ graph can
give an order $Q^{k/2-1}$ contribution.  This means that at NNLO the $Q_r^3$ and
$Q_r^4$ radiation pion graphs need to be considered. In this Appendix we begin by
reviewing the power counting for pions. The order $Q_r^3$ radiation pion
calculation\cite{ms2} is summarized. We then explain how to determine which
radiation pion graphs may give an order $Q$ contribution.  Finally, the order
$Q_r^4$ radiation pion graphs which contribute to nucleon-nucleon scattering are
examined and their order $Q$ contribution is evaluated.

\subsection{Power counting review}

For  $NN$ scattering at NLO the relevant terms in the action are
\begin{eqnarray} \label{oL}
 S &=& \int\!\! dt d^3\!x \: N^\dagger \Big(i \partial_t +\frac{\nabla^2}{2M}
   \Big) N + \pi^\dagger (\partial_t^2 - \nabla^2 - m_\pi^2) \pi
   + \frac{g_A}{\sqrt{2}f} (N^\dagger \sigma^i \tau^j N)(\nabla^i\pi^j ) \nn\\
   && - C_0\, {\cal O}_0  - D_2 m_\pi^2\, {\cal O}_0
   + \frac{C_2}{8}\, {\cal O}_2  \,,
\end{eqnarray}
where ${\cal O}_{0,2}$ are the four nucleon operators given in Eq.~(\ref{Lpi}). (In
this section spin and isospin  dependence is suppressed since it is not relevant
for the rescaling arguements.) To make the power counting in this action manifest
it is useful to rescale the coordinates and fields in a manner similar to the
rescaling done in NRQCD\cite{lm,multi,nrqcd,gries}.  The power counting is
facilitated because factors of $p=M v$, and $M$ are made explicit. For the
nucleon-pion Lagrangian parts of this rescaling were carried out in
Ref.~\cite{lm} and further discussed in Ref.~\cite{gh}. We begin by rescaling the
coordinates in a manner appropriate to the potential regime and rescaling the
fields to keep the kinetic terms invariant:
\begin{eqnarray} \label{rp}
   x = \frac{X}{Mv}\,, \quad t = \frac{T}{Mv^2}\,, \quad N(x,t)=(Mv)^{3/2}\,
   N_p(X,T)\,, \quad \pi(x,t) = M v^{3/2}\, \pi_p(X,T) \,.
\end{eqnarray}
The coefficients of four nucleon operators will also be rescaled to take into
account the KSW power counting which is appropriate for large S-wave scattering
lengths.  Using the PDS \cite{ksw2} or the OS scheme \cite{ms0,ms1} and taking
$\mu = M v$ gives
\begin{eqnarray}
  C_0(M v) = \frac{4\pi}{M^2 v}\: \tilde{C_0} \,,\quad
  C_2(M v) = \frac{4\pi}{M^3 v^2}\:\tilde{C_2} \,, \quad
  D_2(M v) = \frac{4\pi}{M^3 v^2}\:\tilde{D_2} \,,
\end{eqnarray}
where $\tilde{C_0}$, $\tilde{C_2}$, $\tilde{D_2}$ are order $v^0$. This gives the
following rescaled action for the potential regime
\begin{eqnarray} \label{rLp}
 S_p &=& \int\!\! dT d^3X \ N_p^\dagger \Big(i \partial_T -\frac
   {\nabla_{\!X}^2}{2} \Big) N_p + \pi_p^\dagger \bigg[ v^2 \partial_T^2 -
   \nabla_{\!X}^2 -\Big(\frac{m_\pi}{Mv}\Big)^2 \bigg] \pi_p +
   (4\pi)\, \tilde{C_0}\, [N_p^\dagger N_p]^2 \nn \\
  &&+ (4\pi)\, M v\:  \bigg\{\, \tilde{D_2}
  \Big(\frac{m_\pi}{Mv}\Big)^2\, [N_p^\dagger N_p]^2
   + \tilde{C_2} [N_p^\dagger N_p^\dagger  N_p {\tensor{\nabla}_{\!X}}^{\,2} N_p]
   + h.c. \bigg\} \\
   && + \sqrt{4\pi}\ \sqrt{\frac{Mg_A^2}{8\pi f^2}}\:  \sqrt{Mv}\
   (N_p^\dagger \sigma^i \tau^jN_p) (\nabla_{\!X}^i  \pi_p^j ) \nn \,.
\end{eqnarray}
Eq.~(\ref{rLp}) reproduces some familiar features of the power counting. In the
nucleon kinetic term the $\partial_T$ and $\nabla_{\!X}^{\,2}$ terms are the
same order.  In the potential pion kinetic term the $\partial_T^{\,2}$ term is
down by $v^2$ and is therefore treated perturbatively.  Furthermore, the
$\nabla_{\!X}^2$ and $m_\pi^2$ terms are the same size for $v=m_\pi/M \simeq
0.15$  \cite{lm}. Thus, $p\sim m_\pi$ is the natural power counting scale
when calculating graphs with only potential pions.  The $\tilde C_0$ interaction
term is the same size as the nucleon kinetic terms and therefore must be treated
non-perturbatively. Each potential loop gives a factor of $1/(4\pi)$ which will
cancel against factors of $(4\pi)$ multiplying interactions terms like $\tilde
C_0$. Insertions of $\tilde C_2$ or $\tilde D_2$ are suppressed by
$Mv/\Lambda=m_\pi/\Lambda \sim 1/2$ and are therefore treated perturbatively.
Finally, we see that the exchange of a potential pion involves the insertion of
two $NN\pi$ vertices and is suppressed by $Mv/\Lambda_{NN}=m_\pi/\Lambda_{NN}=
0.47$ where $\Lambda_{NN} = (8\pi f^2 )/(M g_A^2) \simeq 300\,{\rm MeV}$.

In the radiation regime the time coordinate has the same scaling as in
Eq.~(\ref{rp}), but the spatial coordinate has a different rescaling.
\begin{eqnarray}  \label{rr}
   x = \frac{X_r}{M v^2} \,, \qquad \pi(x,t) = M v^2\: \pi_r (X_r, T) \,.
\end{eqnarray}
The rescaled radiation pion kinetic term is then
\begin{eqnarray} \label{rSr1}
  S_r = \int\! d^3 X_r\, dT \ \pi_r^\dagger \bigg[ \partial_T^2 -
  \nabla_{\!X_r}^2 - \Big(\frac{m_\pi}{Mv^2}\Big)^2 \bigg] \pi_r \,.
\end{eqnarray}
For radiation pions the derivative terms are the same size as the mass term for
a different value of $v$, namely $v_r=\sqrt{m_\pi/M}$. For $v=v_r$ the radiation
pion energy and momentum are order $m_\pi$.  This $v$ corresponds to nucleon
momenta $p\sim Q_r=\sqrt{M m_\pi}$ which is the pion production threshold.  At
these momenta the power counting for graphs with radiation pions is
straightforward\cite{ms2}. When performing calculations at these momenta the
terms in the $S_p$ action should be scaled up\footnote{\tighten We ignore the
running of the physical $NN\pi$ coupling $g_A(\mu)$ because its $\ln(\mu)$
dependence is down by $Q^2$.} to $\mu \sim M v_r$. The $NN\pi_r$ interaction
term is
\begin{eqnarray} \label{rSr2}
  (4\pi) \frac{g_A}{\sqrt{2}}\:\frac{M v_r}{4\pi f} \: \int\! d^3 X\, dT  \:
   \Big[N_p(X)^\dagger \sigma^i \tau^jN_p(X)\Big]\: \Big[\nabla_{\!X}^i
  \pi_r^j(v X) \Big] \,.
\end{eqnarray}
Since the nucleon and radiation pion fields have a different spatial coordinate
we must perform a multipole expansion \cite{multi} to make the $v_r$ counting
manifest,
\begin{eqnarray}
\nabla_{\!X} \pi_r (v_r X) = v_r (\nabla_{\!X_r} \pi_r)_{X_r=0} + {\cal O}(v_r^2) \,.
\end{eqnarray}
Therefore, a nucleon emitting a radiation pion will not have its three momentum
changed.  From Eq.~(\ref{rSr2}) we see that each radiation pion vertex comes
with\footnote{\tighten Note that since each radiation loop gives a factor of
$1/(4\pi)^2$ we have pulled a $(4\pi)$ out front in the $NN\pi_r$ vertex in
Eq.~(\ref{rSr2}).} a factor of $M v_r^2/(4\pi f)= m_\pi / \Lambda_\chi$. For
evaluating radiation pion graphs we take $p\sim \mu \sim Q_r$ and have the
following power counting rules:

{\tighten
\begin{eqnarray}
& &{\rm radiation\ pion\ propagator} \qquad\qquad\qquad M^2/Q_r^4 \nn \\*
& &{\rm nucleon\ propagator} \qquad\qquad\qquad\qquad\quad  M/Q_r^2 \nn \\*
& &{\rm axial\ pion-nucleon\ coupling} \qquad\qquad\   Q_r^2/M \\*
& &{\rm radiation\ measure} \qquad\qquad\qquad\qquad d^4q \sim Q_r^8/M^4
  \nn\\*
& &{\rm potential\ measure} \qquad\qquad\qquad\qquad  d^4k \sim Q_r^5/M  \nn \,.
\end{eqnarray}\\[-40pt]
}

\noindent At momenta of order $m_\pi$, the mass term in Eq.~(\ref{rSr1}) is
enhanced by $1/v^2$ relative to the kinetic $(\partial_T^2 - \nabla_{\!X_r}^2)$
term.  For $p\ll Q_r$ we see that radiation pions could be integrated out in a
similar fashion to integrating out $W$ bosons for momenta $p\ll M_W$. Matching
onto a low energy theory would absorb radiation contributions into local
operators.  However, this will not be done since the matching gives $m_\pi$
dependence to the coefficients of four nucleon operators, yielding a low energy
theory without a chiral power counting. Instead, radiation pion graphs will be
expanded in $p^2/Q_r^2$ ($\sim Q/M$ for $p\sim m_\pi$), and only the order $Q$
piece of the radiation pion graphs will be included in our calculation.

Finally, consider the soft regime\cite{gries}.  Here the spatial coordinate has
the same scaling as in Eq.~(\ref{rp}), but the time coordinate has a different
rescaling
\begin{eqnarray} \label{rs}
   t = \frac{T_s}{M v} \,, \qquad \pi(x,t) = (M v)\: \pi_s (X, T_s) \,.
\end{eqnarray}
The soft pion action is
\begin{eqnarray} \label{rLs}
  S_s &=&  \int\! d^3 X\, dT_s \ \pi_s^\dagger \bigg[ \partial_{T_s}^2 -
   \nabla_{\!X}^2 -
   \Big(\frac{m_\pi}{Mv}\Big)^2 \bigg] \pi_s \nn\\
  && + (4\pi) \frac{g_A}{\sqrt{2}}\:\frac{M v}{4\pi f} \: \int\! d^3 X\, dT_s\:
   \Big[N_s^\dagger \sigma^i \tau^jN_s\Big]\: \Big[\nabla_{\!X}^i  \pi_s^j \Big]
   \,,
\end{eqnarray}
where $N_s(T_s,X)= N_p(T,X)$.  With this rescaling the nucleon action is
\begin{eqnarray} \label{rLsN}
  \int\! d^3X dT_s\: N_s^\dagger \Big( i \partial_{T_s} -
  v \frac{\nabla_{\!X}^2}{2} \Big) N_s \,.
\end{eqnarray}
Therefore, when a nucleon appears in a soft loop the kinetic energy term is
treated perturbatively making the propagator static. From Eq.~(\ref{rLs}) we see
that the power counting of soft loops is simplest for $v\sim m_\pi/M$ or $p\sim
m_\pi$. Unfortunately, this makes the soft pion modes appear at the same energy
and momentum as the radiation pion modes (i.e. $\sim m_\pi$). Therefore,
calculating with radiation pions at $p\sim Q_r$ and soft pions at $p\sim m_\pi$
may result in double counting.  This problem can be avoided by using $v=v_r$ for
both radiation and soft pions and then scaling down to $v\sim m_\pi/M$. An
explicit example of this procedure is worked out in Ref.~\cite{ms2}. Examples of
soft diagrams are shown in Fig.~\ref{fig_soft}.  These diagrams are order $Q^2$
(even when dressed with $C_0$ bubbles) and therefore will not be discussed
further.

\subsection{T\lowercase{he order \uppercase{$Q$} part of the order
$\uppercase{Q}_r^3$ radiation pion graphs}}

\begin{figure}[!t]
  \centerline{\epsfysize=7.5truecm \epsfbox{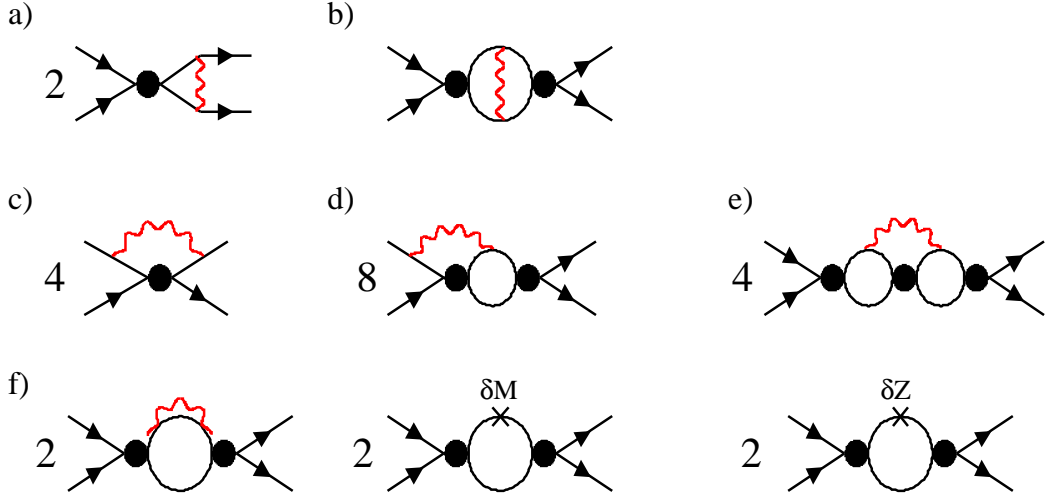}  }
{\tighten \caption[1]{Leading order radiation pion graphs for $NN$ scattering.
The wavy lines are radiation pions and $\delta M$, $\delta Z$ are the mass and
field renormalization counterterms.  There is a further field renormalization
contribution that is included in Eq.~(\ref{fans}).} \label{Qr3} }
\end{figure}

The order $Q_r^3/(M^3 \Lambda_\chi^2)$ radiation pion graphs shown in
Fig.~\ref{Qr3} were calculated\footnote{\tighten The graphs in Fig.~\ref{Qr3}a,b and the field 
renormalization are affected by performing the spin and isospin traces in $n$ 
dimensions, so a) and b) in Eq.~(\ref{gr1}) differ from Ref.~\cite{ms2}. 
However, the sum of graphs in Eq.~(\ref{fans}) is unaffected.} in \hbox{Ref}.~\cite{ms2}.
It is instructive to look at the result of evaluating some of these 
diagrams:
\begin{eqnarray}\label{gr1}
a)&=&
  -3 i {\cal A}_{-1} { g_A^2 m_\pi^2\over (4 \pi f)^2} \bigg[ {1 \over \epsilon} -
  {5\over 3} - {\rm ln}\Big({m_\pi^2 \over {\overline{\mu}}^2}\Big) \bigg] \,,  \\
b) &=& [{\cal A}_{-1}]\,^2\, {g_A^2 M m_\pi^2 \over (4 \pi f)^2}\, \Bigg\{
   {3\,p \over 4 \pi}\, \bigg[ {1 \over \epsilon} + {1\over 3} -
   2\,{\rm ln}\, 2 - {\rm ln}\Big({m_\pi^2 \over {\overline{\mu}}^2}
   \Big) - {\rm ln}\Big({-p^2 \over {\overline{\mu}}^2}\Big) \bigg]  \nn \\
 &&\qquad\qquad\qquad\qquad +{i\sqrt{M m_\pi} \over 4\sqrt{\pi}}\,I_1 \Big({E\over m_\pi}
   \Big) \Bigg\} \,, \nn \\
c) &=& {ig_A^2 \over \sqrt{\pi} f^2}  \Big({m_\pi \over M}\Big)^{3/2}
    I_2 \Big( {E \over m_\pi} \Big)\,, \nn
\end{eqnarray}
where $\bar \mu^2 = \pi e^{-\gamma_E}\mu^2$, and $I_1$ and $I_2$ are
hypergeometric functions given in Ref.~\cite{ms2}. The $1/\epsilon$ poles are
cancelled by insertions of a $D_2 m_\pi^2$ counterterm.  The leading order
amplitude ${\cal A}_{-1} \sim 1/(Mp)$, so we see that Eq.~(\ref{gr1}) has terms
proportional to
\begin{eqnarray}
\left({m_\pi \over M}\right)^{3/2} \!\!\!\!\!\!, \qquad {m_\pi^2 \over M p} \qquad
{\rm and} \qquad {m_\pi^{5/2} \over M^{1/2} p^2}\,.
\end{eqnarray}
For $p \sim Q_r$ these terms scale as $Q_r^3/M^3$, as anticipated by the power
counting. At $p \sim m_\pi \sim Q$, these terms scale like $(Q/M)^{3/2}, Q/M$,
and $(Q/M)^{1/2}$ respectively. The graphs which give rise to the $Q^{1/2}(Q)$
corrections have two (one) external bubble sums. By external bubble sums we mean
bubble sums that do not appear inside radiation loops. External bubble sums go
like $1/p$, which scales like $1/Q_r$ at $p \sim Q_r$ but $1/Q$ at $p \sim
m_\pi$.  So for each external bubble sum, the graph picks up an additional
$Q_r/Q \sim Q^{-1/2}$ upon scaling from $p \sim Q_r$ to $p \sim m_\pi$.
Terms which scale like $Q^{1/2}$ at $p\sim m_\pi$ are actually larger than NNLO
in the $Q$ counting. The $Q^{1/2}$ contributions come from graphs $b), e)$ and
$f)$, and cancel when these graphs are added together.

In the $^1S_0$ channel the sum of all $Q_r^3$ graphs in Fig.~\ref{Qr3} is
\cite{ms2}:
\begin{eqnarray} \label{fans}
 i\,{\cal A}_3^{rad}&=& 6i\, [{\cal A}_{-1}]^2 {g_A^2 m_\pi^2 \over
 (4 \pi f)^2}\, \left( {1\over C_0^{(^1S_0)}} -{1\over C_0^{(^3S_1)}}\right)\,
 \bigg[\frac13 +{\rm ln}\Big({ {\mu}^2\over m_\pi^2 }\Big) \bigg]   \nn \\
 && +i\, [{\cal A}_{-1}]^2 \left( {1\over C_0^{(^1S_0)}} -{1\over C_0^{(^3S_1)}}
 \right)^2  {g_A^2 \over \sqrt{\pi} f^2} \Big({m_\pi \over M} \Big)^{3/2}
 I_2 \Big( {E \over m_\pi} \Big)\,,
\end{eqnarray}
where the $\ln(\mu)$ dependence in Eq.~(\ref{fans}) is cancelled by a
$\ln(\mu)$ in $D_2^{(^1S_0)}(\mu)$. The sum of the $Q_r^3$ diagrams turns out to
be much smaller than anticipated by the power counting. For $p \sim Q_r$, the first
term is suppressed by a factor of $\sim 1/Q_r[1/a^{(^1S_0)}-1/a^{(^3S_1)}]$,
the second by $\sim 1/Q_r^2[1/a^{(^1S_0)}-1/a^{(^3S_1)}]^2$. This suppression
occurs because the radiation pions couple to a charge of Wigner's $SU(4)$
symmetry\cite{Wigner}, which is a symmetry of the leading order Lagrangian in
the limit $a^{(^1S_0)}, a^{(^3S_1)} \rightarrow \infty$ (or $a^{(^1S_0)} =
a^{(^3S_1)}$)\cite{msw}. The order $Q_r^3$ radiation pion graphs are therefore
a small correction to the S-wave scattering amplitude.  Furthermore, to order
$Q$ the $Q_r^3$ graphs simply give an additional contribution to the $\zeta_3$
constant that appears in Eqs.~(\ref{zeta1S0}),
\begin{eqnarray} \label{z33}
 \zeta_3^{(3)} = - 6 \, {g_A^2 \over (4 \pi f)^2}\,
 \left( {1\over C_0^{(^1S_0)}} -{1\over C_0^{(^3S_1)}}\right) \bigg[{1\over 3}
 + {\rm ln}\Big( {\mu^2 \over m_\pi^2 }\Big)\bigg]  \,.
\end{eqnarray}
The $\mu$ dependence in $\zeta_3^{(3)}$ is cancelled by $\mu$ dependence in
$D_2^{(^1S_0)}$. The result in the $^3S_1$ channel is obtained from
Eqs.~(\ref{fans}) and (\ref{z33}) by switching the $^1S_0$ and $^3S_1$ labels.

\subsection{S\lowercase{caling radiation contributions from} $Q_{\lowercase{r}}$
\lowercase{to $m_\pi$} }

Since we are interested in the power counting for $p\sim m_\pi$ it is important
to know how big a radiation pion graph may get when $p$ is lowered from $Q_r$ to
$m_\pi$. The $Q_r^3$ graphs have pieces that scale as $Q^{1/2}, Q$, $Q^{3/2},
\ldots$, for $p \sim m_\pi$ as discussed in the previous section.  In order to
know which radiation pion graphs to include at a given order in the KSW power
counting, we must know the size of the leading term in the $Q$ expansion of a
$Q_r^k$ graph for $p \sim m_\pi$. In this section we will prove that an order
$Q_r^k$ calculation is sufficient to determine the order $Q^{k/2-1}$ result.

To see this first consider the $Q$ expansion of $p \, {\rm cot}\, \delta$ in the
$^1S_0$ channel:
\begin{eqnarray}
p \, {\rm cot}\, \delta &=& i p + {4 \pi \over M} {1 \over {\cal A}} \nn \\
  &=& i p + {4 \pi \over M} {1 \over {\cal A}_{-1}}
  - {4 \pi \over M} {{\cal A}_0 \over  [{\cal A}_{-1}]^2}
  - {4 \pi \over M} \left( {{\cal A}_1 \over  [{\cal A}_{-1}]^2} -
  {{\cal A}_0^2 \over  [{\cal A}_{-1}]^3} \right) \nn \\*
  &&- {4 \pi \over M} \left( {{\cal A}_2 \over [{\cal A}_{-1}]^2} - {2 {\cal A}_0
  {\cal A}_1 \over [{\cal A}_{-1}]^3}+{{\cal A}_0^3 \over [{\cal A}_{-1}]^4} \right)+
  \dots \,.
\end{eqnarray}
$p\, {\rm cot}\, \delta$ is real and an analytic function of $p^2$ near $p=0$.
This will be true order by order in $Q$ so:
\begin{eqnarray}
{{\cal A}_0 \over [{\cal A}_{-1}]^2} = f_0 \, &\Rightarrow& \,
    {\cal A}_0 = f_0  [{\cal A}_{-1}]^2  \,, \\
{{\cal A}_1 \over  [{\cal A}_{-1}]^2} - {[{\cal A}_0]^2 \over  [{\cal A}_{-1}]^3}
    = f_1 \, &\Rightarrow& \, {\cal A}_1 = f_1 [{\cal A}_{-1}]^2+f_0^2 [{\cal A}_{-1}]^3
    \,, \nn \\
{{\cal A}_2 \over  [{\cal A}_{-1}]^2} - {2 {\cal A}_0 {\cal A}_1 \over
    [{\cal A}_{-1}]^3}+{[{\cal A}_0]^3 \over [{\cal A}_{-1}]^4} = f_2 \, &\Rightarrow& \,
    {\cal A}_2 = f_2  [{\cal A}_{-1}]^2+2 f_0 f_1 [{\cal A}_{-1}]^3
    + f_0^3  [{\cal A}_{-1}]^4 \nn \,,
\end{eqnarray}
where the $f_k$ are real functions of $p$ which are analytic about $p^2 = 0$.
The general form of a higher order amplitude is powers of ${\cal A}_{-1}$
multiplied by functions of $p$.  The crucial point is that the function
multiplying the $[{\cal A}_{-1}]^2$ is the only new contribution. The
coefficient of $[{\cal A}_{-1}]^m$, $m > 2$, is determined by lower order
amplitudes. The graphs giving the $m>2$ contributions are ``$C_0$ reducible''
by which we mean that they fall apart when cut at an ${\cal A}_{-1}$
vertex.

This generalizes to the $Q_r$ expansion of radiation pion graphs, the only
difference being that the radiation pion contribution starts out at $Q_r^3$,
while the potential pion starts out at $Q^0$. A $Q_r^k$ radiation pion
correction to the amplitude will be of the form:
\begin{eqnarray} \label{Qrndec}
  {\cal A}_{k} = [{\cal A}_{-1}]^2 f_{k,2} + [{\cal A}_{-1}]^3 f_{k,3} +
  \ldots + [{\cal A}_{-1}]^{k-1} f_{k,k-1} .
\end{eqnarray}
Again, the $f_{k,m}$ are real and analytic about $p^2 = 0$ and all the $f_{k,m}$
except for $f_{k,2}$ will be determined from lower order amplitudes.  Since
${\cal A}_k \sim Q_r^k$ and ${\cal A}_{-1} \sim 1/(M p)$, $f_{k,2} \sim Q_r^{k+2}$
for $p \sim Q_r$. To understand how $f_{k,2}$ scales with $Q$ as $p$ is lowered
to $m_\pi$, note that without loss of generality, $f_{k,2}$ can be written as
\begin{eqnarray}
  f_{k,2} = { ({\sqrt{M m_\pi}})^{k+2} \over \Lambda_\chi^2 \
  {\overline \Lambda}^{\:k-2} }\ {\hat f}_{k,2} \left({p \over \sqrt{M m_\pi}},
  \ldots \right)  \,,
\end{eqnarray}
where the ellipses denote momentum dependence that involves scales other than
$Q_r$, and ${\overline \Lambda}=\Lambda_\chi$, $\Lambda$, or $M$. For $p\sim
m_\pi$ the ellipse denote dependence on the dimensionless variables $p/m_\pi$,
$a p$, and $p/\,{\overline \Lambda}$. For $p \sim m_\pi$, ${p/\sqrt{M m_\pi}}
\sim (Q/M)^{1/2}$ and the function ${\hat f}_{k,2}$ can be expanded in its
first argument:
\begin{eqnarray} \label{scr}
  {\cal A}_{-1}^2 f_{k,2} = {\cal A}_{-1}^2 { ({\sqrt{M m_\pi}})^{k+2} \over
  \Lambda_\chi^2 \ {\overline \Lambda}^{\:k-2} } {\hat f}_{k,2}
  \left( 0, \ldots \right)\left[1 + O \left( {Q \over M} \right)^{1/2} \right] \,.
\end{eqnarray}
Therefore, the new contribution at $Q_r^k$ scales like $Q^{k/2-1}$ (plus
subleading terms) for $p \sim m_\pi$.  This is consistent with the result of
the $Q_r^3$ calculation, where the largest contributions from individual graphs
scaled as $Q^{1/2}$.  A cancellation between graphs resulted in this
contribution vanishing.  The remaining terms scale as $Q$, $Q^{3/2}, \ldots$.

\begin{figure}[!t]
  \centerline{\epsfysize=2.3truecm \epsfbox{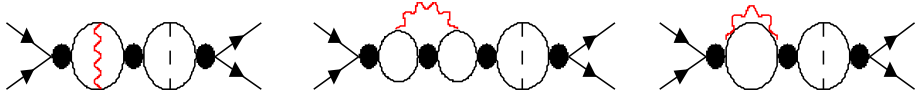}  }
 {\caption[1]{Example of order $Q_r^4$ graphs that have three external bubble
sums.} \label{Qr4a0} }
\end{figure}

Next we consider contributions to the amplitude from $C_0$ reducible graphs.
If a $C_0$ reducible graph is obtained by joining $j$ $C_0$ irreducible graphs
where the $j$'th graph scales as $Q^{\alpha_j}$ at $p \sim m_\pi$, then the $C_0$ reducible graph scales as
\begin{eqnarray}
Q^{\,j-1+\sum_{i=1}^j \alpha_i} \,.
\end{eqnarray}
For example, the order $Q_r^4$ graphs in Fig.~\ref{Qr4a0} are each obtained
by joining a $Q^0$ potential pion graph with a $Q_r^3$ radiation pion
graph. The radiation graphs scale as $Q^{1/2}$ for $p \sim m_\pi$ so the
individual graphs in Fig.[14] scale as $Q^{3/2}$ for $p \sim m_\pi$. No $C_0$
irreducible graphs give an order $Q$ contribution.

Since $Q^{k/2-1} = Q$ for $k=4$, the $Q_r^4$ radiation pion graphs can
have a contribution that is NNLO for $p \sim m_\pi$.  This calculation is
taken up in the next section. Note that a calculation of the order $Q_r^5$
graphs would be necessary to determine the order $Q^{3/2}$ terms.

\subsection{T\lowercase{he order \uppercase{$Q$} part of the order
$\uppercase{Q}_r^4$ radiation pion graphs}}

The order $Q_r^4$ radiation pion contributions come from graphs that
have one radiation pion, an arbitrary number of $C_0$'s, and one
insertion of a $C_2 p^2$, $D_2 m_\pi^2$, or $G_2$ operator or one
potential pion.  The coefficient $G_2$ multiplies a four-nucleon
operator that couples to the axial pion current,
\begin{eqnarray}
  {\cal L} = \frac{i}{2}\, G_2\, [ N^T P^{(s)}_i N ]^\dagger\, [ N^T P^{(s)}_i
 \sigma_j (\xi\partial_j\xi^\dagger - \xi^\dagger\partial_j\xi) N] +h.c. \,.
\end{eqnarray}
Note that due to the hermitian conjugate this operator is the same for
$s=\!^1S_0$ and $s=\!^3S_1$. Power counting these graphs gives $Q_r^4/(M^3
\Lambda_\chi^2 \Lambda)$, i.e. they are suppressed by $Q_r/\Lambda$ relative to
the leading radiation pion graphs in Fig.\ \ref{Qr3}. Note that $Q_r = 360
\,{\rm MeV}$, so for $\Lambda < 360\,{\rm MeV}$, the $Q_r/\Lambda$ expansion
does not converge.   If this is the case then the radiation pion contribution
is not calculable. This is true of radiation contributions even when we scale
down to $p\sim m_\pi$. To make the radiation contributions calculable we must
have a power counting for the pure potential contributions that works for
$p\sim Q_r$. One possible resolution is to ignore radiation pion contributions
since at low momenta the radiation pions can be integrated out. However, this
makes the coefficients of four nucleon operators depend on $m_\pi$ in a
non-trivial way. We will proceed by computing the radiation contribution which
is formally order $Q$ in both S-wave channels even though the size of the spin
triplet potential diagrams in section \ref{s3s1} indicate that a modification of
the power counting is likely necessary to obtain a convergent expansion in this
channel.  Since the $C_2 p^2$ and $D_2 m_\pi^2$ operators and potential pion
exchange do not respect Wigner symmetry, there will be no suppression by factors
of $1/(a\,Q_r)$ at this order.

For calculational purposes it is useful to define \emph{offshell}
amplitudes for S-wave transitions, $N(p_1) N(p_2) \to N(p_3) N(p_4)$,
induced by 4-nucleon operators.  These amplitudes are equal to a sum of
Feynman diagrams where the equations of motion have not been used. They
can be treated as vertices and inserted inside loop graphs, which
greatly reduces the number of order $Q_r^4$ diagrams. The offshell
order $1/Q$ amplitude in $n$ dimensions is
\begin{eqnarray}  \label{offC0}
 \nn \\[-15pt]
i\,{\cal A}^{(-1)} &=& \quad \begin{picture}(15,10)(1,1)
      \put(1,3){\line(1,1){10}} \put(1,3){\line(1,-1){10}}
      \put(1,3){\line(-1,1){10}} \put(1,3){\line(-1,-1){10}}
      \put(-4,15){\mbox{\footnotesize $C_0$}}
  \end{picture} + \quad\
  \begin{picture}(15,10)(1,1)
      \put(1,3){\line(-1,1){10}} \put(1,3){\line(-1,-1){10}}
      \put(11,3){\circle{20}}
      \put(21,3){\line(1,1){10}} \put(21,3){\line(1,-1){10}}
      \put(-6,18){\mbox{\footnotesize $C_0$}}
      \put(14,18){\mbox{\footnotesize $C_0$}}
  \end{picture} \qquad +\, \ldots  \ = \qquad \begin{picture}(15,10)(1,1)
      \put(4,3){\line(1,1){10}} \put(4,3){\line(1,-1){10}} \put(1,3){\circle*{8}}
      \put(-2,3){\line(-1,1){10}} \put(-2,3){\line(-1,-1){10}}
  \end{picture} \ \
  =\ -\frac{4\pi i}{M} \:\frac1{\gamma-\tau (-M\bar E-i\epsilon)^{n/2-1}}\,,
\end{eqnarray}
where
\begin{eqnarray}
  \gamma = \frac{4\pi}{MC_0(\mu)} + \mu \,, \qquad
  \tau = - \, { \Gamma(1-n/2) \over (4\pi)^{n/2-1} } \Big(\frac{\mu}{2}\Big)^{3-n}\,,
\end{eqnarray}
and $\bar E$ is the center of mass energy
\begin{eqnarray}
   \bar E = E_1 + E_2 - { ({\vec p}_1+{\vec p}_2)^2 \over 4 M}
    = E_3 + E_4 - { ({\vec p}_3+{\vec p}_4)^2 \over 4 M} \,.
\end{eqnarray}
At order $Q^0$ the NN amplitude has contributions from the four nucleon
operators $C_2$, $D_2$, and $C_0^{(0)}$.  In $n$ dimensions the offshell
amplitude for $C_2$ graphs is
\begin{eqnarray} \label{offC2}
\nn \\[-15pt] i\,{\cal A}_{C_2} &=& \quad
\begin{picture}(15,10)(1,1)
      \put(1,3){\line(1,1){10}} \put(1,3){\line(1,-1){10}}
      \put(1,3){\line(-1,1){10}} \put(1,3){\line(-1,-1){10}}
      \put(-4,15){\mbox{\footnotesize $C_2$}}
  \end{picture} + \quad\
  \begin{picture}(15,10)(1,1)
      \put(-5,3){\line(-1,1){10}} \put(-5,3){\line(-1,-1){10}}
      \put(11,3){\circle{20}}
      \put(21,3){\line(1,1){10}} \put(21,3){\line(1,-1){10}}
      \put(-2,3){\circle*{8}}
      \put(14,18){\mbox{\footnotesize $C_2$}}
  \end{picture} \qquad + \quad\
  \begin{picture}(15,10)(1,1)
      \put(1,3){\line(-1,1){10}} \put(1,3){\line(-1,-1){10}}
      \put(11,3){\circle{20}}
      \put(27,3){\line(1,1){10}} \put(27,3){\line(1,-1){10}}
      \put(-6,18){\mbox{\footnotesize $C_2$}}
      \put(24,3){\circle*{8}}
  \end{picture}  \qquad + \quad\
  \begin{picture}(15,10)(1,1)
      \put(-5,3){\line(-1,1){10}} \put(-5,3){\line(-1,-1){10}}
      \put(11,3){\circle{20}}
      \put(31,3){\circle{20}}
      \put(47,3){\line(1,1){10}} \put(47,3){\line(1,-1){10}}
      \put(-2,3){\circle*{8}}  \put(44,3){\circle*{8}}
      \put(14,18){\mbox{\footnotesize $C_2$}}
  \end{picture}  \\[15pt]
  &=& -i\, \frac{C_2}{(C_0)^2} \: M \bar E\ \Big[{\cal A}^{(-1)}\,\Big]^2
      +i\, \frac{C_2}{C_0}\: \bigg\{ \frac{ ({\vec p}_1-{\vec p}_2)^2 +
      ({\vec p}_3-{\vec p}_4)^2}{8} - M \bar E \bigg\}\  {\cal A}^{(-1)}  \,. \nn
\end{eqnarray}
Note that $C_2(\mu)/C_0(\mu)$ and therefore the offshell $C_2$ amplitude are
$\mu$ dependent.  The onshell amplitude is $\mu$ independent using the order
$1/Q^2$ part of the beta function in Eq.~(\ref{C0RGE}) since the term
proportional to $C_2/C_0$ vanishes by the equations of motion.  The offshell
amplitude that includes the graphs with $D_2$ or $C_0^{(0)}$ vertices is
\begin{eqnarray} \label{offD2}
\nn \\*[-10pt]
 i\,{\cal A}_{D_2} &=& \quad
  \begin{picture}(15,10)(1,1)
      \put(1,3){\line(1,1){10}} \put(1,3){\line(1,-1){10}}
      \put(1,3){\line(-1,1){10}} \put(1,3){\line(-1,-1){10}}
      \put(-12,20){\mbox{\footnotesize $D_2$,$C_0^{(0)}$}}
  \end{picture} + \quad\
  \begin{picture}(15,10)(1,1)
      \put(-5,3){\line(-1,1){10}} \put(-5,3){\line(-1,-1){10}}
      \put(11,3){\circle{20}}
      \put(21,3){\line(1,1){10}} \put(21,3){\line(1,-1){10}}
      \put(-2,3){\circle*{8}}
      \put(6,22){\mbox{\footnotesize $D_2$,$C_0^{(0)}$}}
  \end{picture} \qquad + \quad\
  \begin{picture}(15,10)(1,1)
      \put(1,3){\line(-1,1){10}} \put(1,3){\line(-1,-1){10}}
      \put(11,3){\circle{20}}
      \put(27,3){\line(1,1){10}} \put(27,3){\line(1,-1){10}}
      \put(-14,22){\mbox{\footnotesize $D_2$,$C_0^{(0)}$}}
      \put(24,3){\circle*{8}}
  \end{picture}  \qquad + \quad\
  \begin{picture}(15,10)(1,1)
      \put(-5,3){\line(-1,1){10}} \put(-5,3){\line(-1,-1){10}}
      \put(11,3){\circle{20}}
      \put(31,3){\circle{20}}
      \put(47,3){\line(1,1){10}} \put(47,3){\line(1,-1){10}}
      \put(-2,3){\circle*{8}}  \put(44,3){\circle*{8}}
      \put(6,22){\mbox{\footnotesize $D_2$,$C_0^{(0)}$}}
  \end{picture}  \phantom{xxxxxxxxxxxxxx} \nn\\[15pt]
 &=& -i\, \bigg[ \frac{D_2}{(C_0)^2}\: m_\pi^2 +
      \frac{C_0^{(0)}}{(C_0)^2} \bigg]  \ \Big[{\cal A}^{(-1)}\,\Big]^2 \,,
\end{eqnarray}

\begin{figure}[!t]
  \centerline{\epsfysize=7.5truecm \epsfbox{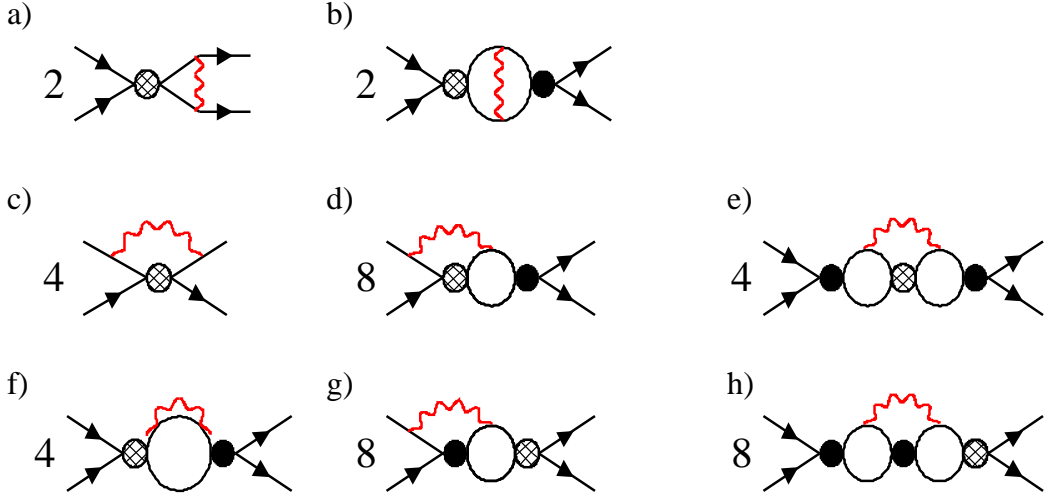}  }
 {\tighten
\caption[1]{Order $Q_r^4$ radiation pion graphs with insertions of $C_2$ and
$D_2$. The solid lines are nucleons and the wavy lines are radiation pions.
The black circle denotes the $C_0$ bubble sum, $i{\cal A}^{(-1)}$, and the
hatched circle denotes an insertion of $i{\cal A}_{C_2}$ or $i{\cal
A}_{D_2}$ given in Eqs.~(\ref{offC2}) and (\ref{offD2}).} \label{Qr4a} }
\end{figure}
The order $Q_r^4$ radiation pion graphs with insertions of $i{\cal A}_{C_2}$ or
$i{\cal A}_{D_2}$ are shown in Fig.~\ref{Qr4a}.  We find that graphs with
insertions of $i{\cal A}_{D_2}$ give contributions that are order $Q^2$ or
higher. The graphs which have an $i{\cal A}_{D_2}$ external to the radiation
loop (Fig.\ref{Qr4a}b,f,h) have no order $Q$ contribution because the same
cancellation that occurs in Fig.~\ref{Qr3}b,e,f occurs here. Of the remaining
graphs only Fig.~\ref{Qr4a}e with two external bubble sums can have an order
$Q$ contribution, however inside the radiation pion loop $i{\cal A}_{D_2}\sim
m_\pi^2/Q_r^2 \sim Q$ (not $Q^0$) so this graph is order $Q^2$.

With insertions of $i{\cal A}_{C_2}$, the only graphs in Fig.~\ref{Qr4a} which
do not give an order $Q$ contribution are a) and c). Diagrams d) and g) give a
non-zero order $Q$ contribution even though they have only one external bubble
sum.  The order $Q$ contribution comes from the $\mu$ dependent part of
$i{\cal A}_{C_2}$.  Since these $Q_r^4$ graphs have one
external bubble sum they are expected to be $\sim Q^{3/2}$.  However, with
$(\vec p_1-\vec p_2)^2 \sim Q_r^2$ or $M\bar E\sim Q_r^2$, and $\mu\sim Q$ the
$\mu$ dependent part of $i{\cal A}_{C_2}$ is order $Q_r/\mu\sim 1/\sqrt{Q}$.
This extra factor\footnote{\tighten Note that enhancements by factors of $\mu$
do not effect the proof in section C3 since the amplitude at a given order in
$Q_r$ is $\mu$ independent.} makes these graphs order $Q$. For the $^1S_0$ channel,
the sum of the order $Q$ contributions from the $C_2$ radiation graphs in
Fig.~\ref{Qr4a} is
\begin{eqnarray} \label{AradC2}
  i{\cal A}^{rad}_{C_2} = \frac{i g_A^2}{f^2} \Big[{\cal A}_{-1}^{(^1S_0)}
  \Big]^2 \,{ M m_\pi^2 (m_\pi - \mu) \over 4\pi } \Bigg\{
  \frac{C_2^{(^1S_0)}}{[C_0^{(^1S_0)}]^2} + \frac{C_2^{(^3S_1)}}
  {[C_0^{(^3S_1)}]^2} -{ C_2^{(^1S_0)} +C_2^{(^3S_1)} \over
  C_0^{(^1S_0)} C_0^{(^3S_1)} }
\Bigg\} \,.
\end{eqnarray}
The corresponding amplitude in the $^3S_1$ channel is obtained by exchanging
the $^1S_0$ and $^3S_1$ labels in Eq.~(\ref{AradC2}).

\begin{figure}[!t]
  \centerline{\epsfysize=5.5truecm \epsfbox{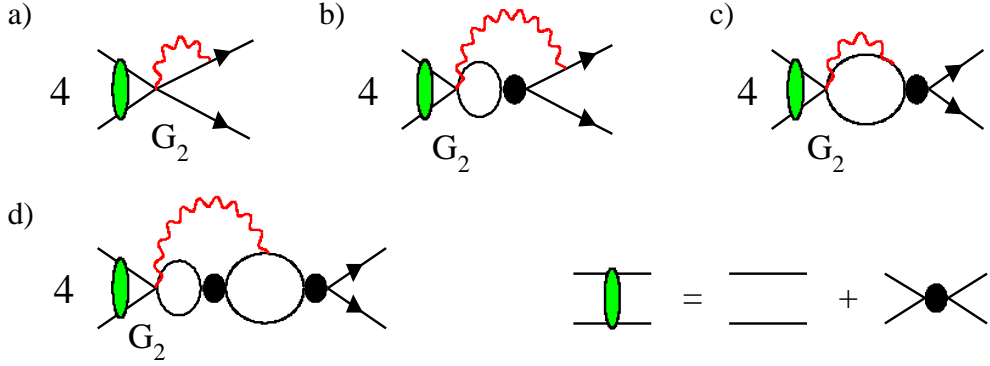}  }
 {\tighten
\caption[1]{Order $Q_r^4$ radiation pion graphs with insertions of $G_2$. }
\label{Qr4b} }
\end{figure}
The result in Eq.~(\ref{AradC2}) is $\mu$ dependent. For the term proportional
to $m_\pi^2$ the $\mu$ dependence is cancelled by a radiation contribution to
the beta function for $D_2(\mu)$.  The $\mu$ dependence of the $m_\pi^3$ term
is cancelled by $G_2(\mu)$.  To calculate the PDS beta function for $G_2(\mu)$
we consider graphs with $G_2$ dressed with $C_0^{(^1S_0)}$ bubbles on one side
and $C_0^{(^3S_1)}$ bubbles on the other.  We also consider graphs with a $C_2$
vertex next to a $NN\pi$ vertex:
\begin{eqnarray}
  \begin{picture}(40,30)(1,1)
   \put(1,3){\line(1,1){20}} \put(1,3){\line(1,-1){20}}
   \put(1,3){\line(-1,1){20}} \put(1,3){\line(-1,-1){20}}
      \put(-9,25){\mbox{ $C_2$}}
     \multiput(13,-9)(15,10){3}{\line(3,2){10}}
  \end{picture} \nn \\[-15pt]
\end{eqnarray}
When the derivatives in ${\cal O}_2$ act on the nucleons on the right there is
a piece in which the numerator cancels the propagator exactly. This piece has
the same form as a $NNNN\pi$ vertex and contributes to the beta function for
$G_2$ when dressed with $C_0$ bubbles. We find
\begin{eqnarray}
  \beta_{G_2} &=& {\mu M \over 4\pi}\, G_2 (C_0^{(^1S_0)}+C_0^{(^3S_1)})
  \nn \\
  &&  + {g_A \mu M^2 \over 4\pi}\Big[ C_0^{(^1S_0)} C_2^{(^1S_0)} +
    C_0^{(^3S_1)} C_2^{(^3S_1)} - C_0^{(^3S_1)} C_2^{(^1S_0)} -
    C_0^{(^1S_0)} C_2^{(^3S_1)} \Big] \,,
\end{eqnarray}
which has the solution
\begin{eqnarray} \label{G2soln}
  G_2(\mu) = \kappa_G \frac{M C_0^{(^1S_0)}}{4\pi}
  \frac{M C_0^{(^3S_1)}}{4\pi} + g_A M\, [C_2^{(^1S_0)}+C_2^{(^3S_1)}]
  \,,
\end{eqnarray}
with $\kappa_G$ the constant of integration.  Eq.~(\ref{G2soln}) gives
$G_2(\mu)\sim 1/\mu^2$. Because $G_2$ is the coefficient of the four nucleon
coupling to the axial current it has the same renormalization group equation as
the weak axial four nucleon operator\footnote{In Ref.~\cite{ChenWk} $G_2$ was
denoted by $L_{1,A}$.} considered in Ref.~\cite{ChenWk}. Using the scaling of
$G_2$ to power count the diagrams in Fig.~\ref{Qr4b} we find that they are
order $Q_r^4$ radiation pion graphs.
For the $^1S_0$ channel, the sum of the order $Q$ part of the diagrams
in Fig.~\ref{Qr4b} is
\begin{eqnarray} \label{AradG2}
  i\,{\cal A}_{G_2}^{rad} = \frac{i g_A}{f^2} [ A_{-1}^{(^1S_0)} ]^2
  \frac{G_2}{C_0^{(^1S_0)} C_0^{(^3S_1)}} {m_\pi^2 (m_\pi-\mu)
  \over 4\pi } \,.
\end{eqnarray}
The result in the $^3S_1$ channel is obtained by interchanging the labels
$^1S_0$ and $^3S_1$.  Using Eq.~(\ref{G2soln}) we see that the $\mu$ dependence
of the $m_\pi^3$ term in Eq.~(\ref{AradC2}) is cancelled by the $m_\pi^3$ term
in Eq.~(\ref{AradG2}).  The $\mu$ dependence of the $m_\pi^2$ term in
Eq.~(\ref{AradG2}) is again cancelled by $D_2(\mu)$.

\begin{figure}[!t]
  \centerline{\epsfysize=5.5truecm \epsfbox{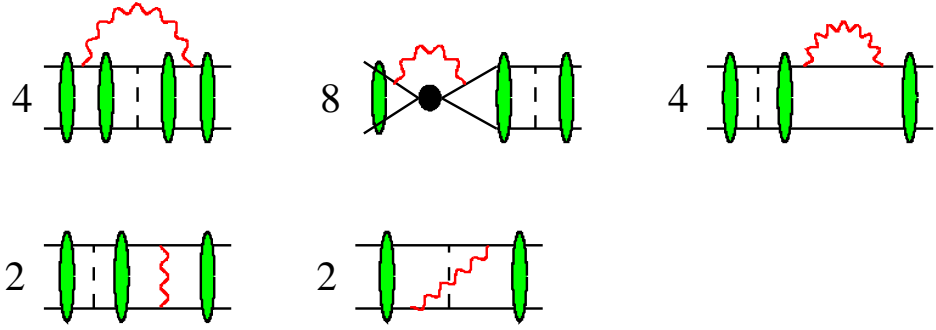}  }
 {\tighten
\caption[1]{Order $Q_r^4$ radiation pion graphs with one potential pion. }
\label{Qr4c} }
\end{figure}

\begin{figure}[!t]
  \centerline{\epsfysize=2.5truecm \epsfbox{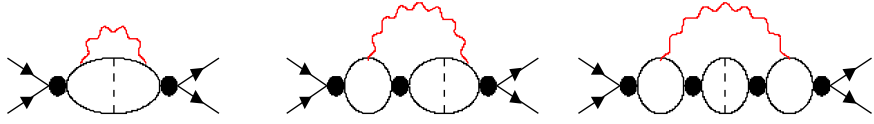}  }
 {\tighten
  \caption[1]{The three order $Q_r^4$ radiation pion graphs with one
  potential pion that give an order $Q$ contribution for $p\sim m_\pi$. }
  \label{Qr4d} }
\end{figure}

\begin{figure}[!t]
  \centerline{\epsfysize=2.5truecm \epsfbox{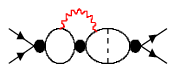}  }
 {\tighten
  \caption[1]{An order $Q_r^4$ radiation pion graphs with one
  potential pion that does not give an order $Q$ contribution. } \label{Qr4e} }
\end{figure}

The final order $Q_r^4$ diagrams that we must consider are those with
one potential pion, one radiation pion and an arbitrary number of
$C_0$'s shown in Fig.~\ref{Qr4c}.
Here we find that only graphs with two external bubble sums can give an
order $Q$ contribution. Graphs in Fig.~\ref{Qr4c} with three external
bubbles sums are $C_0$ reducible and do not give order $Q$ contributions.
Furthermore, of all the diagrams in Fig.~\ref{Qr4c} with two external bubble
sums only the three shown in Fig.~\ref{Qr4d} give an order $Q$ contribution.
These are the graphs in which the potential pion exchange is inside the
radiation pion loop.  This ensures that all potential loop momenta in the graph
see the scale $Q_r$ which is necessary for the graph to give an order $Q$
contribution.  So, for example, we find that the graph in Fig.~\ref{Qr4e} does
not give an order $Q$ contribution. The diagrams in Fig.~\ref{Qr4d} look
somewhat daunting since they involve a three, four, and five loop calculation.
Nevertheless, their order $Q$ contribution can be evaluated analytically.
Adding up the the order $Q$ part of the diagrams in Fig.~\ref{Qr4d} gives
\begin{eqnarray} \label{Aradppi}
  i\,{\cal A}^{rad}_{\pi} =i\, \frac53\, \Big(\frac{g_A^2}{2f^2}\Big)^2
  \, \frac{M m_\pi^3}{4\pi} \Big( \frac{M A_{-1}}{4\pi} \Big)^2 \,.
\end{eqnarray}
It is interesting that the order $Q$ part of these graphs is not multiplied by
a nontrivial function of $p/m_\pi$.  At one higher order, $Q^{3/2}$, the
diagrams in Fig.~\ref{Qr4c} will give a result which involves a function of
$p/m_\pi$.

The results in Eqs.~(\ref{AradC2},\ref{AradG2},\ref{Aradppi}) give the complete
order $Q$ contribution from order $Q_r^4$ graphs.  It is interesting to note
that all contributions are equal to a constant times $[{\cal A}_{-1}]^2$.
Therefore they simply give an additional contribution to the constant $\zeta_3$
that appears in Eqs.~(\ref{zeta1S0}) and (\ref{zeta3S1}),
\begin{eqnarray} \label{z34}
  \zeta_3^{(4)} &=&  -\frac53 \, \Big(\frac{M}{4\pi}\Big)^2\:
  \Big(\frac{g_A^2}{2 f^2}\Big)^2\:  \frac{M m_\pi}{4\pi} \\
  &&- \frac{(m_\pi-\mu)g_A}{4\pi f^2} \Bigg\{
  \frac{G_2 - M g_A (C_2^{(^1S_0)}+C_2^{(^3S_1)})}{C_0^{(^1S_0)}
  C_0^{(^3S_1)}} + M g_A \Bigg(\: \frac{C_2^{(^1S_0)}}{[C_0^{(^1S_0)}]^2}
  +\frac{C_2^{(^3S_1)}}{[C_0^{(^3S_1)}]^2}\: \Bigg) \Bigg\} \,. \nn
\end{eqnarray}
The result for $\zeta_3^{(4)}$ is the same in the $^3S_1$ channel.  The $\mu$
dependence in $\zeta_3^{(4)}$ is cancelled by $\mu$ dependence in $D_2$.

The complete order $Q$ contribution from radiation pion graphs is the sum of
Eqs.~(\ref{z33}) and (\ref{z34}):
\begin{eqnarray} \label{z3rad}
   \zeta_3^{rad} = \zeta_3^{(3)} + \zeta_3^{(4)} \,.
\end{eqnarray}
The order $Q$ radiation pion contribution to the $D_2^{(^1S_0)}$ beta function
in Eq.~(\ref{C0RGE}) is
\begin{eqnarray} \label{bd2rad}
  \beta_{D_2}^{rad} &=& 12 \frac{g_A^2}{(4\pi f)^2} \Big(
  \frac{1}{C_0^{(^1S_0)}}-\frac{1}{C_0^{(^3S_1)}} \Big)
  \Big[\,C_0^{(^1S_0)}\Big]^2 \\*
  && - \frac{\mu g_A}{4\pi f^2} \Bigg\{
  \frac{G_2 - M g_A (C_2^{(^1S_0)}+C_2^{(^3S_1)})}{C_0^{(^1S_0)}
  C_0^{(^3S_1)}} + M g_A \Bigg(\: \frac{C_2^{(^1S_0)}}{[C_0^{(^1S_0)}]^2}
  +\frac{C_2^{(^3S_1)}}{[C_0^{(^3S_1)}]^2}\: \Bigg) \Bigg\}
  \Big[\,C_0^{(^1S_0)}\Big]^2 \,. \nn
\end{eqnarray}
The contribution to the $D_2^{(^3S_1)}$ beta function in Eq.~(\ref{C0RGE2}) is
obtained by switching the spin singlet and triplet labels.

For a consistent radiation pion calculation at momenta $p\sim Q_r$ it would be
necessary to keep all powers of $p/Q_r$ in computing the $Q_r^4$ diagrams.  If
the $Q_r/\Lambda$ expansion were convergent then the magnitude of this
radiation contribution would be small; being down by $Q_r^5$ relative to the
leading order amplitude.  For $Q_r/\Lambda > 1$ it is necessary to modify the
power counting for the potential diagrams to increase the scale $\Lambda$
before the radiation pion power counting will yield a convergent series.  Since
diagrams with one radiation pion are suppressed by $m_\pi^2/\Lambda_\chi^2$
relative to any leading order amplitude these contributions are always likely
to contribute at the few percent level.

{\tighten

} 


\begin{references}

\bibitem{dgh} J.F. Donoghue, E. Golowich and B.R. Holstein, {\it Dynamics of the
Standard Model} (Cambridge University Press,, 1992).

\bibitem{mj} E. Jenkins and A. Manohar, UCSD-PTH-91-30; E. Jenkins and
A. Manohar, Phys. Lett. {\bf B255} (1991) 558, {\em ibid} {\bf B 259} (1991)
353.

\bibitem{bkm} V. Bernard, N. Kaiser, and U. Meissner, Phys.\ Rev.\ Lett.\
{\bf 67} (1991) 1515; Phys.\ Lett.\ B {\bf 319} (1993) 269; Int. J. Mod. Phys.
E4 (1995) 193.

\bibitem {weinberg} S. Weinberg, Phys.\ Lett.\ {\bf B251} (1990) 288; Nucl.\
Phys.\ {\bf B363} (1991) 3;

\bibitem{orefs}  C. Ordonez and U. van Kolck, Phys. Lett. {\bf B291} (1992) 459;
C.  Ordonez, L. Ray and U. van Kolck, Phys. Rev. Lett. {\bf 72} (1994) 1982;  U.
van Kolck, Phys. Rev. {\bf C49} (1994) 2932;  G.P. Lepage, nucl-th/9706029; T-S.
Park, K. Kubodera, D-P.  Min and M. Rho, Nucl. Phys. {\bf A646} (1999) 83; D.R.
Phillips and T.D. Cohen, nucl-th/9906091. See the review, U. van Kolck,
nucl-th/9902015 and references therein.

\bibitem{ork}
C.~Ordonez, L.~Ray, and U.~van Kolck,
\newblock Phys. Rev. {\bf C53}, (1996) 2086,
\newblock 

\bibitem{Ulf} E. Epelbaum, W. Glockle, and U.G. Meissner, nucl-th/9910064,

\bibitem{ksw1}
D.~B. Kaplan, M.~J. Savage, and M.~B. Wise,
\newblock Phys. Lett. {\bf B424} (1998) 390,
\newblock 

\bibitem{ksw2}
D.~B. Kaplan, M.~J. Savage, and M.~B. Wise,
\newblock Nucl. Phys. {\bf B534} (1998) 329,
\newblock 

\bibitem{bira} U. van Kolck, hep-ph/9711222,

\bibitem{kaiser}  N. Kaiser, R. Brockmann, and W. Weise, Nucl. Phys. {\bf A625}
(1997) 758.

\bibitem{oNLO}
E.~Epelbaum and U.-G. Meissner, nucl-th/9903046;
\newblock 
X.~Kong and F.~Ravndal, hep-ph/9903523;
\newblock 
\newblock Phys. Lett. {\bf B450}, (1999) 320;
\newblock 
X.~Kong and F.~Ravndal, nucl-th/9902064;
\newblock 
nucl-th/9904066;
\newblock 
D.~B. Kaplan, M.~J. Savage, and M.~B. Wise, Phys. Rev. {\bf C59} (1999) 617;
\newblock 
J.-W. Chen et.al., \newblock Nucl. Phys. {\bf A644} (1998) 221;
\newblock 
M.~J. Savage, K.~A. Scaldeferri, and M.~B. Wise, nucl-th/9811029;
\newblock 
J.-W. Chen, G.~Rupak, and M.~J. Savage, nucl-th/9905002;
\newblock 
J.-W. Chen, H.~W. Griesshammer, M.~J. Savage, and R.~P. Springer, Nucl. Phys.
{\bf A644}, 245 (1998);
\newblock 
J.-W. Chen, nucl-th/9810021;
\newblock 
D.~B. Kaplan, M.~J. Savage, R.~P. Springer, and M.~B. Wise, Phys. Lett. {\bf
B449} (1999) 1;
\newblock 
M.~J. Savage and R.~P. Springer, Nucl. Phys. {\bf A644} (1998) 235;
\newblock 
P.F. Bedaque and H.W. Griesshammer, nucl-th/9907077,

\bibitem{ChenWk} Malcolm Butler and Jiunn-Wei Chen, nucl-th/9905059,

\bibitem{martin} M. J. Savage, nucl-th/9905009, Proceedings of the
INT Workshop on Nuclear Physics with Effective Field Theory, Seattle, WA 25-26
Feb 1999.

\bibitem{binger} M.~Binger,  nucl-th/9901012,
\newblock 

\bibitem{rupak1} G.~Rupak and N.~Shoresh,  nucl-th/9902077,
\newblock 

\bibitem{rupak2} G.~Rupak and N.~Shoresh,  nucl-th/9906077,


\bibitem{msconf} T. Mehen and I.~W. Stewart, nucl-th/9906010,

\bibitem{fms} S. Fleming, T. Mehen amd I. Stewart, nucl-th/9906056.

\bibitem{Nij} V.G.J. Stoks, et.al., Phys. Rev. {\bf C48} (1993) 792;  V.G.J.
Stoks et.al., Phys. Rev. {\bf C49} (1994) 2950, nucl-th/9406039.  (cf.
http://nn-online.sci.kun.nl/NN/)

\bibitem{ms1} T.~Mehen and I.~W. Stewart, Phys. Rev. {\bf C59} (1999) 2365,
\newblock 

\bibitem{Wigner} E.~Wigner,  Phys. Rev. {\bf 51}, 106, 947 (1937).

\bibitem{msw} T.~Mehen, I.~W. Stewart, and M.~B. Wise, hep-ph/9902370,
\newblock 

\bibitem{gegelia} J. Gegelia, nucl-th/9802038.

\bibitem{ms0} T.~Mehen and I.~W. Stewart, Phys. Lett. {\bf B445}, (1999) 378,
\newblock 

\bibitem{birse} M.C.~Birse, J.A.~McGovern, and K.G.~Richardson, hep-ph/9807302.

\bibitem{pths} I.~W. Stewart, Ph.D. thesis, hep-ph/9907448.

\bibitem{nopi} J.-W. Chen, G.~Rupak, and M.~J. Savage, Nucl.Phys. {\bf A653}
(1999) 386,
\newblock 

\bibitem{ms2} T.~Mehen and I.~W. Stewart, nucl-th/9901064,
\newblock 

\bibitem{sk} J. Steele and D.B. Kaplan, nucl-th/9905027.

\bibitem{ch} T.D. Cohen and J.M. Hansen, Phys. Rev. {\bf C59} (1999) 13;
nucl-th/9901065.

\bibitem{kaplanpc} D.B. Kaplan, {\em private communication},

\bibitem{Stapp} H.P. Stapp, T.J. Ypsilantis, and N. Metropolis, Phys. Rev. {\bf 105}
(1957) 302,

\bibitem{kaiser2} N. Kaiser, S. Gerstendorfer, and W. Weise, Nucl. Phys. {\bf A637}
(1998) 395,

\bibitem{nopi2} J.W. Chen, G.~Rupak, and M.~J. Savage, nucl-th/9905002,

\bibitem{resd} D.R.~Phillips, G.~Rupak, and  M.J.~Savage, nucl-th/9908054,

\bibitem{lm} M. Luke and A. Manohar, Phys.\ Rev.\ {\bf D55} (1997) 4129,

\bibitem{3bdy}
P.F.~Bedaque, H.W.~Hammer and U.~van Kolck,
Nucl.\ Phys.\ {\bf A646}, (1999) 444;
P.F.~Bedaque, H.W.~Hammer and U.~van Kolck,
nucl-th/9906032.
P.F.~Bedaque and U.~van Kolck,
Phys.\ Lett.\ {\bf B428}, (1998) 221;
P.F.~Bedaque, H.W.~Hammer and U.~van Kolck,
Phys.\ Rev.\ {\bf C58}, (1998) R641.

\bibitem{gp} P.F.~Bedaque and H.W.~Griesshammer,
nucl-th/9907077.

\bibitem{intbyparts} F.V. Tkachov, Phys. Lett. {\bf B100} (1981) 65;
K.G. Chetyrkin and F.V. Tkachov, Nucl. Phys. {\bf B192} (1981) 159,

\bibitem{tensdec} G. Passarino and M. Veltman, Nucl. Phys. {\bf B160} (1979) 151;
G. Weiglein, R. Scharf, and M. Bohm, Nucl. Phys. {\bf B416} (1994) 606,

\bibitem{Tarasov} O.V. Tarasov, Phys. Rev. {\bf D54} (1996) 6479; Nucl. Phys.
{\bf B502} (1997) 455.

\bibitem{mertig} R. Mertig and R. Scharf, Comput. Phys. Commun. {\bf 111} (1998) 265.

\bibitem{looprev} R. Harlander and M. Steinhauser, hep-ph/9812357,

\bibitem{braaten} E.~Braaten and A.~Nieto,
Phys.\ Rev.\ {\bf D51}, (1995) 6990.

\bibitem{Rajan} A.K. Rajantie, Nucl. Phys. {\bf B480} (1996) 729.

\bibitem{beneke} M. Beneke and V.A. Smirnov, Nucl. Phys. {\bf B522} (1998) 321,

\bibitem{NRQCD0} W.E. Caswell and G.P. Lepage, Phys.\ Lett.\ {\bf B167} (1986)
437; G.T. Bodwin, E. Braaten, and G.P. Lepage, Phys.\ Rev.\ {\bf D55} (1997) 1125.

\bibitem{multi} B. Grinstein and I. Rothstein, Phys. Rev. {\bf D 57} (1998) 78;
P. Labelle, Phys.\ Rev.\ {\bf D58} (1998) 093013; M. Luke, A.V. Manohar, and
I.Z. Rothstein, hep-ph/9910209,

\bibitem{nrqcd}
M. Luke and M.J. Savage, Phys.\ Rev.\ {\bf D57} (1998) 413;

\bibitem{gries} H.W. Griesshammer, Phys. Rev. {\bf D58} (1998) 094027 and
hep-ph/9810235,

\bibitem{gh} H.W. Griesshammer, hep-ph/9804251,


\end{references}
\end{document}